\documentclass[twocolumn,aps,superscriptaddress,showpacs,floatfix]{revtex4}

\usepackage{graphicx}
\usepackage{dcolumn}
\usepackage{bm}
\usepackage{amsmath,amssymb}
\usepackage{booktabs}
\usepackage{array}
\usepackage{multirow}
\usepackage{makecell,rotating,diagbox}
\usepackage{tabularx}
\usepackage{longtable}     
\usepackage{gensymb}
\usepackage[section]{placeins}

\usepackage[caption=false]{subfig}     
\usepackage[percent]{overpic}          
\usepackage[colorlinks=true,linkcolor=blue,citecolor=blue,urlcolor=blue]{hyperref}

\usepackage[normalem]{ulem}
\usepackage[draft]{changes}

\usepackage{lineno}

\newcounter{panel}

\usepackage[colorlinks=true,linkcolor=blue,citecolor=blue,urlcolor=blue]{hyperref}

\begin{document}
\title{Laser-assisted $\alpha$ decay of actinide nuclei in bichromatic fields}

\author{You-Tian Zou}
\affiliation{College of Science, National University of Defense Technology, 410073 Changsha, People’s Republic of China}
\author{Tong-Pu Yu}
\email{tongpu@nudt.edu.cn }

\begin{abstract}
Actinide nuclei provide a suitable platform for studying the laser-assisted nuclear $\alpha$ decay, with potential applications in nuclear transmutation, nuclear radiotherapy, and nuclear battery regulation. In the present work, we develop a deformed one-parameter model to quantitatively study the influence of ultra-intense laser fields on the $\alpha$ decay of actinide nuclei. Our calculations show that the $\alpha$-decay half-lives of these nuclei can be altered to some finite extent under laser intensities anticipated at near-future laser facilities. Furthermore, we found that, from the perspective of the nucleus, the laser field’s effect on $\alpha$ decay is governed by the nuclear shell structure and decay energy. The $\alpha$-emitting nuclei with lower decay energies and located farther from neutron shell closures are more susceptible to the laser fields. From the perspective of the laser driver, we proposed a bichromatic laser scheme to enhance the effects of laser fields on $\alpha$ tunneling of actinide nuclei. With appropriate phase conditions and amplitude ratios, it is shown that a fundamental–second-harmonic ($\omega$–$2\omega$) bichromatic field can increase the time-averaged modification by one to two orders of magnitude. 
\end{abstract}

\maketitle

\section{Introduction}
\setlength{\parskip}{1.0pt}
The $\alpha$-decay rate of nuclei is a crucial parameter characterizing the stability and structural properties of nuclear matter \cite{RevModPhys.72.733,PhysRevC.105.024327,PhysRevC.93.034316, PhysRevC.101.034307,PhysRevC.111.034330,Deng_2025,Deng_2024}, and it is commonly regarded as a different intrinsic constant of each radioactive nuclide. However, the actual decay rate may be influenced by changes under external physical or chemical environments\cite{PhysRevC.90.054619,PhysRevC.92.024301,wan2016alpha,PhysRevC.105.024307}. This characteristic holds promising potential for applications in nuclear waste transmutation \cite{CHWASZCZEWSKI200387,GUDOWSKI2000169c,KURNIAWAN2022108736}, nuclear battery regulation \cite{KRAUSE201251,prelas2014review,doi:10.1504/IJNEST.2023.135375}, astrophysical nucleosynthesis \cite{NatureCo,PhysRevC.68.015804,OrAstr}, and the synthesis of superheavy elements \cite{Oganessian_2015,Oganessian_2017}. Since Becquerel’s early observation of radioactivity in uranium cooled to liquid-air temperature ($-192\ ^{\circ}\mathrm{C}$) \cite{inbook}, a great deal of the studies have explored the influence of temperature, pressure, electromagnetic field, and gravitational acceleration on nuclear radioactive processes \cite{Zhou_2011}. However, no measurable variation in decay rates has been observed under such conditions.  This is primarily because such perturbations are too weak to significantly modify the tunneling barrier, leaving a fundamental challenge to conventional $\alpha$-decay control strategies.

The development of high-intensity laser technology has provided a uniquely extreme environment and an operable platform for the investigation of laser-nucleus interaction \cite{Fucb}. Particularly, improvements in chirped pulse amplification (CPA) techniques \cite{STRICKLAND1985447} have enabled the generation of laser fields with broader frequency ranges, higher laser intensities, and shorter pulse durations \cite{RevModPhys.78.309}—advancements that were recognized with the Nobel Prize in Physics in 2018. Currently, the laser peak field intensities have reached $10^{23}\ \mathrm{W/cm^{2}}$ \cite{Yoon:21}, the corresponding electric field strength is equivalent to the Coulomb field strength formed by the atomic nucleus at a distance of about 10 fm. Secondly, the ponderomotive energy of a proton in a laser field under the current laser intensity has exceeded several MeV, reaching the energy magnitudes of nuclear physics. Meanwhile, ongoing plans for next-generation petawatt-class laser systems \cite{Colin2, Radier,Yu-Tong} expect that the laser intensities will be further increased by one to two orders of magnitude beyond current levels in the future, which has further motivated the exploration of the laser-driven or laser-assisted nuclear phenomena \cite{ELI, 10.1117/12.2671369, GB, Xiaoppnp}.

Indeed, recent theoretical works have reported that the high-intense lasers can influence the nuclear dynamics such as nuclear $\alpha$ decay \cite{PhysRevC99, PhysRevC102, BAI201823, Misicu2013, Misicurizea, Po212, Cheng123, XqGamow, CHENGoddA, wang2025alpha, PhysRevC1103}, proton radioactivity \cite{Misicu_2019,PhysRevC.105.024312,PhysRevC.106.064610}, two-proton raioactivity \cite{Zou_2024}, cluster radioactivity \cite{w6wq-mj9b}, nuclear fussion \cite{PhysRevC.102.011601,PhysRevC.105.064615} and some low-energy nuclear excitation \cite{EPJA93Mo,PhysRevLett.130.112501,PhysRevC.110.L051601,Mzg}. Experimental progress has begun to substantiate these possibilities. For instance, Feng $et\ al.$ presented femtosecond pumping of isomeric nuclear states of $^{83}$Kr by the Coulomb excitation of ions with the quivering electrons induced by laser fields \cite{Feng}. Moreover, Shvyd$\rm{k^{'}}$o $et\ al.$ reported the resonant x-ray excitation of the $^{45}\mathrm{Sc}$ isomeric state by using the 12.4-keV laser pulses \cite{45Sc}, reducing the transition energy uncertainty by two orders of magnitude. Together, these developments establish high–intensity lasers as a powerful emerging tool for manipulating nuclear processes, underscoring both their theoretical importance and their potential for transformative applications.

Actinide nuclei, which predominantly decay via the $\alpha$ channel, provide natural platforms for the study of laser-assisted $\alpha$ decay. Besides that, such control has potential applications in nuclear transmutation (e.g., $^{239}$Pu and $^{237}$Np~\cite{CHWASZCZEWSKI200387,GUDOWSKI2000169c,KURNIAWAN2022108736}), nuclear radiotherapy (e.g., $^{225}$Ac and $^{227}$Th~\cite{NY1,NY2}), and nuclear battery regulation (e.g., $^{238}$Pu and $^{244}$Cm~\cite{KRAUSE201251,prelas2014review,doi:10.1504/IJNEST.2023.135375}), etc. The goal of the current paper is to quantitatively study the influence of ultra-intense laser fields on the $\alpha$ decay of actinide nuclei. Given that multipolar deformation effects and the intrinsic orientation dependence of actinide nuclei significantly influence the study of $\alpha$ decay, the nuclear deformation must be considered in this research. To achieve this goal, we first develop a deformed one-parameter model, in which the double-folding form obtains the deformed Coulomb potential. The numerical results show that the $\alpha$ decay-rate of those nuclei can be altered by non-negligible amounts, for example, on the order of 0.01$\%$ - 0.1$\%$ in intense laser fields available in the forthcoming years. Furthermore, we found that the influence of a laser field on $\alpha$ decay is determined by two aspects. From the perspective of the nucleus, the laser field’s effect on $\alpha$ decay is governed by the nuclear shell structure and decay energy. The $\alpha$-emitting nuclei with lower decay energies and located farther from neutron shell closures are more susceptible to the laser fields. Based on the robustness of the shell closure in the laser field, we also predicted that $N=142$ may be a possible deformed neutron sub-shell. From the perspective of the laser driver, we proposed a bichromatic laser scheme to enhance the influence of laser fields on the $\alpha$ tunneling. With appropriate phase conditions and amplitude ratios, it is shown that a fundamental–second-harmonic ($\omega$–$2\omega$) bichromatic field can increase the time-averaged modification by one to two orders of magnitude.

This article is organized as follows. A brief introduction of the theoretical framework is presented in Section \ref{section 2}. Detailed numerical results and discussion are given in Section \ref{section 3}. Section \ref{section 4} is a simple summary. 

\section{Theoretical framework}
\label{section 2}
\subsection{The deformed one-parameter model}
In the theoretical framework of the one-parameter model \cite{Tavares_2005,Medeiros_2006,Tavares_2012,Zou_2021,Qi_2025}, the decay constant $\lambda$ equals the product of three quantities
\begin{equation}
\lambda = P_{f}\ \nu_{0}\ P_{t},
\label{(1)}
\end{equation}
with $ P_{f} = e^{-G_{ov}}$ and $P_{t}= e^{-G_{se}}$. Here, $P_{f}$ is the formation probability of an $\alpha$ particle at the nuclear surface, $P_{t}$ is the tunneling probability through the external separation barrier region, i.e., the Coulomb-plus-centrifugal potential barrier. The assault frequency $\nu_{0}$ is related to the oscillation frequency $\omega_n$ by  \cite{PhysRevC.81.064309,PhysRevC.84.027303,Liu_2020} 
\begin{equation}
\nu_{0}=\frac{\omega_n}{2\pi}=\frac{(2n_{r}+l+\frac{3}{2})\hbar}{2 \pi \mu_{0} R_{n}^2}=\frac{(G_0+\frac{3}{2})\hbar}{1.2\pi \mu_0 R_{0}^{2}},
\end{equation}
where $\hbar$ is the reduced Planck constant. $\mu_0 = \frac{m_{\alpha} m_{d}}{m_{\alpha}+ m_{d}}$ is the final reduced mass of $\alpha$-daughter nucleus system with $m_{\alpha}$ and $m_{d}$ being the atomic mass of $\alpha$ particle and daughter nucleus, respectively. The relationship of $R^{2}_n = \frac{3}{5}R^{2}_0$ is used here for the nucleus RMS radius \cite{PhysRevC.62.044610}. $G_0 = 2n_{r}+l$ represents the main quantum number with $n_{r}$ and $l$ being the radial quantum number and the angular momentum quantum number, respectively. For $\alpha$ decay, $G_0$ is given by \cite{PhysRevC.69.024614} %
\begin{eqnarray}
G_0=2n_{r}+l=\left\{\begin{array}{ll}
18, & N \leq 82, \\[0.25pt]  \\
20, &82 < N \leq 126,\\[0.25pt]  \\
22, & N > 126 .
\end{array}\right.
\end{eqnarray}
The Gamow factors $G_{i}\ (i = ov, se)$ can be evaluated by the Wentzel-Kramers-Brillouin (WKB) integral approximation and expressed as
\begin{equation}
G_{ov}=\int_{R_{in}}^{R_{c}} 2 k (r) d r , 
\label{eq4}
\end{equation}
\begin{equation}
G_{se}= \int_{R_{c}}^{R_{out}} 2k (r) d r  \ , 
\label{eq5}
\end{equation}
where $k(r) = \sqrt{\frac{2\mu(r)}{\hbar ^2} \left[V(r) - Q_{\alpha}\right]}$ is the wave number. $r$ is the centeral distance between the $\alpha$ particle and daughter nucleus. $\mu(r)$ is the reduced mass of nuclear fission system and $V(r)$ is the potential barrier suffered by the $\alpha$ particle. The classical turning points, denoted as $R_{in}$ and $R_{out}$, can be calculated using the equation $V(r) = Q_{\alpha}$. The $\alpha$-decay energy $Q_{\alpha}$ is obtained by 
\begin{equation}
\label{Q}
Q_{\alpha} = \Delta M_{p} - (\Delta M_{d} + \Delta M_{\alpha}) \ \rm{MeV} \ ,
\end{equation}
where $\Delta M_{p}$, $\Delta M_{d}$, and $\Delta M_{\alpha}$ are the mass excess of parent nucleus, daughter nucleus, and $\alpha$ particle, respectively.

\begin{figure}[t]
\vspace{-10pt} 
\centering
	\includegraphics[width=\linewidth]{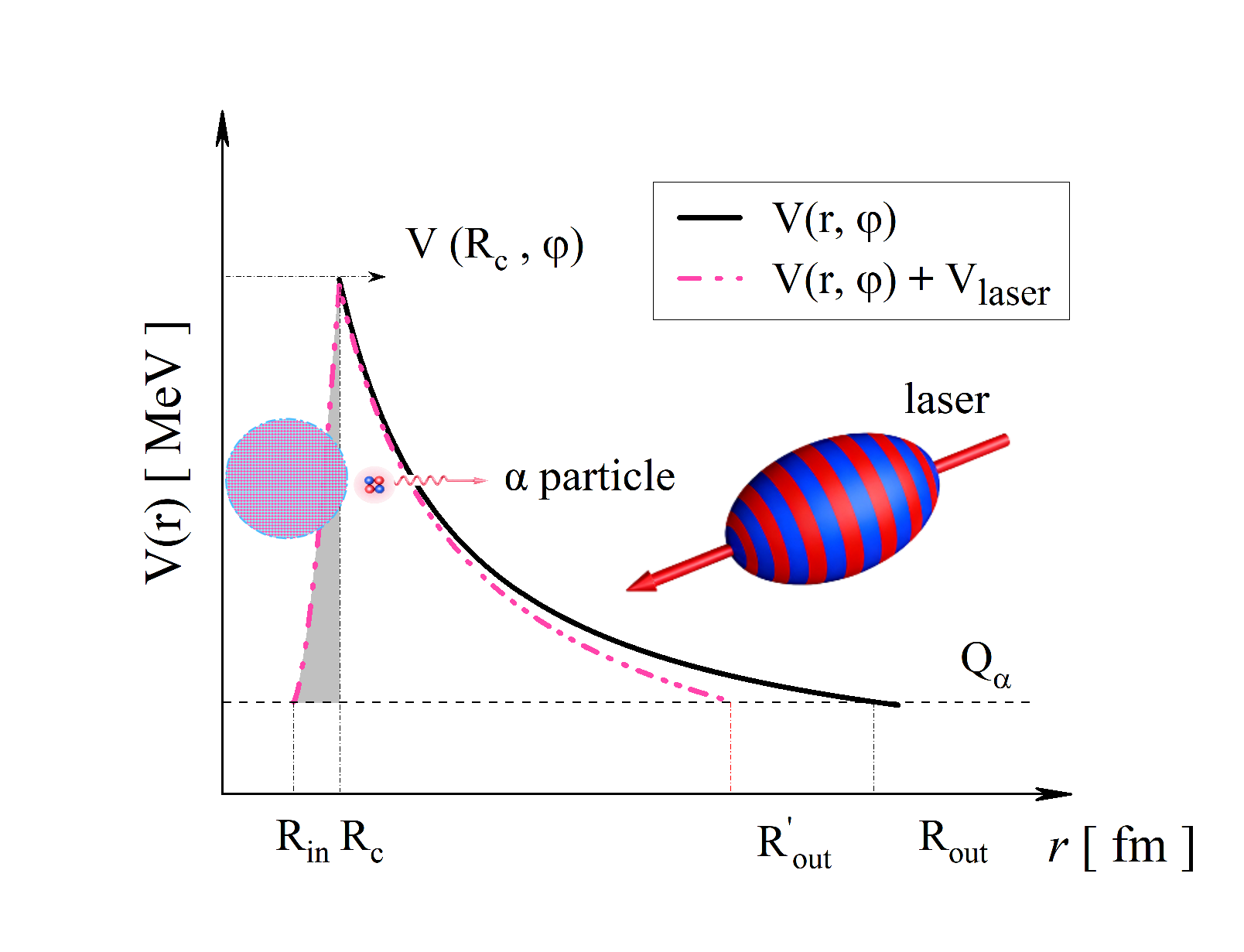}
	\caption{Schematic diagram of the total potential with laser field (red dotted line) and without laser field (black line).}
	\label{Fig:1}
\end{figure}

The nuclear deformation effects can significantly influence the directional quantum tunneling paths. Integrating the deformation effects into the one-parameter model, the penetration factor $P$ can be expressed as 
\begin{equation}
P_i = \frac{1}{2} \int_{0}^{\pi} P_{\varphi} \sin(\varphi)\ d\varphi \ ,
\end{equation}
with $P_{\varphi} = \exp\left[ -2\int k(r, \varphi) \, dr \right]$. Here, $\varphi$ is the deformed angle that denotes the emission direction from the symmetry axis for symmetric deformed nuclei. In the deformed one-parameter model, the total interacting potential $V(r,\theta)$ and reduced mass $\mu(r)$ between the emitted $\alpha$ particle and daughter nucleus is divided into two parts, i.e., overlapping region for $R_{in} < r  < R_c$ and separated region for $R_{out} > r  >  R_c$, respectively, as shown in Fig.\,\ref{Fig:1}. In the overlapping region, the reduced mass and interaction potential are dealt with in the exponential form, i.e., $\mu(r)=\mu_0 \left[(r-R_{in})/(R_c-R_{in})\right]^k$ and $V(r,\ \varphi)= \left[V(R_c,\ \varphi) -Q_{\alpha}\right]\left[(r-R_{in})(R_c-R_{in})\right]^q + Q_{\alpha}$, respectively. Here, the adjustable prarmeter of model is defiend as g =$\left(1+\frac{k+q}{2}\right)^{-1}$ from the integrate of Gamow factor $G_{ov}$ in Eq(\,\ref{eq4}). $R_{in} = R_p - R_{\alpha}$ is the difference between the radius of the parent nucleus and the $\alpha$ particle. The radius of the nucleus can be expressed as
\begin{equation}
 R_{i} = 1.28A_{i}^{1/3}-0.76+0.8A_{i}^{-1/3}\ ,\ i = p,\ d,\ \alpha,
\end{equation}
where $i = p,\ d,\ \alpha$ denote the parent nucleus, daughter nucleus and $\alpha$ particle, respectively. $R_c = R_T(\varphi) + R_{\alpha}$ is the separation configuation at which the $\alpha$ particle arrives at the surface of deforemd daughter nucleus. Here, $R_T(\varphi)$ is the radius of the deformed daughter nucleus, which can be expressed as
\begin{equation}
R_{T}(\varphi) =R_{d} \cdot \left[1+\sum_{\lambda} \beta_{\lambda} Y_{\lambda}^{(0)}(\varphi) \right] (\lambda = 2,\ 4,\ 6)\ ,
\end{equation} 
where $\beta_{2}$, $\beta_{4}$, $\beta_{6}$ denote the quadrupole, hexadecapole, and hexacontatetrapole deformation of the nuclear ground state, respectively. $Y_{\lambda}^{(0)}(\varphi)$ is the shperical harmonics function. 

In the separated region, $V(r,\varphi)$ consists of the deformed Coulomb potential $V_{C}(r,\varphi)$ and the centrifugal potential $V_{l}(r)$. The deformed Coulomb potential can be obtained in the double-folding from and expressed as \cite{PhysRevC.61.044607}
\begin{equation}
V_{C}(\vec{r},\ \varphi)=\iint \frac{\rho_{1}(\vec{r}_{1}) \rho_{2}(\vec{r}_{2})}{|\vec{r}+\vec{r}_{2}-\vec{r}_{1}|} \, dr_{1} \, dr_{2}, 
\end{equation}
where $\vec{r}_1$ and $\vec{r}_2$ are the radius vectors in the charge distributions of the daughter nucleus and $\alpha$ particle, respectively. $\rho_1(\vec{r}_1)$ and  $\rho_2(\vec{r}_2)$ denote the density distribution of the daughter nucleus and $\alpha$ particle. Simplified appropriately by the Fourier transform, the deformed Coulomb potential can be approximated as \cite{Gui_2022}
\begin{equation}
V_{C}(\vec{r},\varphi)=V_{C}^{(0)}(\vec{r},\varphi)+V_{C}^{(1)}(\vec{r},\varphi)+V_{C}^{(2)}(\vec{r},\varphi) \ .
\end{equation}
As for the centrifugal potential $V_{l}(r)$, since $l\ (l+1)$ $\to$ $(l+\frac{1}{2})^2$ is a necessary correction for one-dimensional problems, we choose it as the Langer modified form in this work. Thus it can be expressed as \cite{PhysRevC.97.044322}
\begin{equation}
V_{l}(r)=\frac{\hbar^{2}(l+\frac{1}{2})^{2}}{2\mu r^{2}},
\end{equation}
where $l$ is the angular momentum carried by the emitted $\alpha$ particle. According to the spin-parity selection rule, the angular momentum satisfies the following conditions
\begin{eqnarray}
|I_{p}-I_{d}| \leq l \leq I_{p} + I_{d} \ ,  \ \ \    \frac{\pi_{p}}{\pi_{d}} = (-1)^{l} \ ,
\end{eqnarray}
where $I_{p},\pi_{p}$ and $I_{d}, \pi_{d}$ represent the spin and parity values of the parent and daughter nuclei, respectively.

\begin{figure}[h]
\vspace{-10pt} 
\centering
	\includegraphics[width=\linewidth]{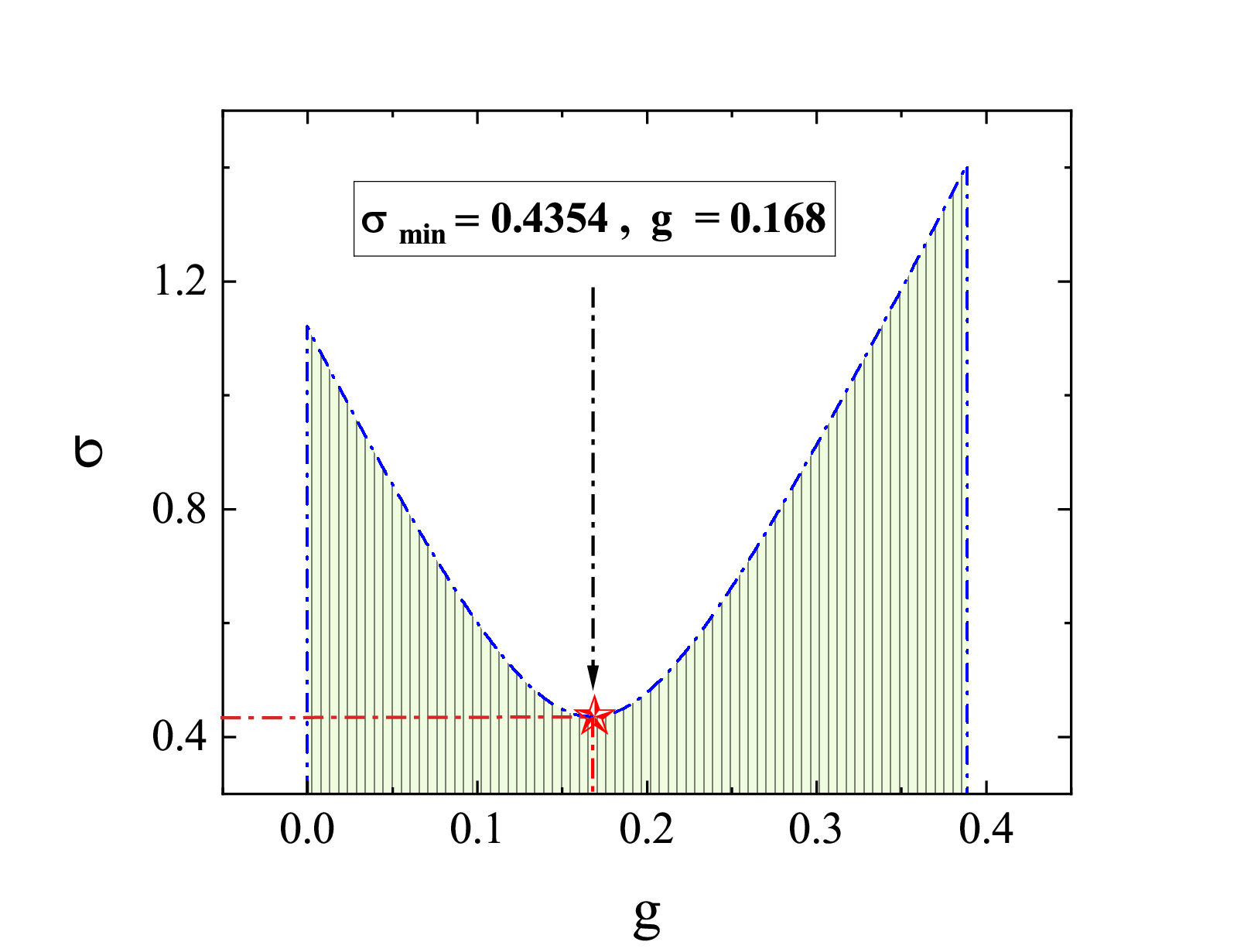}
	\caption{The relationship between the standard deviation $\sigma$ and the adjustable parameter $\rm{g}$ in the deformed one-parameter model.}
	\label{fig 2}
\end{figure}
\vspace{-5pt}
\subsection{$\alpha$ decay under the Gaussian laser field}
For the $\alpha$ decay under the laser field, an electric-dipole term is introduced to describe the effect of the laser field on $\alpha$ decay, which is expressed as \cite{PhysRevC102,PhysRevC99,BAI201823,Misicu2013}
 \begin{equation}
V_i(\vec{r}, t, \theta) = -Z_{\rm{eff}}\ \vec{r}\ \cdot \vec{E}(t) = -Z_{\rm{eff}}\ r \ E(t)\ \rm{\cos \theta} \ ,
\end{equation}
where $\theta$ is the angle between the vector $\vec{r}$ and $\vec{E}$. $Z_{\rm{eff}}$ =$\frac{2A - 4Z}{A + 4}$ is the effective charge describing the tendency of the laser electric field to separate the $\alpha$ particle and residual daughter nucleus. $A$ and $Z$ are the mass number and proton number of the daughter nucleus, respectively. $E(t) = E_{0} f(t) \rm{sin}(\omega t)$ is the laser electric field with $f(t)= {\exp}\left({-t^{2}/{2\sigma_E}^2}\right)$ and $\omega$ being the pulse temporal profile and the angular frequency, and $\sigma_E = \tau /(2\sqrt{2\ln2})$ with $\tau = m T_0$ being the pulse width of the laser electric field amplitude. $T_0 = \frac{2\pi}{\omega}$ is the laser period, and $m$ is the number of the laser period, respectively. The peak of the laser electric field $E_{0}$ depends on the laser peak intensity $I_{0}$ and expressed as \cite{Misicu_2019}
 \begin{equation}
E_{0}[\mathrm{V} / \mathrm{cm}]=\left(\frac{2 I_{0}}{c_{0} \epsilon_{0}}\right)^{1 / 2}=27.44\left(I_{0}\left[\mathrm{~W} / \mathrm{cm}^{2}\right]\right)^{1 / 2}
\end{equation}
where $c_{0}$ and $\epsilon_{0}$ are the speed of light in vacuum and the permittivity of free space, respectively.
\begin{figure*}[t]
  \centering
  \subfloat{%
    \begin{overpic}[width=0.5\linewidth]{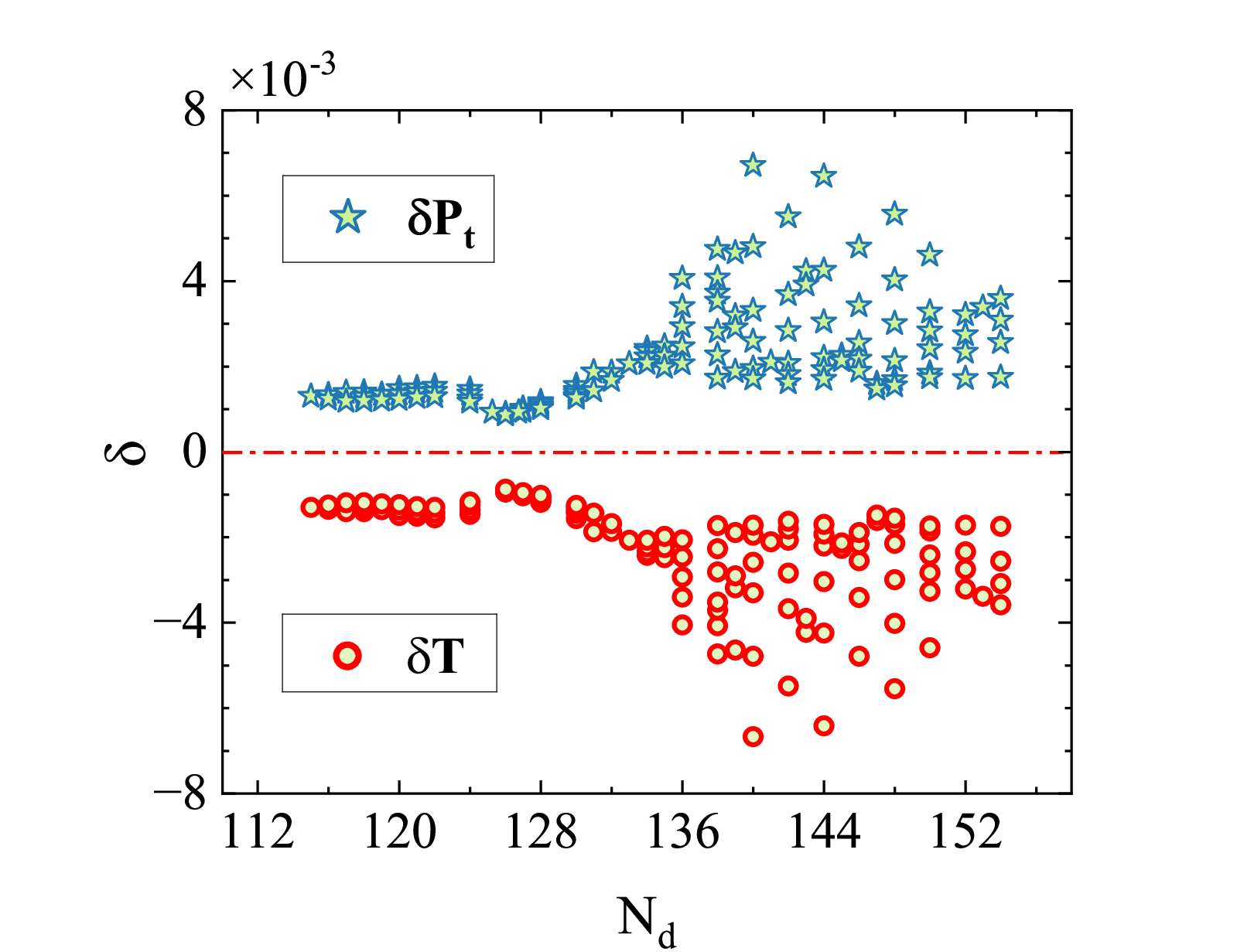}
      \put(19,64){\large \fontsize{10pt}{8pt}\selectfont \bfseries (a)} 
    \end{overpic}
    \label{fig:3a}}
  \hspace{-0.06\linewidth}
  \subfloat{%
    \begin{overpic}[width=0.5\linewidth]{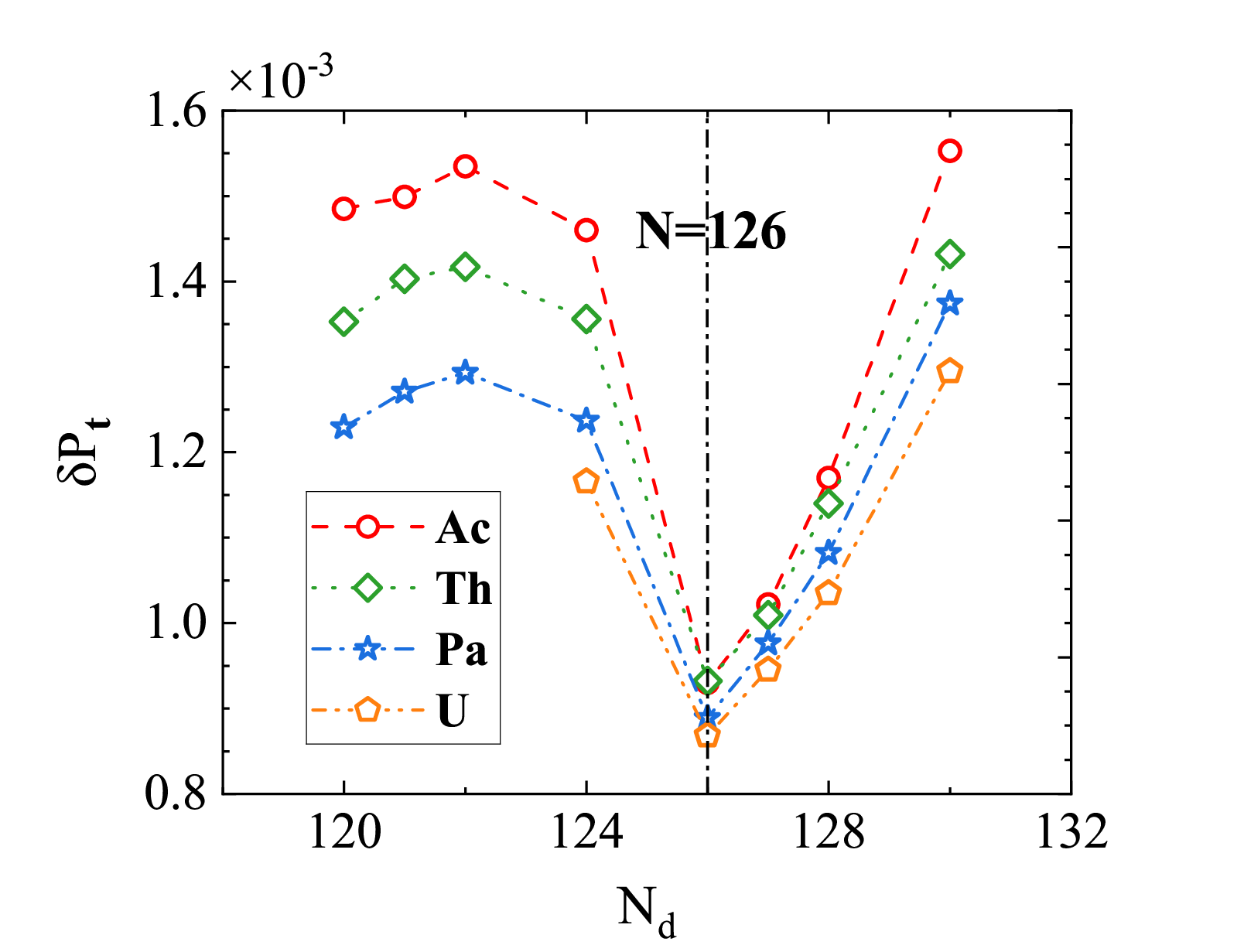}
      \put(19,64){\large \fontsize{10pt}{8pt}\selectfont \bfseries (b)}
    \end{overpic}
    \label{fig:3b}}
  \\[-9pt]

  \subfloat{%
    \begin{overpic}[width=0.5\linewidth]{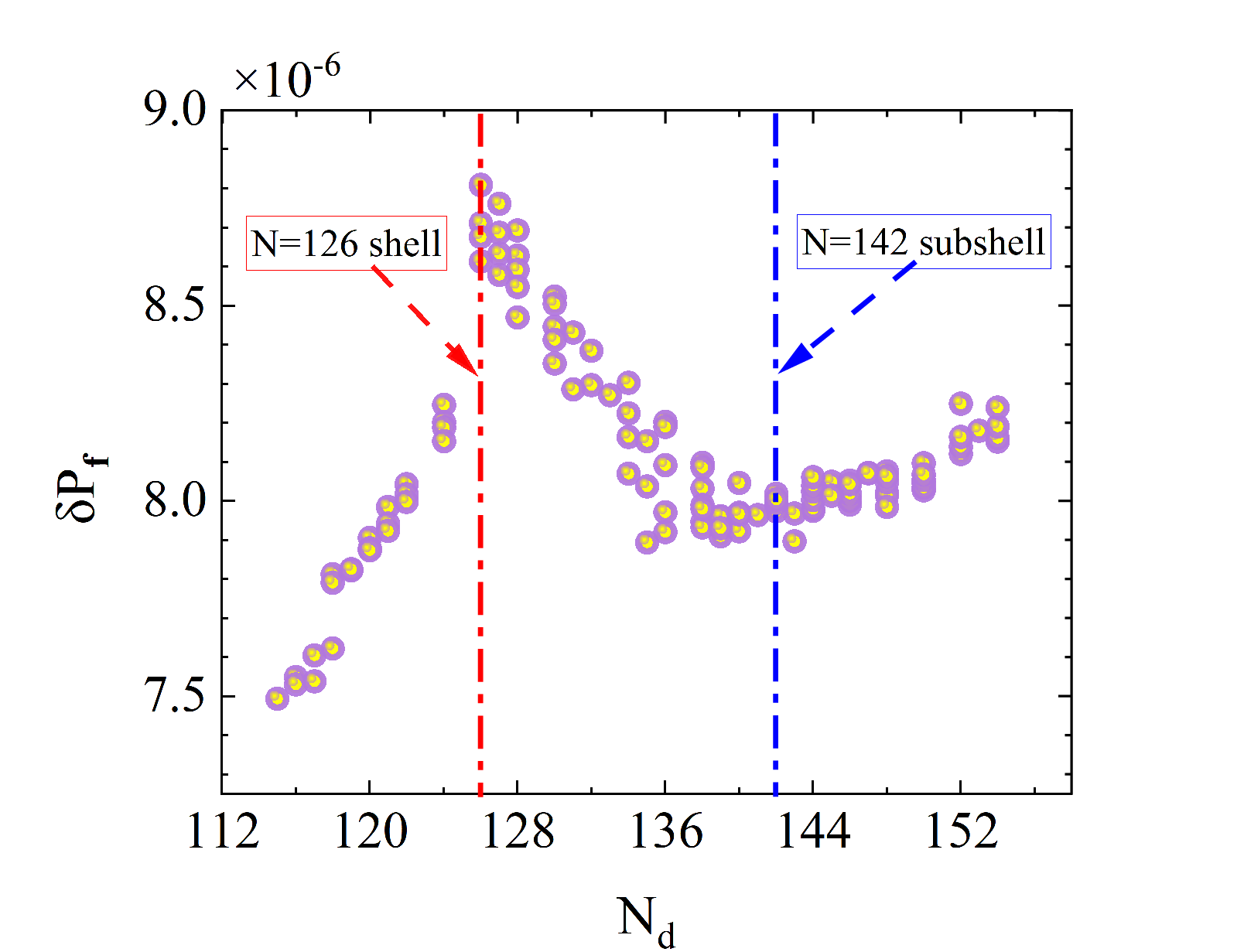}
      \put(19,64){\large \fontsize{10pt}{8pt}\selectfont \bfseries (c)}
    \end{overpic}
    \label{fig:3c}}
  \hspace{-0.06\linewidth}
  \subfloat{%
    \begin{overpic}[width=0.5\linewidth]{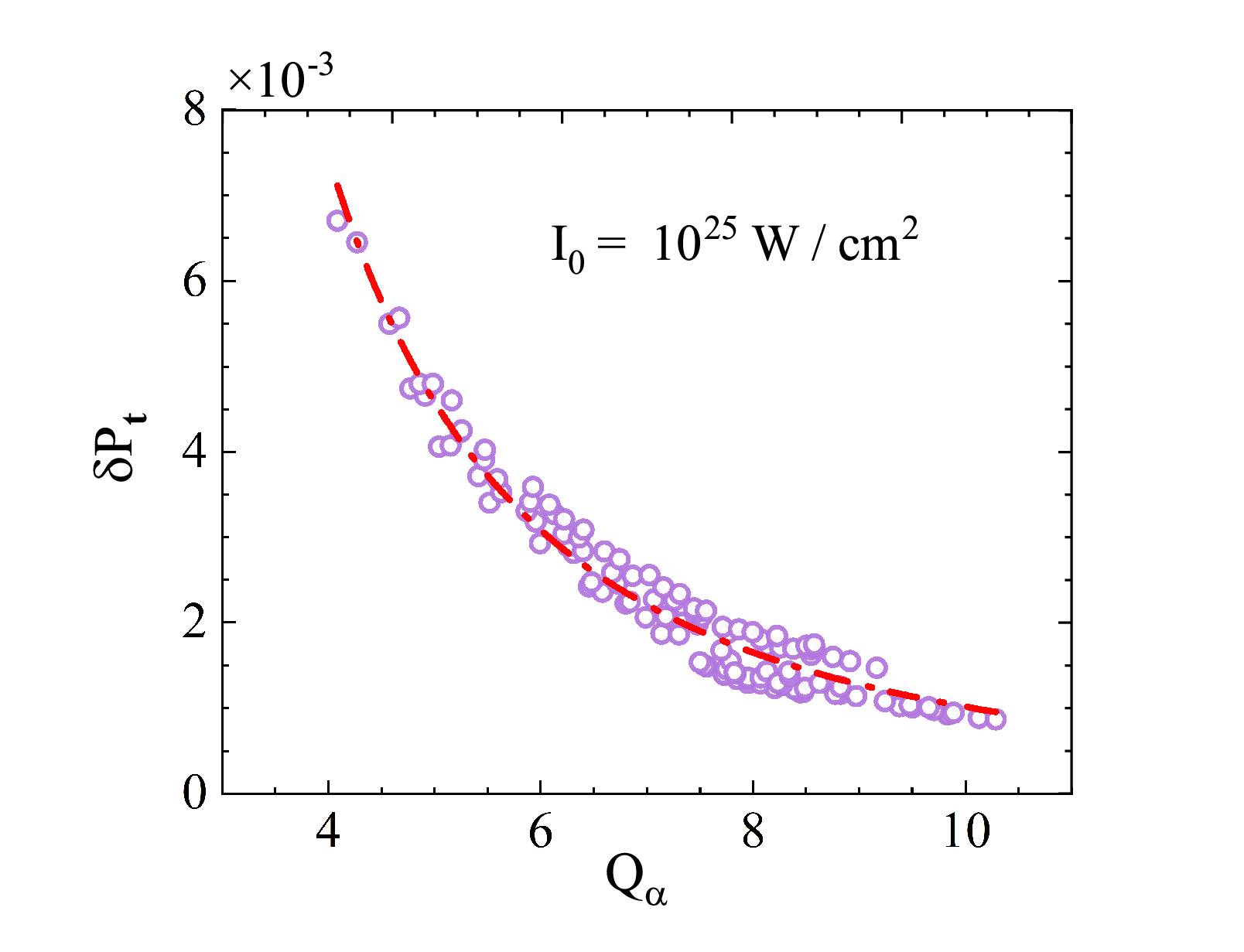}
      \put(19,64){\large \fontsize{10pt}{8pt}\selectfont \bfseries (d)}
    \end{overpic}
    \label{fig:3d}}
  \caption{(a)The instantaneous rate of the change of tunneling probability $\delta P_t$ and half-lives $\delta T$ for the different actinide nuclei. (b)The influence of the shell effect on the $\delta P_t$ for the isotopic nuclei. (c)The influence of the shell effect on the $\delta P_f$ for different actinide nuclei. (d)The relationship between $\delta P_t$ and the $\alpha$ decay energy $Q_{\alpha}$ for different actinide nuclei at the laser intensity $I_0 = 1.0 \times 10^{25}\ \rm{W/cm^2}$.}
  \label{fig:3}
\end{figure*}
Furthermore, the $\alpha$-decay energy $Q_{\alpha}$ can also be affected by the laser field, the change of energy equals the work of the $\alpha$ particle accelerated along the radius of the daughter nucleus. Thus, the decay energy under the laser field can be rewritten as \cite{Cheng123}
\begin{equation}
 Q^{*}_{\alpha} = Q_{\alpha} + 2e E(t) R_{T}(\varphi)\rm{\cos\theta}.
\end{equation}

\section{RESULTS AND DISCUSSIONS}
\label{section 3}
\subsection{ Laser-induced instantaneous modifications to the $\alpha$-decay half-lives}
In the deformed one-parameter model employed in this work, the only adjustable parameter $\rm{g}$ of the model is obtained by fitting the experimental half-lives of 118 favored $\alpha$ emitters utilizing the least-squares method. Experimental data for parity, spin, $\alpha$-decay energy, and half-lives are taken from AME2020 \cite{Wang_2021} and NUBASE2020 \cite{Kondev_2021}. The quadrupole $\beta_2$, hexadecapole $\beta_4$, and hexacontetrapole $\beta_6$ deformation parameters of the daughter nucleus are derived from the FRDM(2012) \cite{MOLLER20161}. The standard deviation $\sigma$ is defined as 
\begin{equation}
 \sigma=\sqrt{\frac{1}{N} \sum_{i=1}^{N}\left(\log _{10} T_{1 / 2}^{\text {cal.} i}-\log _{10} T_{1 / 2}^{\text {exp.} i}\right)^{2}} \ ,
\end{equation}
where $\log_{10} T_{1 / 2}^{\text {cal.} i}$ and $\log_{10} T_{1 / 2}^{\text {exp.} i}$ represent the logarithms of the theoretical and experimental half-lives for the $i$-th nucleus, respectively. The relationship between $\sigma$ and $\rm{g}$ is shown in Fig. \ref{fig 2}. From this figure, one can see that the minimal value of $\sigma_{\text{min}} = 0.435$ with $\rm{g} = 0.168$, indicating that the theoretical predictions of the half-lives are in good agreement with the experimental data, within a factor of 2.7, for these 118 actinide nuclei. Notably, the recently synthesized actinide nucleus $^{210}\rm{Pa}$, measured at the newly constructed China Accelerator Facility for Superheavy Elements, has a reported $\alpha$-particle energy of $E_{\alpha} = 8.258\ \rm{MeV}$ and half-life of $T_{1/2\alpha} = 6.0_{-1.1}^{+1.5}$ ms \cite{210Pa}. This is well reproduced in our prediction, where the half-life is calculated to be 5.047 ms (See the appendix \ref{sec:appendix}).

\begin{figure*}[t]
  \centering
  \subfloat{%
    \begin{overpic}[width=0.5\linewidth]{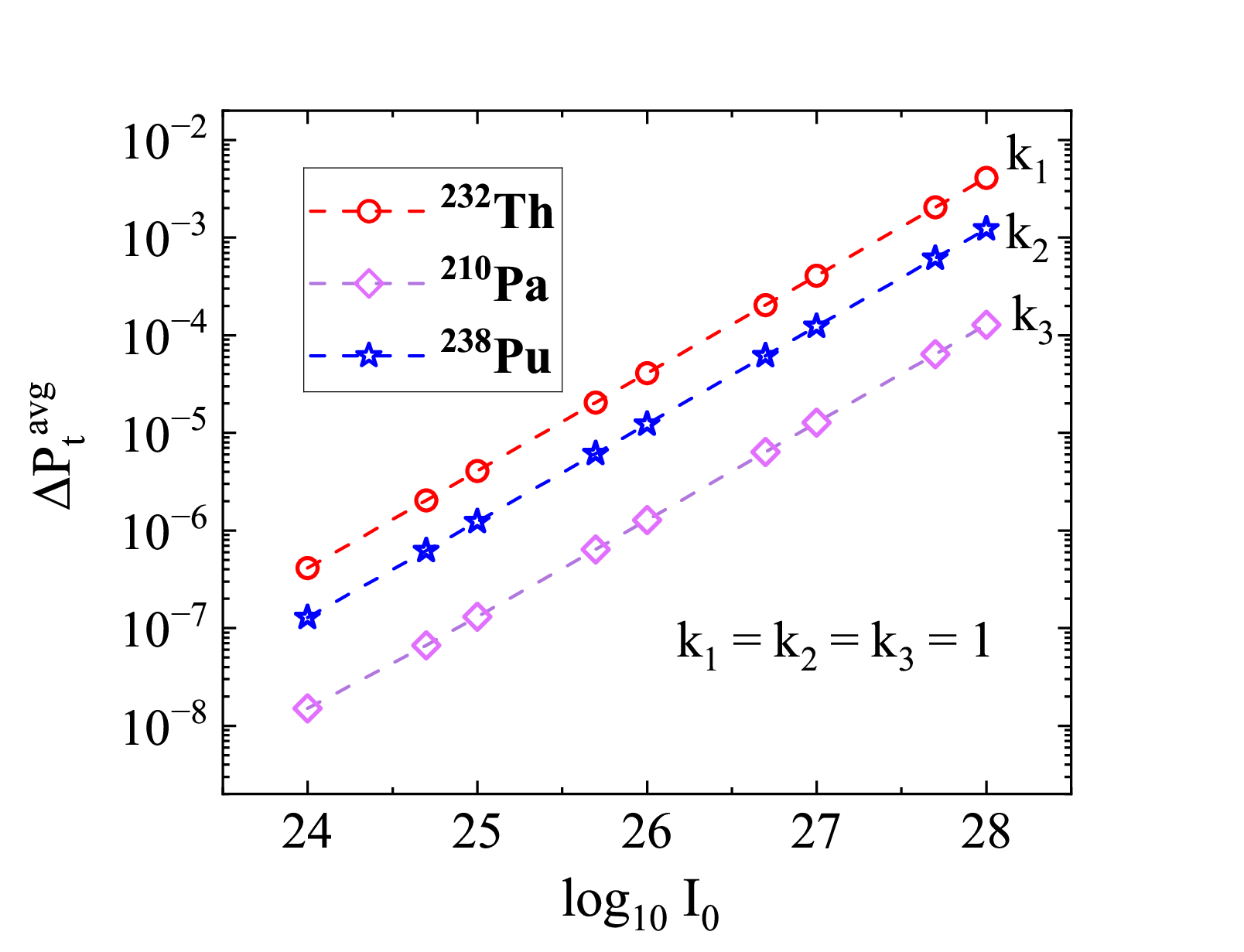}
      \put(19,64){\large \fontsize{10pt}{8pt}\selectfont \bfseries (a)} 
    \end{overpic}
    \label{fig:4a}}
  \hspace{-0.06\linewidth}
  \subfloat{%
    \begin{overpic}[width=0.5\linewidth]{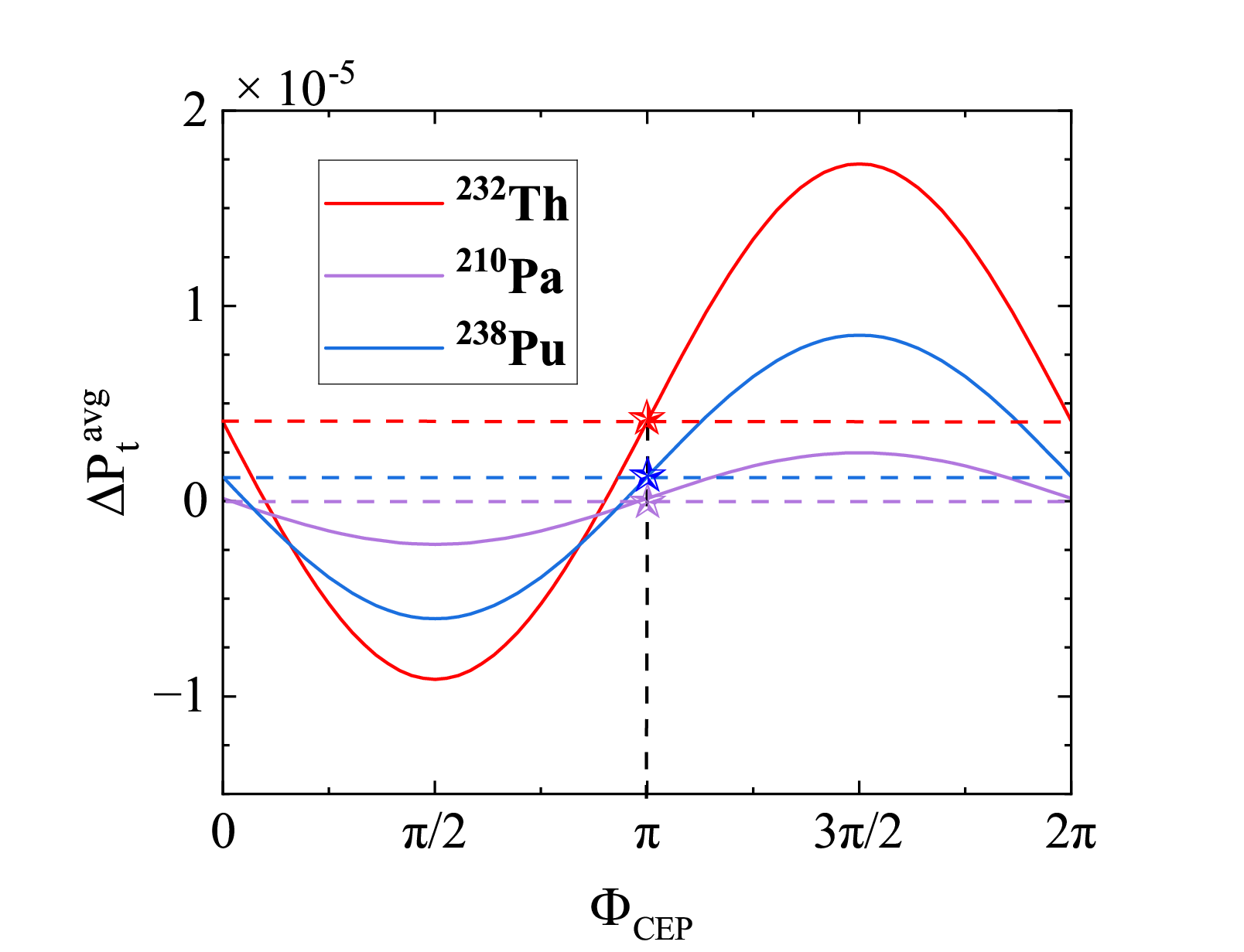}
      \put(19,64){\large \fontsize{10pt}{8pt}\selectfont \bfseries (b)}
    \end{overpic}
    \label{fig:4b}}
  \\[-12pt]
  
  \subfloat{%
    \begin{overpic}[width=0.5\linewidth]{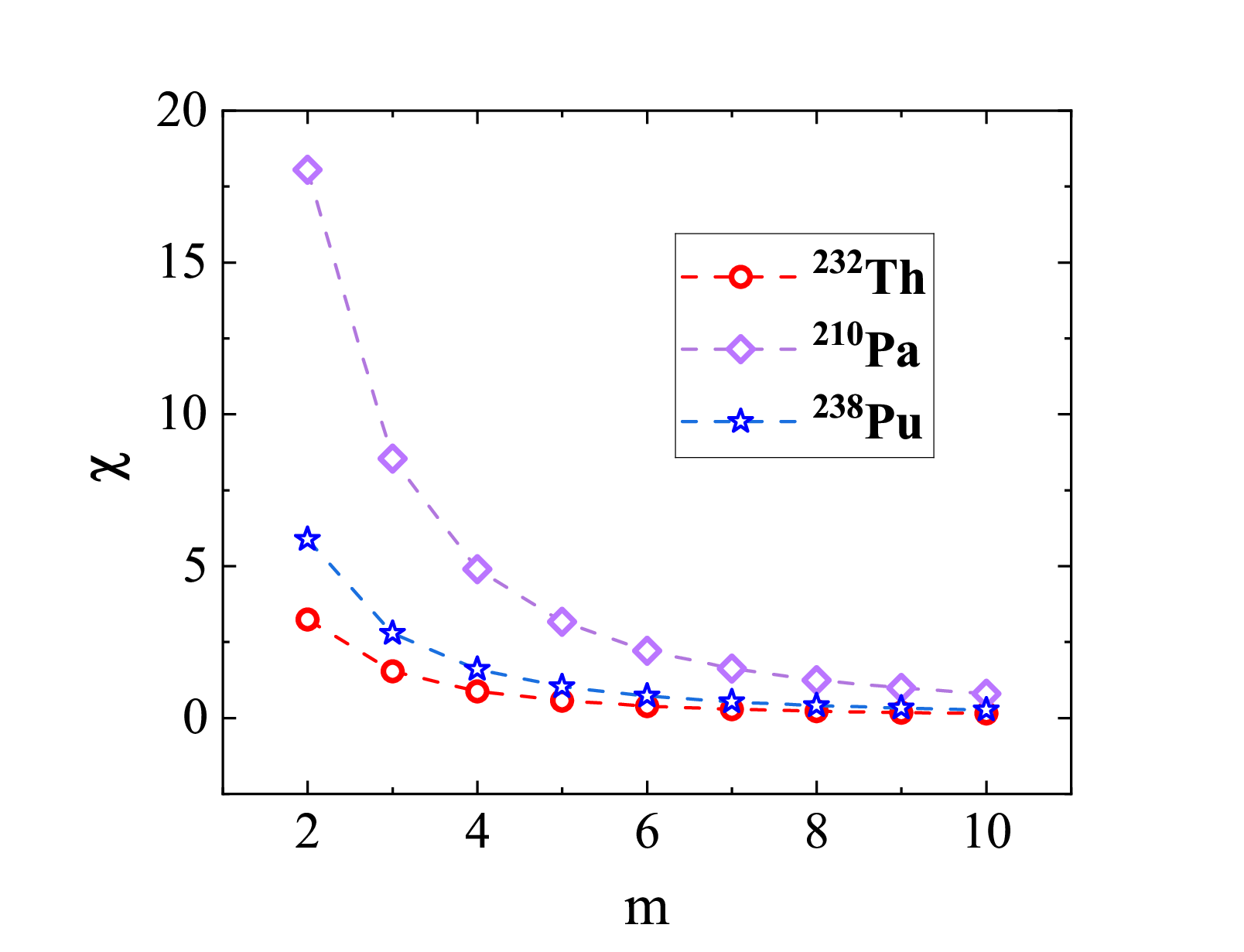}
      \put(19,64){\large \fontsize{10pt}{8pt}\selectfont \bfseries (c)} 
    \end{overpic}
    \label{fig:4c}}
  \hspace{-0.06\linewidth}
  \subfloat{%
    \begin{overpic}[width=0.5\linewidth]{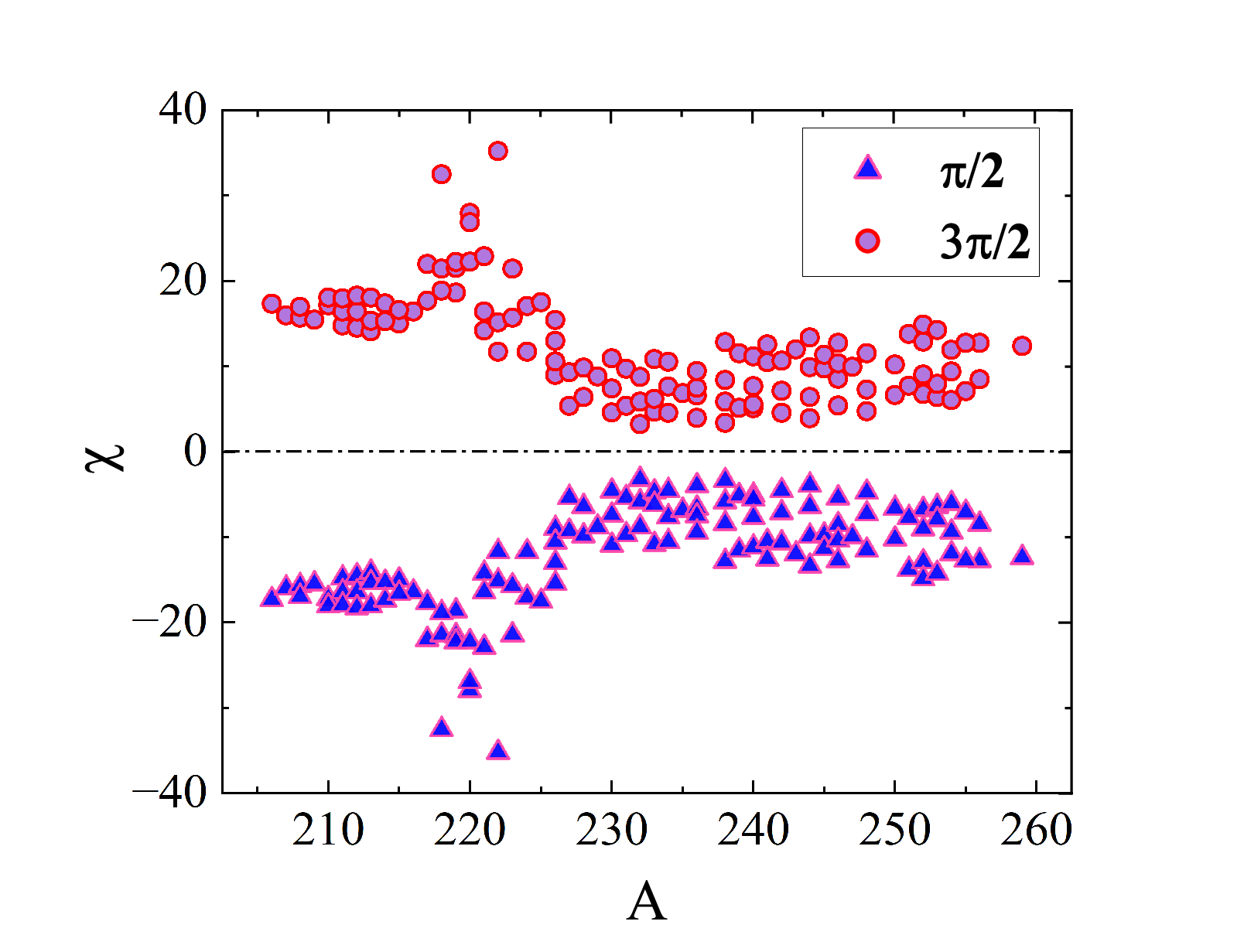}
      \put(19,64){\large \fontsize{10pt}{8pt}\selectfont \bfseries(d)}
    \end{overpic}
    \label{fig:4d}}
  \caption{(a)The time-averaged modifications to the tunneling probability as a function of the laser intensity for three elements $^{232}$Th, $^{210}$Pa, and $^{238}$Pu, respectively. (b)The effects of the carrier envelope phase $\Phi_{\text{CEP}}$ on the $\Delta P_t^{\text{avg}}$ in the case of the number of the laser period $m = 2$. The horizontal underline with different colors marks the values of the $\Delta P_t^{\text{avg}}(\Phi_{\text{CEP}} = 0)$ for different nuclei. (c)The correction factor $\chi$ as a function of the $m$ for the $\Phi_{\text{CEP}} = \frac{3\pi}{2}$. (d)The correction factor $\chi$ as a function of mass number A for different actinide nuclei.}
  \label{fig:4}
\end{figure*}

Since tunneling is highly sensitive to the interaction potential between the emitted $\alpha$ particle and the residual nucleus, and the existence of the laser field modifies this potential, as shown in Fig.~\ref{Fig:1}. It is reasonable to expect that there will be some effects on the $\alpha$ decay processes. Furthermore, the process of emitting an $\alpha$ particle penetrating through the potential barrier can be regarded as the quasistatic approximation since the tunneling time $t_{\tau} \sim \frac{1}{\nu_0} = 10^{-21}\ \rm{s}$ is much smaller than the optical cycle of strong lasers. This has been widely employed in the tunneling ionization of atoms in strong-field atomic physics \cite{PhysRevA.54.R2551,PhysRevA.63.011404}. Based on the quasistatic approximation, we systematically investigate the effects of ultra-intense laser fields on the $\alpha$ decay of the actinide nuclei within $89 \leq Z \leq 103$.  To quantitatively describe the laser-induced modifications on the $\alpha$ decay, we define the instantaneous change rate of the formation probability $\delta P_f$, the tunneling probability $\delta P_t$, and the half-lives $\delta T$ as
\begin{equation}
 \delta P_i = \frac{P_i(E,\ \theta) - P_i(E=0,\ \theta)}{P_i(E=0,\ \theta)}\ ,
\end{equation}
\begin{equation}
 \delta T = \frac{T(E,\ \theta) - T(E=0,\ \theta)}{T(E=0,\ \theta)} \ \ ,
\end{equation}
where $E \neq 0$ and $E = 0$ denote the field-assisted and field-free $\alpha$ decay cases, respectively. Here, we focus on the forward direction of $\theta = 0$, i.e., maximizing the exploration of the impact of laser fields on the $\alpha$ decay. Table \ref{T1} in the appendix \ref{sec:appendix} shows the numerical results of $\delta P_f$, $\delta P_t$, and $\delta T$. For clarity, the $\alpha$-emitting parent nucleus, decay energy, deformation parameter, experimental half-lives, and theoretical ones in logarithmic form have also been listed. The laser wavelength is 800\ nm, and the laser intensity is assumed to be $10^{25}\ \rm{W/cm^2}$, which is expected to be achievable in the forthcoming years \cite{MREL123}. One sees that the $\alpha$ decay half-lives can be altered by $0.01\%$ to $0.1\%$ for these actinide nuclei, and this modification on half-life is mainly induced by the changes in the tunneling probability, i.e., $\delta P_f \ll \delta P_t \approx -\delta T$. A more intuitive symmetry between $\delta P_t$ and $-\delta T$ is shown in Fig.~\ref{fig:3}\subref{fig:3a}. From this figure, we see that a distinct turning point for $\delta P_t$ and $\delta T$ occurs around the neutron number $N = 126$. This phenomenon may be associated with the shell effects of the magic number nucleus. To verify this conjecture, we present the $\delta P_t$ and $\delta P_f$ as a function of the neutron number of the daughter nucleus in Figs.~\ref{fig:3}\subref{fig:3b} and ~\ref{fig:3}\subref{fig:3c}, respectively. From these figures, it is evident that both $\delta P_t$ and $\delta P_f$ are significantly influenced by the shell effects at $N = 126$, demonstrating the robustness of the $N = 126$ shell closure effect in intense laser fields. Additionally, we observe a subtle increase in $\delta P_f$ at $N = 142$, similar to the behavior of $\delta P_f$ at $N = 126$, suggesting a possible deformed neutron sub-shell closure at $N = 142$. This inference aligns with predictions of neutron sub-shell closures at $N = 142$ for field-free cases \cite{PhysRevC.87.054324,PhysRevC.111.024322}.

In addition to the shell structure, which influences the modulation of the $\alpha$-decay rate under the laser field, the decay energy may also affect the modifications induced by lasers. To explore the effect of decay energy, we draw the relationship between $\delta P_t$ and the decay energy $Q_{\alpha}$ for different parent nuclei in Fig~\ref{fig:3}\subref{fig:3d}. From this figure, one can see that as the $Q_{\alpha}$ decreases, the $\delta P_t$ increases. This result can be explained by the fact that the low-energy $\alpha$ decay typically involves higher Coulomb potential barriers and longer tunneling paths $ R_{c}\to R_{out}$, allowing the laser field to exert a more significant effect. Therefore, the intrinsic nuclear properties of long-lived actinide nuclides such as $^{232}\rm{Th}$, $^{238}\rm{U}$, $^{238}\rm{Pu}$, and $^{239}\rm{Pu}$ naturally confer an advantage, making their $\alpha$-decay rates more sensitive to modulation by the lasers.

\subsection{Time-averaged modifications to the tunneling probability in few-cycle monochrome fields}
The instantaneous rate of change of tunneling probability oscillates back and forth within the laser time envelope, so that most of the instantaneous components are cancelled out. To capture the overall effect of the single laser pulse on nuclear $\alpha$ decay, it is essential to evaluate the time-averaged modification to the tunneling probability. Here, we define an averaged rate of change in tunneling probability of $\alpha$ decay, denoted as $\Delta P_t^{\text{avg}}$, to estimate this influence, which is expressed as
\vspace{-2pt} 
\begin{equation}
 \Delta P_t^{avg} = \frac{1}{2\tau} \int_{-\tau}^{\tau} \delta P_t\ (t)\ \mathrm{d}t\ .\\ \
\end{equation}
Figure~\ref{fig:4}\subref{fig:4a} shows the dependence of $\Delta P_t^{\mathrm{avg}}$ on the laser intensity $I_0$ for three representative actinide nuclei, $^{232}$Th, $^{210}$Pa, and $^{238}$Pu. The pulse width is fixed at $\tau=2T_0$, typical of few-cycle pulses. A linear trend is observed between $\log_{10}\Delta P_t^{\text{avg}}$ and $\log_{10}I_0$ with a slope $k_i = 1$ for all three nuclei, indicating that the time-averaged modification of the tunneling probability is proportional to $I_0$ within the explored range. Consequently, increasing $I_0$ by one decade leads to a one-decade increase in $\Delta P_t^{\text{avg}}$. While this scaling suggests that stronger laser intensity produces proportionally larger modifications, substantially increasing the laser intensity is experimentally greatly challenging.

\begin{figure*}[t]
  \centering
    \subfloat{%
    \begin{overpic}[width=0.5\linewidth]{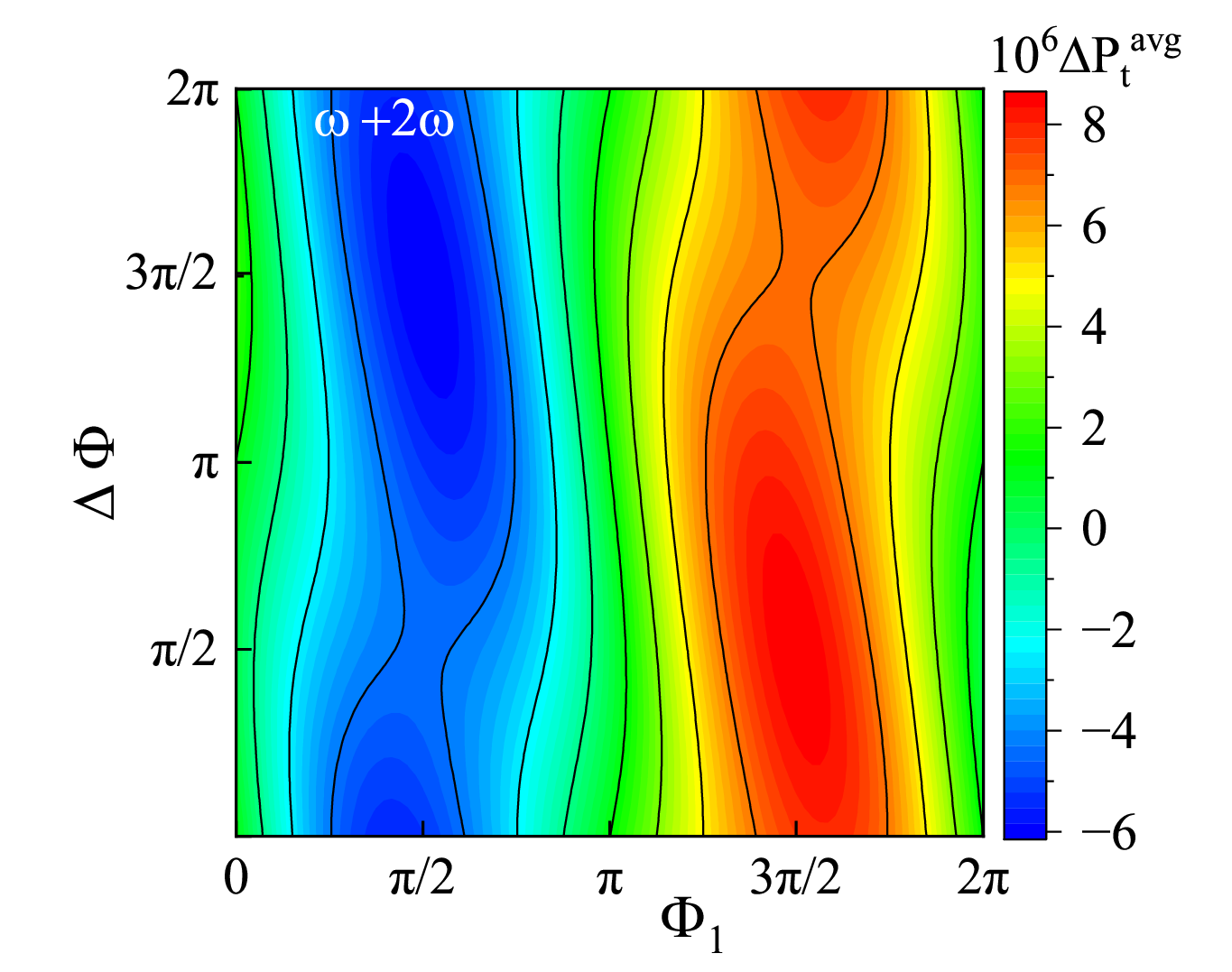}
      \put(19,68){\large \fontsize{10pt}{8pt}\selectfont \bfseries {\color{white}(a)}} 
    \end{overpic}
    \label{fig:5a}}
  \hspace{-0.03\linewidth}
     \subfloat{%
    \begin{overpic}[width=0.5\linewidth]{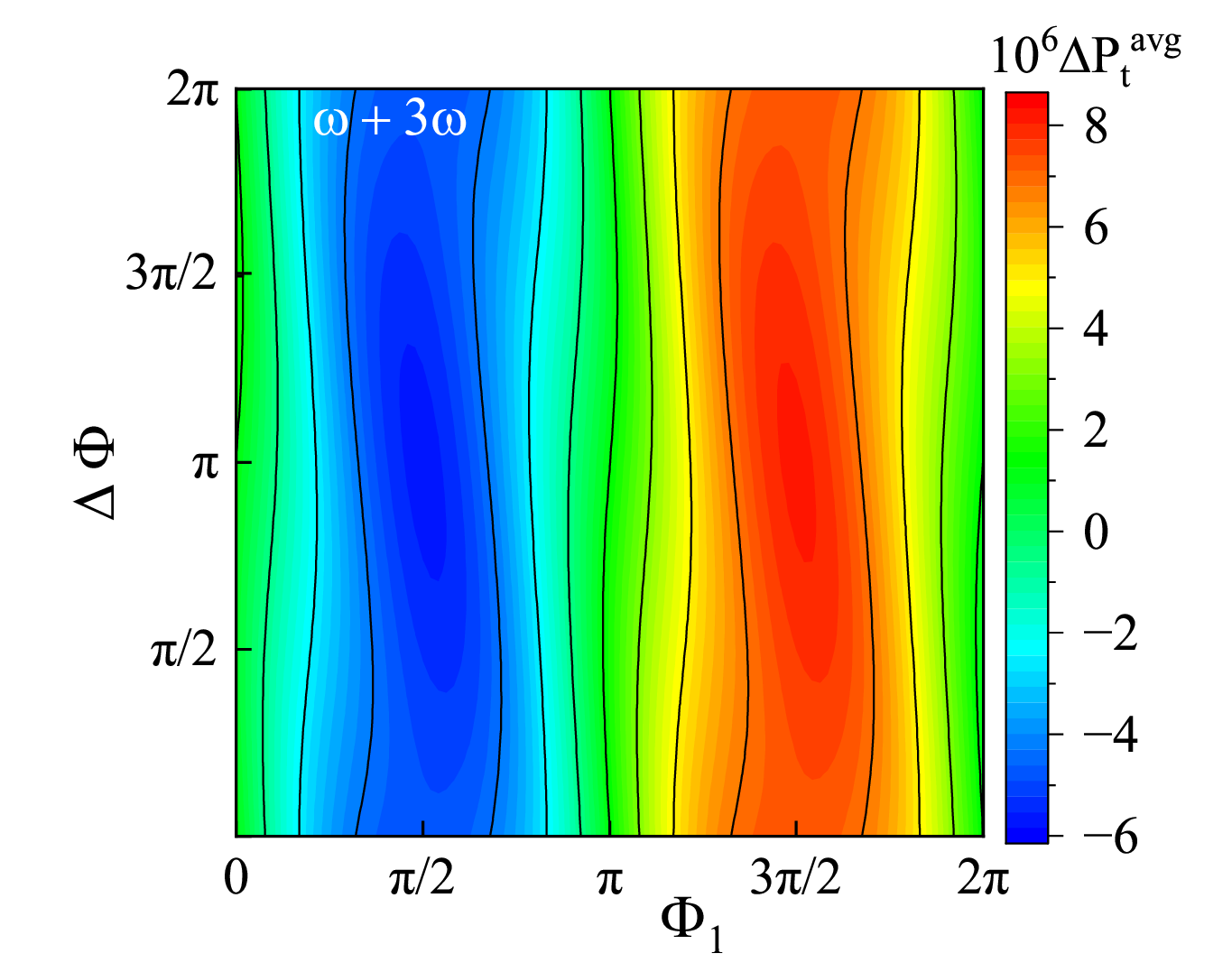}
      \put(19,68){\large \fontsize{10pt}{8pt}\selectfont \bfseries {\color{white}(b)}} 
    \end{overpic}
    \label{fig:5b}}
    \\[-5pt]
  
   \subfloat{%
    \begin{overpic}[width=0.5\linewidth]{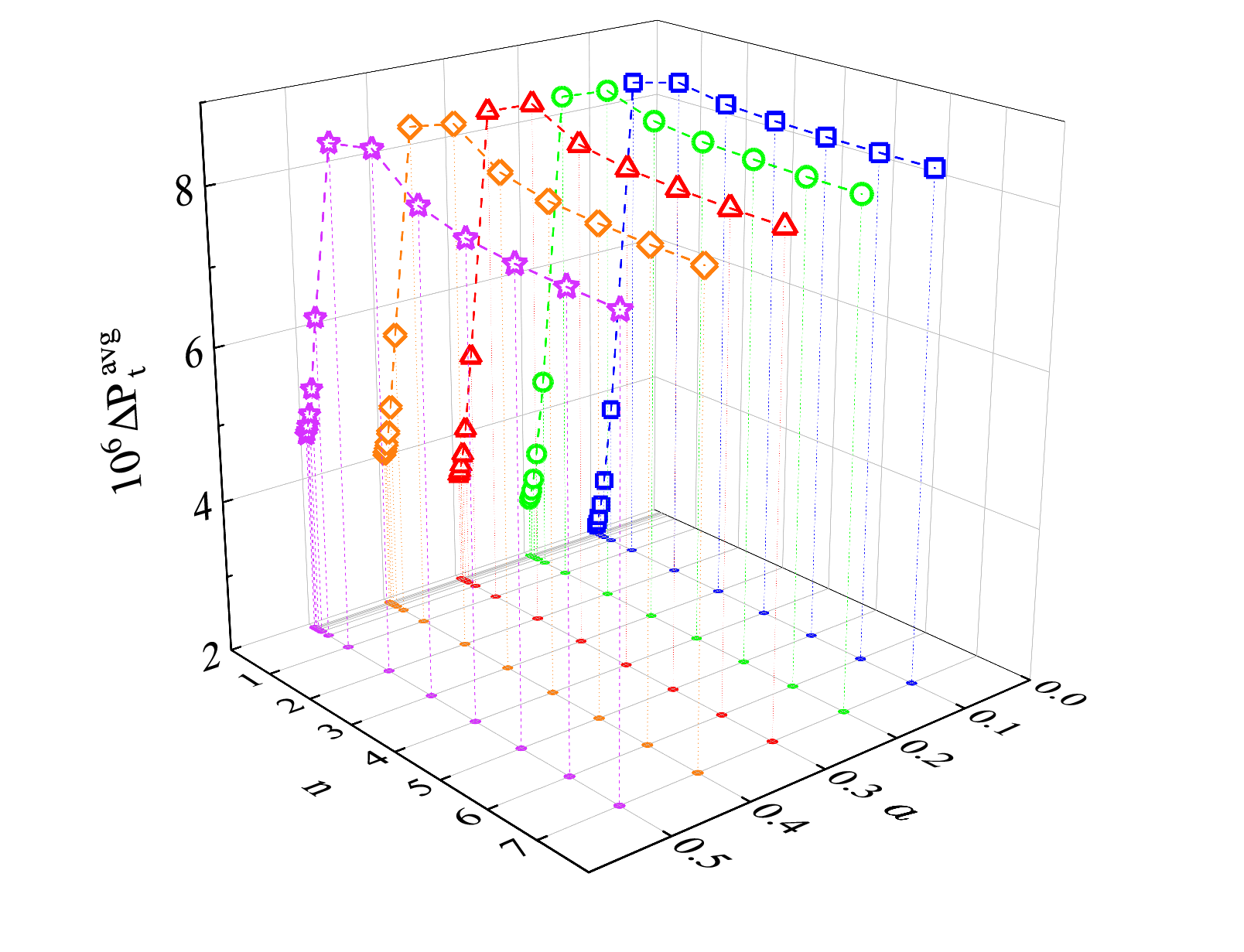}
      \put(18,65){\large \fontsize{10pt}{8pt}\selectfont \bfseries (c)} 
    \end{overpic}
    \label{fig:5c}}
  \hspace{-0.03\linewidth}
     \subfloat{%
    \begin{overpic}[width=0.5\linewidth]{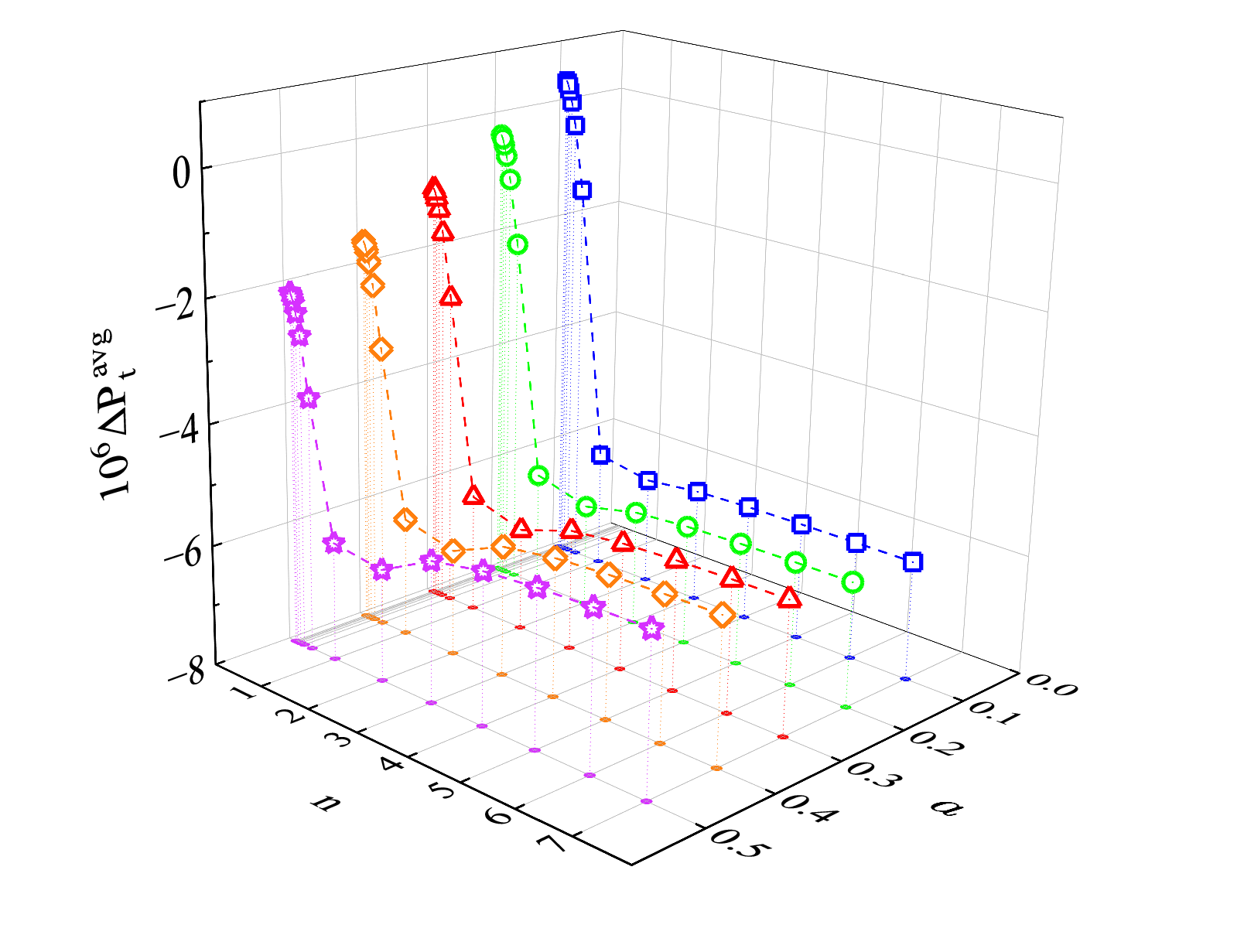}
      \put(20,65){\large \fontsize{10pt}{8pt}\selectfont \bfseries (d)} 
    \end{overpic}
    \label{fig:5d}} 
  \caption{(a)-(b)The phase scan heatmaps of the $\Delta P_t^{\text{avg}}$ for $^{238}$Pu under the conditions of $n = 2$ and $n = 3$, with the amplitude ratio $a =0.5$ and the pulse width $\tau =2 T_0$. (c)-(d)The effects of the different frequency ratio $n$ and the amplitude ratio $a$ of the Gaussian bichromatic laser pulses on $\Delta P_t^{\text{avg}}$ for $^{238}$Pu under the maximal and minimal phase modulation conditions, respectively.}
  \label{fig:5}
\end{figure*}

To explore alternative control strategies, previous studies have employed nonstandard temporal waveforms and polarization states. Qi \emph{et al.} proposed an elliptically polarized scheme~\cite{PhysRevC99}, in which the analysis focused on the angle $\theta$ between the laser electric field and the decay axis, without considering the frequency detuning. Mi\c{s}icu and Rizea theoretically demonstrated that short rectangular pulses containing an odd number of half-cycles could increase the decay rate of proton radioactivity by three orders of magnitude~\cite{Misicu_2019}. However, removing an even number of half-cycles from the pulse shape appears physically implausible. Cheng \emph{et al.} investigated frequency-chirped, temporally asymmetric pulses and reported an increase in $\Delta P_t^{\text{avg}}$ by up to two orders of magnitude~\cite{PhysRevC.105.024312}. In this study, we examine the effect of the carrier–envelope phase $\Phi_{\mathrm{CEP}}$ on $\Delta P_t^{\text{avg}}$. The results are shown in Fig.~\ref{fig:4}\subref{fig:4b}. From this figure, it can be seen that for an initial Gaussian waveform of $f(t)\sin(\omega t)$, $\Delta P_t^{\mathrm{avg}}$ is suppressed for $0<\Phi_{\mathrm{CEP}}<\pi$ and enhanced for $\pi<\Phi_{\mathrm{CEP}}<2\pi$, with a minimum at $\Phi_{\mathrm{CEP}}=\pi/2$ and a maximum at $\Phi_{\mathrm{CEP}}=3\pi/2$. For the form of $f(t)\cos(\omega t)$, the pattern is simply shifted by $\pi/2$, the strongest suppression occurs at $\Phi_{\mathrm{CEP}}=0$, and the strongest enhancement occurs at $\Phi_{\mathrm{CEP}}=\pi$, respectively. This emphasizes that the modulation of the time-averaged tunneling probability is dependent on the initial phase of the waveform. To quantify this impact, a correction factor $\chi$ is defined as
\begin{equation}
 \chi = \frac{\Delta P_t^{\text{avg}}(\Phi_{\text{CEP}}) - \Delta P_t^{\text{avg}}(\Phi_{\text{CEP}} = 0)}{\Delta P_t^{\text{avg}}(\Phi_{\text{CEP}} = 0)} \ .
\end{equation}
Figure~\ref{fig:4}\subref{fig:4c} shows the correction factor $\chi$ as a function of the number of laser period $m$ for the carrier envelope phase value $\Phi_{\text{CEP}} = \frac{3\pi}{2}$. A negative correlation is observed between $\chi$ and $m$, indicating that the correction factor decreases as the pulse duration increases. This result suggests that the $\Phi_{\text{CEP}}$ has a significant impact on the time-averaged modification of the tunneling probability in the few-cycle pulse regime. Specifically, for $m = 2$, $\Phi_{\text{CEP}}$ significantly affects the tunneling process, with $\chi$ reaching values on the order of tens. However, as the pulse duration increases (e.g., for $m = 10$), the influence of the $\Phi_{\text{CEP}}$ diminishes, approaching zero, making its effect on the tunneling probability less significant.

After establishing the role of the $\Phi_{\text{CEP}}$ in few-cycle pulses, we calculated the effect of the $\Phi_{\text{CEP}}$ on $\Delta P_t^{\mathrm{avg}}$ for different actinide nuclei in Fig.~\ref{fig:4}\subref{fig:4d}. In this figure, the correction factor $\chi$ is plotted as a function of the mass number of parent nucleus $A$ for the two values of $\Phi_{\text{CEP}} = \frac{\pi}{2}$ and $\Phi_{\text{CEP}} = \frac{3\pi}{2}$, respectively. From the figure, it can be seen that $\chi (\frac{\pi}{2})$ and $\chi (\frac{3\pi}{2})$ exhibit good symmetry around $\chi = 0$ for the different parent nuclei, with the maximum value of $|\chi|$ reaching approximately 35. This indicates that simple adjustments to the carrier-envelope phase can increase or decrease the $\Delta P_t^{\mathrm{avg}}$ by up to a factor of thirty-five times.

\begin{figure*}[t]
  \centering
    
     \subfloat{%
    \begin{overpic}[width=0.5\linewidth]{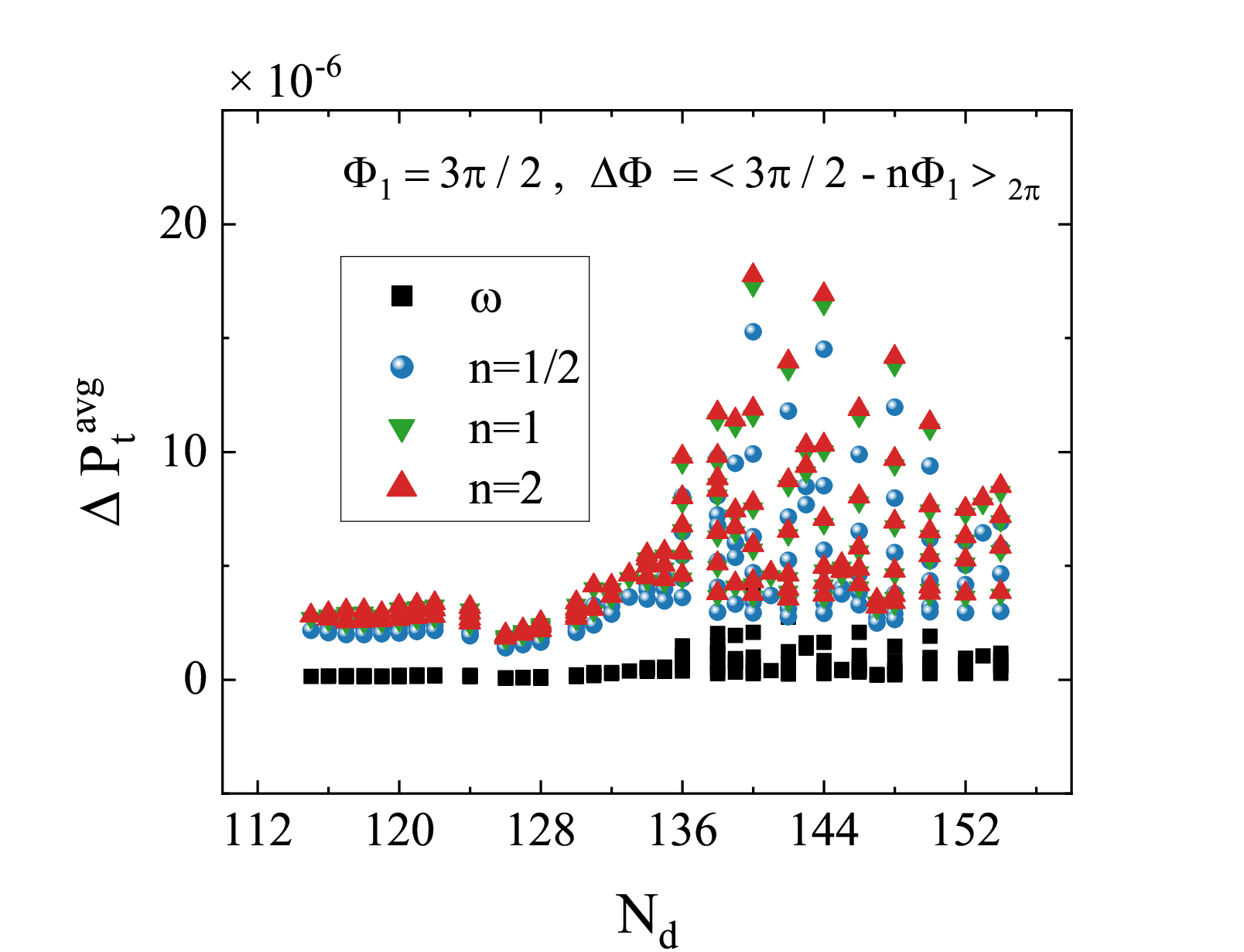}
      \put(18,63){\large \fontsize{10pt}{8pt}\selectfont \bfseries(a)} 
    \end{overpic}
    \label{fig:6a}}
 \hspace{-0.03\linewidth}
  \subfloat{%
    \begin{overpic}[width=0.5\linewidth]{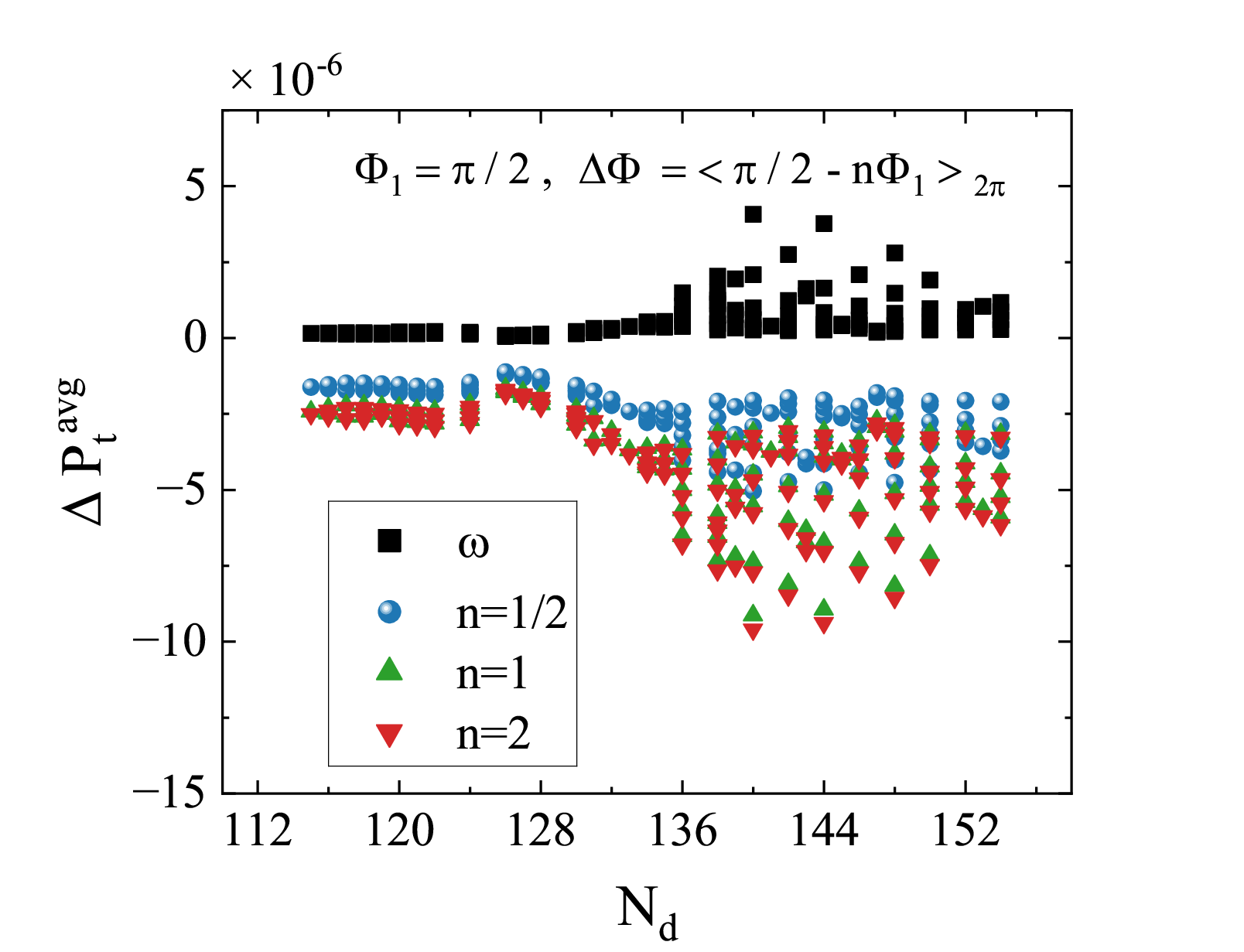}
      \put(18,63){\large \fontsize{10pt}{8pt}\selectfont \bfseries(b)}
    \end{overpic}
    \label{fig:6b}}
    \\[-12pt]
  
  \subfloat{%
    \begin{overpic}[width=0.5\linewidth]{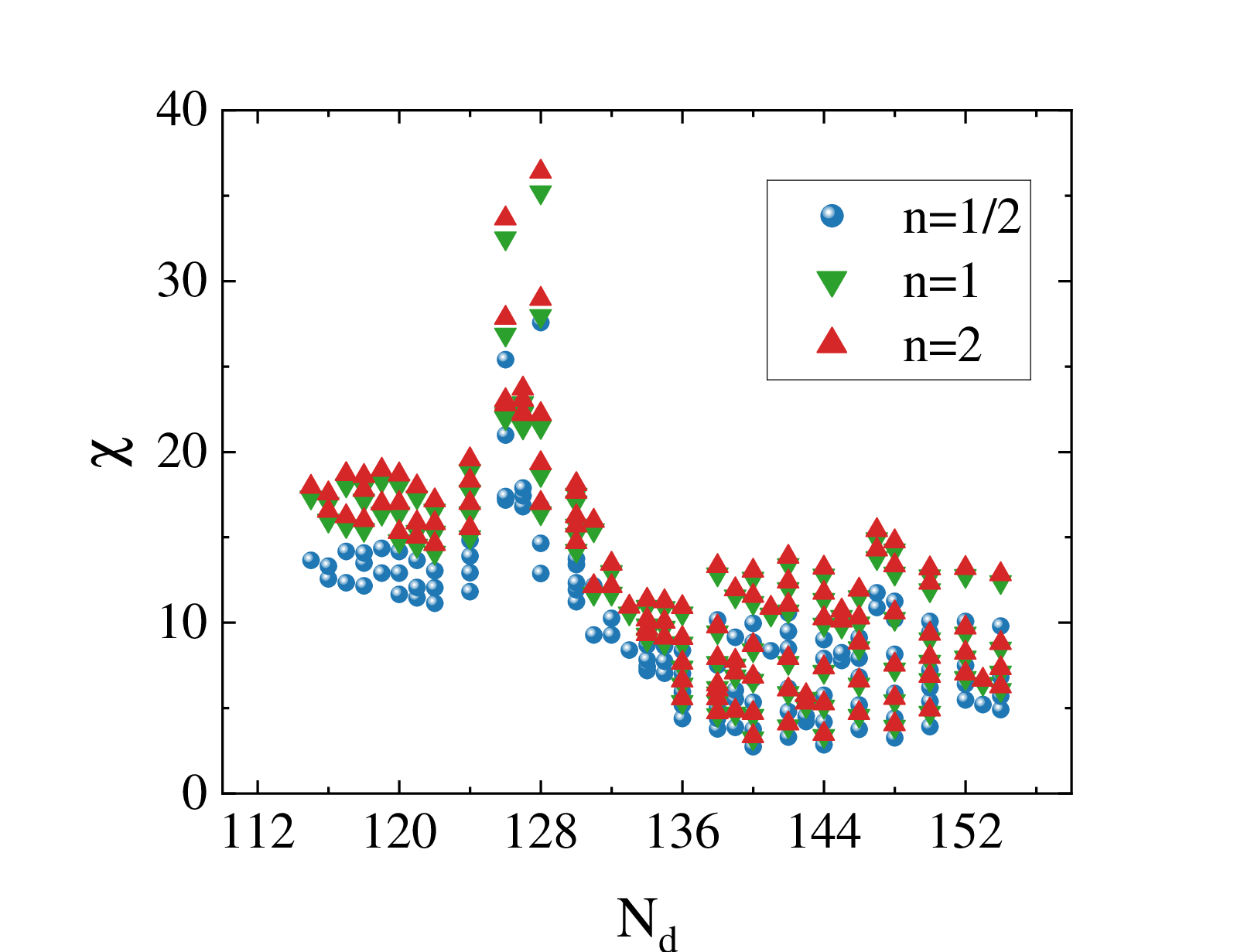} 
      \put(18,63){\large \fontsize{10pt}{8pt}\selectfont \bfseries(c)} 
    \end{overpic}
    \label{fig:6c}}
 \hspace{-0.03\linewidth}
  \subfloat{%
    \begin{overpic}[width=0.5\linewidth]{ 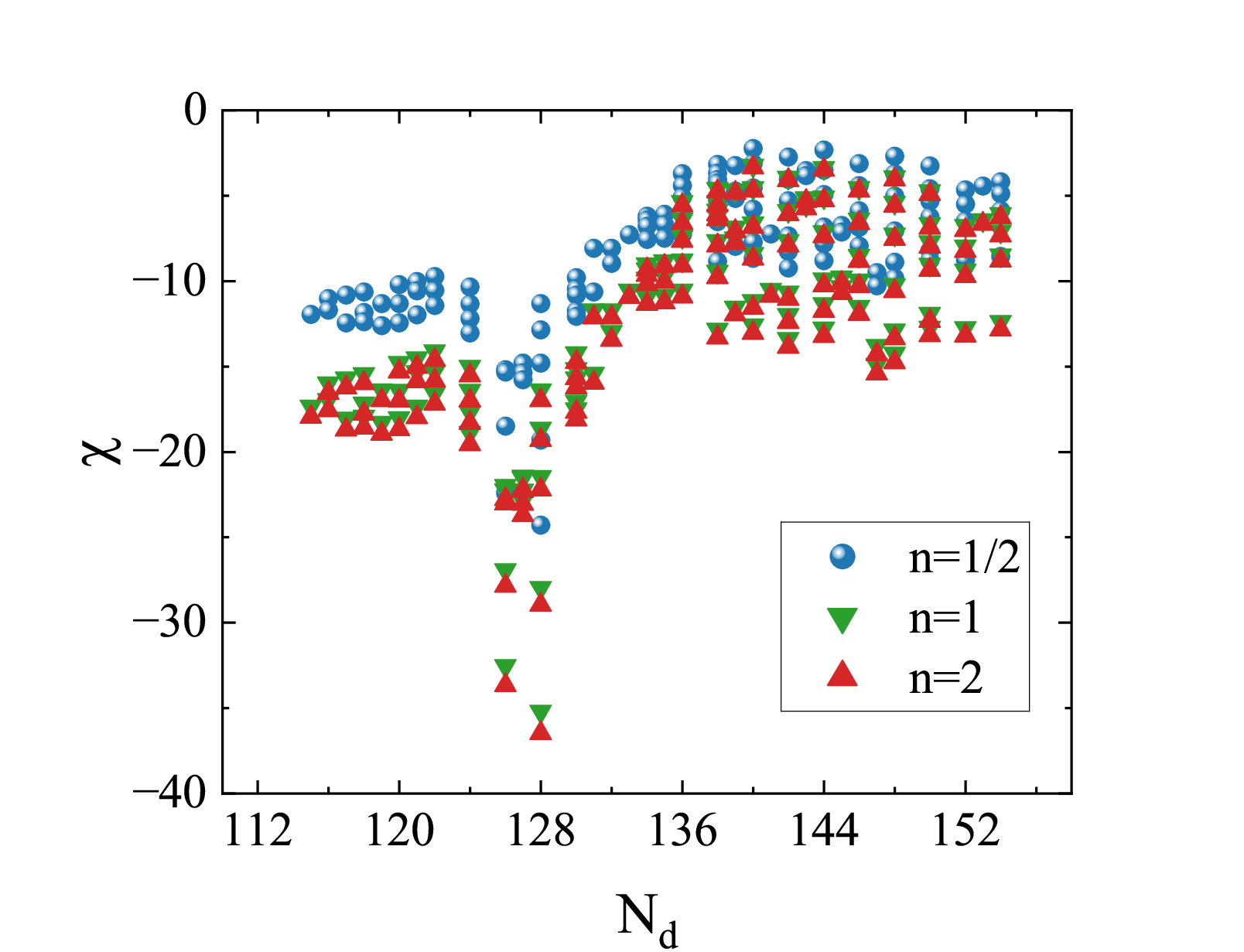}
      \put(18,63){\large \fontsize{10pt}{8pt}\selectfont \bfseries(d)}
    \end{overpic}
    \label{fig:6d}}
    
 \caption{(a)-(b) The effects of the bichromatic laser field on $\Delta P_t^{\text{avg}}$ for different actinide nuclei with varying frequency ratios. For $n = 1/2$, the amplitude ratio is $a = 0.5$, while for $n = 2$, the amplitude ratio is $a = 0.3$. (c)-(d) The growth rate of $\Delta P_t^{\text{avg}}$ under bichromatic fields for different actinide nuclei with varying frequency ratios.}
  \label{fig:6}  
\end{figure*}

\subsection{Laser-assisted $\alpha$ decay in bichromatic fields}
The bichromatic fields introduce additional degrees of freedom for controlling the nuclear $\alpha$-decay process compared to the monochromatic case, including parameters such as the frequency and field amplitude ratio of the two lasers, the phase difference, and the temporal delay. Therefore, it is a newly important aspect in understanding the manipulation of the tunneling process, as these parameters may be varied to optimize or modulate the $\alpha$-decay rate. In this work, we systematically investigated the effects of the bichromatic field on the $\alpha$ decay of actinide nuclei. The linearly polarized Gaussian bichromatic laser field can be expressed as
\begin{equation}
E(t) = \Lambda\exp\left(\frac{-t^2}{2\sigma^2}\right)E_{0}\left[\sin(\omega_1 t + \Phi_1) + a\sin(\omega_2 t + \Phi_2)\right] \ ,
\end{equation}
where $\Lambda = \sqrt{\frac{\varepsilon_\text{{m}}}{\varepsilon_\text{{b}}}}$ is the energy normalization factor with $\varepsilon_\text{{m}}$ and $\varepsilon_\text{{b}}$ being the pulse energy for the monochrome and bichromatic cases, respectively. Here, a key parameter is defined as the \emph{frequency ratio} $n = \omega_2/\omega_1$—i.e., the \emph{harmonic order} when $n>1$ and the \emph{subharmonic ratio} when $0<n<1$, $a$ is the amplitude ratio between the second and first frequency components, and $\Phi_1$ and $\Phi_2$ are the phases for the two frequency components $\omega_1$ and $\omega_2$, respectively. Here, $\Phi_2 = n \Phi_1 + \Delta \Phi$ with $\Delta \Phi$ being the phase difference.

Figure ~\ref{fig:5}\subref{fig:5a} and ~\ref{fig:5}\subref{fig:5b} show the phase scan heatmaps $\Delta P_t^{\text{avg}}(\Phi_1, \Delta \Phi)$ of $^{238}$Pu for the frequency ratio $n = 2$ and $n = 3$ with the amplitude ratio $a=0.5$, respectively. From these figures, one sees that the contour lines appear as nearly vertical stripes, indicating that the response depends primarily on $\Phi_{1}$, while $\Delta\Phi$ provides only minor fine-tuning. When $ \Phi_1=\pi/2 $, $\Delta P_t^{\text{avg}}$ shows a blue valley for suppression, and when $\Phi_1 = 3\pi/2 $, $\Delta P_t^{\text{avg}}$ corresponds to a red ridge for enhancement, which is consistent with the case of $\Phi_{\text{CEP}}$ in the few-cycle monochromatic field-assisted $\alpha$ decay. Furthermore, we also find that the maximal $\Delta P_t^{\text{avg}}$ in bichromatic laser field statified the conditions of $\Phi_1 = \frac{3\pi}{2}$ and $\Delta \Phi = \langle \frac{3\pi}{2} - n \Phi_1 \rangle_{2\pi}$, while the minimal one statified the conditions of $\Phi_1 = \frac{\pi}{2}$ and $\Delta \Phi = \langle \frac{\pi}{2} - n \Phi_1 \rangle_{2\pi}$, where the $\langle x \rangle_{2 \pi}$ denotes reduction of $x$ modulo $2\pi$ onto $[0,2\pi)$. These results highlight the capability of the bichromatic laser fields to precisely control the nuclear $\alpha$-decay rate through the phase modulation. This may provide a potential means for important prospective applications in the refined regulation of nuclear battery power.

Besides the phase modulation, we also investigated the effects of the frequency ratio $n$ and the amplitude ratio $a$ of a linearly polarized Gaussian bichromatic field on $\Delta P_t^{\text{avg}}$ for $^{238}$Pu. The pulse width $\tau$ is the same as used in Figs.~\ref{fig:5}\subref{fig:5a} and \subref{fig:5b}. Differently, for $n<1$, the pulse width $\tau$ should be replaced by the $\tau^{'} = \tau/n$, ensuring that the subharmonic component contains at least a few complete carrier cycles within the pulse. The calculated results are shown in Figs.~\ref{fig:5}\subref{fig:5c} and ~\ref{fig:5}\subref{fig:5d} under the maximal and minimal phase modulation conditions, respectively. From these figures, one can see that for $n < 1$, $|\Delta P_t^{\text{avg}}|$ increases with the increase of $a$ and $n$. For $n > 1$, the maximal modifications occur under the conditions of the frequency ratio $n = 2$ with the best amplitude ratio around  $a = 0.3$. After the peak of $n > 2$, $|\Delta P_t^{\text{avg}}|$ increases with the decrease of $a$ for the large $n$-value. In particular, for a fixed parameter $a$, it can be observed that $|\Delta P_t^{\text{avg}}|$ increases with $n$, reaching a peak, after which it decreases as $n$ increases. For the case of $n = 1$, the bichromatic field reduces to a $\Phi_{\text{CEP}}$-controlled monochrome case. Under energy normalization, the calculated results agree with the $\Phi_{\text{CEP}}$-controlled monochrome case and show no dependence on the amplitude ratio.

Based on the phase modulation rule $(\Phi_1, \Delta \Phi)$ and the relationship between the frequency ratio $n$ and amplitude ratio $a$, we quantitatively investigated the effects of the bichromatic laser field on the $\Delta P_t^{\text{avg}}$ for different actinide nuclei with different frequency ratios under the conditions of the maximum suppression and enhancement, respectively. The results are shown in Figs.~\ref{fig:6}\subref{fig:6a} and~\ref{fig:6}\subref{fig:6b}. From these figures, one can see that, compared with the monochrome field $\omega$, the bichromatic laser field can also make the $\Delta P_t^{\text{avg}}$ more significant. Moreover, the postive phase modulation $(\Phi_1 =\frac{3\pi}{2},\Delta \Phi = \langle \frac{3\pi}{2} - n \Phi_1 \rangle_{2\pi})$ can increase $\Delta P_t^{\text{avg}}$, while the negetive phase modulation $(\Phi_1 =\frac{\pi}{2},\Delta \Phi = \langle \frac{\pi}{2} - n \Phi_1 \rangle_{2\pi})$ can desease $\Delta P_t^{\text{avg}}$. Therefore, the $\Delta P_t^{\text{avg}}$ can be enhanced using a bichromatic laser field with the positive phase relation. To make a more intuitive comparison of monochrome and bichromatic cases, the growth rate of $\Delta P_{t}^{avg}$ with the different frequency ratios is shown in Figs.~\ref{fig:6}\subref{fig:6c} and~\ref{fig:6}\subref{fig:6d}. One can observe that the $\omega + 2\omega$ bichromatic field can further increase $|\Delta P_t^{\text{avg}}|$ compared to the $\Phi_{\text{CEP}}$-controlled monochrome case, and this enhancement is entirely due to the phase modulation after energy normalization. With optimized parameters, the bichromatic modulation on $\alpha$ decay is equivalent to increasing the laser intensity more than one order of magnitude. This result proves that it is feasible to use the bichromatic laser field to improve $\Delta P_t^{\text{avg}}$. Moreover, the bichromatic laser field can precisely speed up or delay the half-life of $\alpha$ decay by phase modulation as required in future applications.
\vspace{-1.5pt}
\section{Summary}
\label{section 4}
In summary, we present a quantitative analysis of laser-assisted nuclear $\alpha$ decay for 120 favored actinide emitters from $^{206}$Ac\ –\ $^{259}$Lr within a deformed one-parameter model. The calculations indicate that intense, yet experimentally foreseeable laser fields can modify the $\alpha$-decay penetrability by $0.01$–$0.1\%$ at intensities anticipated at near-term facilities such as ELI–NP and SEL. From the perspective of the nuclear structure, the $N=126$ shell closure remains robust in the presence of the laser field, and the decay energy $Q_\alpha$ governs the sensitivity of the field-induced modification of the tunneling probability. Based on the robustness of the shell effects, we also predicted that $N=142$ is a possible deformed neutron sub-shell. From the laser-driver standpoint, we analyze the dependence of $\Delta P_t^{\text{avg}}$ on the carrier–envelope phase of few-cycle pulses and on the relative phase of bichromatic fields. The results show that the $\omega$-2$\omega$ bichromatic field with the optimized parameters can amplify the time-averaged tunneling modification by $\sim 35$, which is equivalent to increasing the laser intensity by approximately one to two orders of magnitude.

These findings strengthen the case that intense laser fields can directly influence nuclear dynamics via barrier tunneling, providing a concrete and quantifiable pathway for laser–nucleus coupling beyond atomic-electronic channels. Observation of such effects in the laboratory will, however, hinge on judicious isotope selection—favoring long-lived $\alpha$ emitters with low decay energy in the actinide region—and continued advances in high-intensity, phase-controlled laser technology. Actinide nuclei thus constitute a practical platform for laser control of $\alpha$ decay, with potential applications ranging from tunable nuclear batteries to strategies for radiotoxic-waste transmutation. Accordingly, the present study serves as a theory-driven extrapolation and benchmark for future theoretical and experimental efforts on laser-controlled nuclear $\alpha$ decay.
\begin{acknowledgments}
This work was supported by the National Natural Science Foundation of China (Grant No.12135009, 12375244), the Natural Science Foundation of Hunan Province of China (Grant No.2025JJ30002), and the Hunan Provincial Innovation Foundation for Postgraduate (Grant No.CX20230008).
\end{acknowledgments}

\appendix 
\section{Supplementary Data} 
\label{sec:appendix}         
\onecolumngrid            
{\setlength\LTcapwidth{\textwidth}
\begin{longtable*}{@{\extracolsep{\fill}} l c c c c c c c c c @{}}
\caption{The influence of a laser pulse on the favored $\alpha$ decay of actinide nuclei with a laser intensity of $I_0 = 10^{25}$ W/cm$^2$. The $*$ represent the possible $\alpha$ emitter candidates.}\\
\hline
Nucleus & $Q_{\alpha}$ (MeV) & $\beta_2$ & $\beta_4$ & $\beta_6$ & $\log_{10} T_{1/2}^{\rm exp}$ & $\log_{10} T_{1/2}^{\rm cal}$ & $\delta P_f$ & $\delta P_t$ & $\delta T$ \\
\hline
\addlinespace[0.25em]
\endfirsthead

\multicolumn{10}{l}{\textbf{Table \thetable\ (continued)}}\\
\hline
Nucleus & $Q_{\alpha}$ (MeV) & $\beta_2$ & $\beta_4$ & $\beta_6$ 
& $\log_{10} T_{1/2}^{\rm exp}$ & $\log_{10} T_{1/2}^{\rm cal}$ 
& $\delta P_f$ & $\delta P_t$ & $\delta T$ \\
\hline
\addlinespace[0.25em]
\endhead

\hline
\multicolumn{10}{r}{\textit{Continued on next page}}\\
\endfoot

\bottomrule
\endlastfoot
$^{	206	}$Ac	&	7.953 	&	-0.217 	&	0.006 	&	0.001 	&	-1.602 	&	-1.569 	&	7.494 	$\times$ 10$^{	-6	}$ &	1.300 	$\times$ 10$^{	-3	}$ &	-1.300 	$\times$ 10$^{	-3	}$\\
$^{	207	}$Ac	&	7.849 	&	-0.207 	&	-0.007 	&	0.003 	&	-1.509 	&	-1.231 	&	7.549 	$\times$ 10$^{	-6	}$ &	1.340 	$\times$ 10$^{	-3	}$ &	-1.350 	$\times$ 10$^{	-3	}$\\
$^{	208	}$Ac	&	7.728 	&	-0.197 	&	-0.009 	&	0.003 	&	-1.013 	&	-0.842 	&	7.605 	$\times$ 10$^{	-6	}$ &	1.390 	$\times$ 10$^{	-3	}$ &	-1.400 	$\times$ 10$^{	-3	}$\\
$^{	209	}$Ac	&	7.725 	&	-0.125 	&	0.018 	&	0.008 	&	-1.027 	&	-0.742 	&	7.813 	$\times$ 10$^{	-6	}$ &	1.400 	$\times$ 10$^{	-3	}$ &	-1.400 	$\times$ 10$^{	-3	}$\\
$^{	211	}$Ac	&	7.564 	&	-0.115 	&	0.017 	&	0.008 	&	-0.672 	&	-0.233 	&	7.906 	$\times$ 10$^{	-6	}$ &	1.480 	$\times$ 10$^{	-3	}$ &	-1.490 	$\times$ 10$^{	-3	}$\\
$^{	212	}$Ac	&	7.540 	&	-0.094 	&	0.003 	&	0.000 	&	-0.048 	&	-0.136 	&	7.985 	$\times$ 10$^{	-6	}$ &	1.500 	$\times$ 10$^{	-3	}$ &	-1.500 	$\times$ 10$^{	-3	}$\\
$^{	213	}$Ac	&	7.498 	&	-0.084 	&	0.002 	&	0.000 	&	-0.132 	&	0.000 	&	8.043 	$\times$ 10$^{	-6	}$ &	1.530 	$\times$ 10$^{	-3	}$ &	-1.540 	$\times$ 10$^{	-3	}$\\
$^{	215	}$Ac	&	7.746 	&	0.011 	&	0.000 	&	0.000 	&	-0.767 	&	-0.811 	&	8.246 	$\times$ 10$^{	-6	}$ &	1.460 	$\times$ 10$^{	-3	}$ &	-1.470 	$\times$ 10$^{	-3	}$\\
$^{	217	}$Ac	&	9.831 	&	0.000 	&	0.000 	&	0.000 	&	-7.161 	&	-6.438 	&	8.809 	$\times$ 10$^{	-6	}$ &	9.300 	$\times$ 10$^{	-4	}$ &	-9.380 	$\times$ 10$^{	-4	}$\\
$^{	218	}$Ac	&	9.383 	&	-0.010 	&	0.012 	&	0.000 	&	-6.000 	&	-5.437 	&	8.761 	$\times$ 10$^{	-6	}$ &	1.020 	$\times$ 10$^{	-3	}$ &	-1.030 	$\times$ 10$^{	-3	}$\\
$^{	219	}$Ac	&	8.827 	&	0.000 	&	0.000 	&	0.000 	&	-5.027 	&	-4.068 	&	8.693 	$\times$ 10$^{	-6	}$ &	1.170 	$\times$ 10$^{	-3	}$ &	-1.180 	$\times$ 10$^{	-3	}$\\
$^{	221	}$Ac	&	7.790 	&	0.079 	&	0.054 	&	0.021 	&	-1.284 	&	-1.254 	&	8.524 	$\times$ 10$^{	-6	}$ &	1.550 	$\times$ 10$^{	-3	}$ &	-1.560 	$\times$ 10$^{	-3	}$\\
$^{	222	}$Ac	&	7.138 	&	0.090 	&	0.055 	&	0.012 	&	0.703 	&	0.946 	&	8.432 	$\times$ 10$^{	-6	}$ &	1.870 	$\times$ 10$^{	-3	}$ &	-1.870 	$\times$ 10$^{	-3	}$\\
$^{	227	}$Ac	&	5.042 	&	0.132 	&	0.083 	&	0.008 	&	10.696 	&	10.659 	&	8.203 	$\times$ 10$^{	-6	}$ &	4.060 	$\times$ 10$^{	-3	}$ &	-4.050 	$\times$ 10$^{	-3	}$\\
$^{	208	}$Th	&	8.204 	&	-0.207 	&	-0.007 	&	0.003 	&	-2.620 	&	-1.944 	&	7.530 	$\times$ 10$^{	-6	}$ &	1.240 	$\times$ 10$^{	-3	}$ &	-1.240 	$\times$ 10$^{	-3	}$\\
$^{	210	}$Th	&	8.069 	&	-0.125 	&	0.018 	&	0.008 	&	-1.796 	&	-1.445 	&	7.791 	$\times$ 10$^{	-6	}$ &	1.290 	$\times$ 10$^{	-3	}$ &	-1.300 	$\times$ 10$^{	-3	}$\\
$^{	211	}$Th	&	7.945 	&	-0.125 	&	0.018 	&	0.008 	&	-1.319 	&	-1.080 	&	7.822 	$\times$ 10$^{	-6	}$ &	1.340 	$\times$ 10$^{	-3	}$ &	-1.350 	$\times$ 10$^{	-3	}$\\
$^{	212	}$Th	&	7.958 	&	-0.125 	&	0.018 	&	0.008 	&	-1.499 	&	-1.140 	&	7.879 	$\times$ 10$^{	-6	}$ &	1.350 	$\times$ 10$^{	-3	}$ &	-1.360 	$\times$ 10$^{	-3	}$\\
$^{	213	}$Th	&	7.837 	&	-0.104 	&	0.004 	&	0.009 	&	-0.842 	&	-0.737 	&	7.942 	$\times$ 10$^{	-6	}$ &	1.400 	$\times$ 10$^{	-3	}$ &	-1.410 	$\times$ 10$^{	-3	}$\\
$^{	214	}$Th	&	7.827 	&	-0.084 	&	-0.009 	&	0.001 	&	-1.060 	&	-0.696 	&	8.018 	$\times$ 10$^{	-6	}$ &	1.420 	$\times$ 10$^{	-3	}$ &	-1.420 	$\times$ 10$^{	-3	}$\\
$^{	216	}$Th	&	8.073 	&	-0.053 	&	-0.011 	&	0.001 	&	-1.580 	&	-1.478 	&	8.202 	$\times$ 10$^{	-6	}$ &	1.360 	$\times$ 10$^{	-3	}$ &	-1.360 	$\times$ 10$^{	-3	}$\\
$^{	218	}$Th	&	9.849 	&	0.000 	&	0.000 	&	0.000 	&	-6.914 	&	-6.161 	&	8.712 	$\times$ 10$^{	-6	}$ &	9.320 	$\times$ 10$^{	-4	}$ &	-9.400 	$\times$ 10$^{	-4	}$\\
$^{	219	}$Th	&	9.503 	&	-0.021 	&	0.012 	&	0.000 	&	-5.990 	&	-5.394 	&	8.687 	$\times$ 10$^{	-6	}$ &	1.010 	$\times$ 10$^{	-3	}$ &	-1.020 	$\times$ 10$^{	-3	}$\\
$^{	220	}$Th	&	8.974 	&	0.000 	&	0.000 	&	0.000 	&	-4.991 	&	-4.104 	&	8.629 	$\times$ 10$^{	-6	}$ &	1.140 	$\times$ 10$^{	-3	}$ &	-1.150 	$\times$ 10$^{	-3	}$\\
$^{	222	}$Th	&	8.132 	&	0.078 	&	0.054 	&	0.010 	&	-2.650 	&	-1.918 	&	8.505 	$\times$ 10$^{	-6	}$ &	1.430 	$\times$ 10$^{	-3	}$ &	-1.440 	$\times$ 10$^{	-3	}$\\
$^{	224	}$Th	&	7.299 	&	0.111 	&	0.081 	&	0.026 	&	0.017 	&	0.541 	&	8.385 	$\times$ 10$^{	-6	}$ &	1.860 	$\times$ 10$^{	-3	}$ &	-1.860 	$\times$ 10$^{	-3	}$\\
$^{	226	}$Th	&	6.453 	&	0.122 	&	0.082 	&	0.018 	&	3.265 	&	3.815 	&	8.303 	$\times$ 10$^{	-6	}$ &	2.420 	$\times$ 10$^{	-3	}$ &	-2.420 	$\times$ 10$^{	-3	}$\\
$^{	228	}$Th	&	5.520 	&	0.143 	&	0.084 	&	0.009 	&	7.781 	&	8.286 	&	8.190 	$\times$ 10$^{	-6	}$ &	3.400 	$\times$ 10$^{	-3	}$ &	-3.400 	$\times$ 10$^{	-3	}$\\
$^{	230	}$Th	&	4.770 	&	0.164 	&	0.098 	&	0.010 	&	12.376 	&	12.710 	&	8.098 	$\times$ 10$^{	-6	}$ &	4.740 	$\times$ 10$^{	-3	}$ &	-4.730 	$\times$ 10$^{	-3	}$\\
$^{	232	}$Th	&	4.082 	&	0.174 	&	0.099 	&	0.009 	&	17.645 	&	18.002 	&	8.046 	$\times$ 10$^{	-6	}$ &	6.700 	$\times$ 10$^{	-3	}$ &	-6.670 	$\times$ 10$^{	-3	}$\\
$^{	210	}$Pa$^{*}$	&	8.445 	&	-0.207	&	-0.018	&	0.006	&	-	&	-2.297 	&	7.538 	$\times$ 10$^{	-6	}$ &	1.190 	$\times$ 10$^{	-3	}$ &	-1.190 	$\times$ 10$^{	-3	}$\\      
$^{	211	}$Pa	&	8.475 	&	-0.197 	&	-0.020 	&	0.005 	&	-2.222 	&	-2.382 	&	7.622 	$\times$ 10$^{	-6	}$ &	1.190 	$\times$ 10$^{	-3	}$ &	-1.190 	$\times$ 10$^{	-3	}$\\
$^{	212	}$Pa	&	8.415 	&	-0.125 	&	0.018 	&	0.008 	&	-2.237 	&	-2.126 	&	7.825 	$\times$ 10$^{	-6	}$ &	1.210 	$\times$ 10$^{	-3	}$ &	-1.210 	$\times$ 10$^{	-3	}$\\
$^{	213	}$Pa	&	8.385 	&	-0.125 	&	0.018 	&	0.008 	&	-2.131 	&	-2.058 	&	7.874 	$\times$ 10$^{	-6	}$ &	1.230 	$\times$ 10$^{	-3	}$ &	-1.240 	$\times$ 10$^{	-3	}$\\
$^{	214	}$Pa	&	8.275 	&	-0.115 	&	0.005 	&	0.009 	&	-1.770 	&	-1.731 	&	7.923 	$\times$ 10$^{	-6	}$ &	1.270 	$\times$ 10$^{	-3	}$ &	-1.280 	$\times$ 10$^{	-3	}$\\
$^{	215	}$Pa	&	8.235 	&	-0.094 	&	0.004 	&	0.009 	&	-1.854 	&	-1.602 	&	7.998 	$\times$ 10$^{	-6	}$ &	1.290 	$\times$ 10$^{	-3	}$ &	-1.300 	$\times$ 10$^{	-3	}$\\
$^{	217	}$Pa	&	8.489 	&	-0.063 	&	-0.010 	&	-0.009 	&	-2.420 	&	-2.349 	&	8.188 	$\times$ 10$^{	-6	}$ &	1.240 	$\times$ 10$^{	-3	}$ &	-1.240 	$\times$ 10$^{	-3	}$\\
$^{	219	}$Pa	&	10.124 	&	0.000 	&	0.000 	&	0.000 	&	-7.252 	&	-6.448 	&	8.676 	$\times$ 10$^{	-6	}$ &	8.890 	$\times$ 10$^{	-4	}$ &	-8.970 	$\times$ 10$^{	-4	}$\\
$^{	220	}$Pa	&	9.703 	&	-0.021 	&	0.012 	&	0.000 	&	-6.071 	&	-5.535 	&	8.634 	$\times$ 10$^{	-6	}$ &	9.760 	$\times$ 10$^{	-4	}$ &	-9.840 	$\times$ 10$^{	-4	}$\\
$^{	221	}$Pa	&	9.243 	&	0.000 	&	0.000 	&	0.000 	&	-5.229 	&	-4.447 	&	8.592 	$\times$ 10$^{	-6	}$ &	1.080 	$\times$ 10$^{	-3	}$ &	-1.090 	$\times$ 10$^{	-3	}$\\
$^{	223	}$Pa	&	8.345 	&	0.090 	&	0.055 	&	0.012 	&	-2.276 	&	-2.217 	&	8.446 	$\times$ 10$^{	-6	}$ &	1.370 	$\times$ 10$^{	-3	}$ &	-1.380 	$\times$ 10$^{	-3	}$\\
$^{	226	}$Pa	&	6.987 	&	0.122 	&	0.082 	&	0.018 	&	2.161 	&	2.072 	&	8.271 	$\times$ 10$^{	-6	}$ &	2.060 	$\times$ 10$^{	-3	}$ &	-2.070 	$\times$ 10$^{	-3	}$\\
$^{	227	}$Pa	&	6.580 	&	0.133 	&	0.083 	&	0.019 	&	3.431 	&	3.651 	&	8.225 	$\times$ 10$^{	-6	}$ &	2.360 	$\times$ 10$^{	-3	}$ &	-2.360 	$\times$ 10$^{	-3	}$\\
$^{	231	}$Pa	&	5.150 	&	0.164 	&	0.098 	&	0.010 	&	12.013 	&	10.739 	&	8.086 	$\times$ 10$^{	-6	}$ &	4.070 	$\times$ 10$^{	-3	}$ &	-4.070 	$\times$ 10$^{	-3	}$\\
$^{	218	}$U	&	8.775 	&	-0.063 	&	-0.022 	&	-0.008 	&	-3.451 	&	-2.796 	&	8.153 	$\times$ 10$^{	-6	}$ &	1.170 	$\times$ 10$^{	-3	}$ &	-1.170 	$\times$ 10$^{	-3	}$\\  
$^{	220	}$U	&	10.286 	&	0.000 	&	0.000 	&	0.000 	&	-6.180 	&	-6.484 	&	8.614 	$\times$ 10$^{	-6	}$ &	8.680 	$\times$ 10$^{	-4	}$ &	-8.760 	$\times$ 10$^{	-4	}$\\  
$^{	221	}$U	$^{*}$&	9.889 	&	-0.021 	&	0.012 	&	0.000 	&	-	&	-5.639 	&	8.579 	$\times$ 10$^{	-6	}$ &	9.450 	$\times$ 10$^{	-4	}$ &	-9.530 	$\times$ 10$^{	-4	}$\\      
$^{	222	}$U	&	9.478 	&	0.000 	&	0.000 	&	0.000 	&	-5.328 	&	-4.691 	&	8.548 	$\times$ 10$^{	-6	}$ &	1.040 	$\times$ 10$^{	-3	}$ &	-1.040 	$\times$ 10$^{	-3	}$\\  
$^{	224	}$U	&	8.628 	&	0.090 	&	0.055 	&	0.012 	&	-3.402 	&	-2.670 	&	8.413 	$\times$ 10$^{	-6	}$ &	1.300 	$\times$ 10$^{	-3	}$ &	-1.300 	$\times$ 10$^{	-3	}$\\  
$^{	226	}$U	&	7.701 	&	0.111 	&	0.069 	&	0.015 	&	-0.570 	&	0.025 	&	8.297 	$\times$ 10$^{	-6	}$ &	1.680 	$\times$ 10$^{	-3	}$ &	-1.680 	$\times$ 10$^{	-3	}$\\  
$^{	228	}$U	&	6.799 	&	0.144 	&	0.084 	&	0.010 	&	2.748 	&	3.151 	&	8.164 	$\times$ 10$^{	-6	}$ &	2.220 	$\times$ 10$^{	-3	}$ &	-2.230 	$\times$ 10$^{	-3	}$\\  
$^{	229	}$U	&	6.476 	&	0.143 	&	0.084 	&	0.009 	&	4.239 	&	4.512 	&	8.153 	$\times$ 10$^{	-6	}$ &	2.470 	$\times$ 10$^{	-3	}$ &	-2.470 	$\times$ 10$^{	-3	}$\\  
$^{	230	}$U	&	5.992 	&	0.154 	&	0.085 	&	0.010 	&	6.243 	&	6.704 	&	8.092 	$\times$ 10$^{	-6	}$ &	2.930 	$\times$ 10$^{	-3	}$ &	-2.930 	$\times$ 10$^{	-3	}$\\  
$^{	232	}$U	&	5.414 	&	0.174 	&	0.100 	&	0.011 	&	9.337 	&	9.585 	&	8.032 	$\times$ 10$^{	-6	}$ &	3.710 	$\times$ 10$^{	-3	}$ &	-3.710 	$\times$ 10$^{	-3	}$\\  
$^{	233	}$U	&	4.909 	&	0.184 	&	0.113 	&	0.021 	&	12.701 	&	12.534 	&	7.958 	$\times$ 10$^{	-6	}$ &	4.660 	$\times$ 10$^{	-3	}$ &	-4.640 	$\times$ 10$^{	-3	}$\\  
$^{	234	}$U	&	4.858 	&	0.195 	&	0.114 	&	0.022 	&	12.889 	&	12.788 	&	7.970 	$\times$ 10$^{	-6	}$ &	4.800 	$\times$ 10$^{	-3	}$ &	-4.790 	$\times$ 10$^{	-3	}$\\  
$^{	236	}$U	&	4.573 	&	0.205 	&	0.103 	&	0.010 	&	14.869 	&	14.912 	&	7.992 	$\times$ 10$^{	-6	}$ &	5.500 	$\times$ 10$^{	-3	}$ &	-5.480 	$\times$ 10$^{	-3	}$\\  
$^{	238	}$U	&	4.270 	&	0.226 	&	0.095 	&	-0.001 	&	17.149 	&	16.257 	&	7.975 	$\times$ 10$^{	-6	}$ &	6.450 	$\times$ 10$^{	-3	}$ &	-6.420 	$\times$ 10$^{	-3	}$\\  
$^{	223	}$Np	&	9.655 	&	0.055 	&	0.027 	&	0.005 	&	-5.602 	&	-4.828 	&	8.470 	$\times$ 10$^{	-6	}$ &	1.010 	$\times$ 10$^{	-3	}$ &	-1.020 	$\times$ 10$^{	-3	}$\\
$^{	225	}$Np	&	8.825 	&	0.100 	&	0.056 	&	0.013 	&	-2.220 	&	-2.895 	&	8.352 	$\times$ 10$^{	-6	}$ &	1.250 	$\times$ 10$^{	-3	}$ &	-1.260 	$\times$ 10$^{	-3	}$\\
$^{	226	}$Np	&	8.335 	&	0.110 	&	0.068 	&	0.014 	&	-1.456 	&	-1.583 	&	8.286 	$\times$ 10$^{	-6	}$ &	1.430 	$\times$ 10$^{	-3	}$ &	-1.430 	$\times$ 10$^{	-3	}$\\
$^{	233	}$Np	&	5.628 	&	0.185 	&	0.126 	&	0.024 	&	8.492 	&	8.530 	&	7.947 	$\times$ 10$^{	-6	}$ &	3.520 	$\times$ 10$^{	-3	}$ &	-3.520 	$\times$ 10$^{	-3	}$\\
$^{	230	}$Pu	&	7.178 	&	0.143 	&	0.084 	&	0.009 	&	2.021 	&	2.488 	&	8.070 	$\times$ 10$^{	-6	}$ &	2.070 	$\times$ 10$^{	-3	}$ &	-2.070 	$\times$ 10$^{	-3	}$\\
$^{	231	}$Pu	&	6.839 	&	0.153 	&	0.085 	&	-0.001 	&	3.582 	&	3.811 	&	8.037 	$\times$ 10$^{	-6	}$ &	2.240 	$\times$ 10$^{	-3	}$ &	-2.240 	$\times$ 10$^{	-3	}$\\
$^{	232	}$Pu	&	6.716 	&	0.186 	&	0.126 	&	0.035 	&	4.005 	&	3.672 	&	7.971 	$\times$ 10$^{	-6	}$ &	2.460 	$\times$ 10$^{	-3	}$ &	-2.460 	$\times$ 10$^{	-3	}$\\
$^{	234	}$Pu	&	6.310 	&	0.185 	&	0.126 	&	0.024 	&	5.723 	&	5.521 	&	7.990 	$\times$ 10$^{	-6	}$ &	2.820 	$\times$ 10$^{	-3	}$ &	-2.820 	$\times$ 10$^{	-3	}$\\
$^{	235	}$Pu	&	5.951 	&	0.195 	&	0.114 	&	0.022 	&	7.723 	&	7.309 	&	7.961 	$\times$ 10$^{	-6	}$ &	3.180 	$\times$ 10$^{	-3	}$ &	-3.180 	$\times$ 10$^{	-3	}$\\
$^{	236	}$Pu	&	5.867 	&	0.206 	&	0.116 	&	0.013 	&	7.955 	&	7.711 	&	7.964 	$\times$ 10$^{	-6	}$ &	3.310 	$\times$ 10$^{	-3	}$ &	-3.300 	$\times$ 10$^{	-3	}$\\
$^{	238	}$Pu	&	5.593 	&	0.215 	&	0.106 	&	0.001 	&	9.442 	&	9.261 	&	7.987 	$\times$ 10$^{	-6	}$ &	3.680 	$\times$ 10$^{	-3	}$ &	-3.680 	$\times$ 10$^{	-3	}$\\
$^{	239	}$Pu	&	5.244 	&	0.215 	&	0.106 	&	0.001 	&	11.881 	&	11.320 	&	7.968 	$\times$ 10$^{	-6	}$ &	4.230 	$\times$ 10$^{	-3	}$ &	-4.220 	$\times$ 10$^{	-3	}$\\
$^{	240	}$Pu	&	5.256 	&	0.226 	&	0.108 	&	-0.009 	&	11.316 	&	11.207 	&	7.980 	$\times$ 10$^{	-6	}$ &	4.250 	$\times$ 10$^{	-3	}$ &	-4.240 	$\times$ 10$^{	-3	}$\\
$^{	242	}$Pu	&	4.984 	&	0.236 	&	0.098 	&	-0.021 	&	13.073 	&	13.008 	&	7.991 	$\times$ 10$^{	-6	}$ &	4.800 	$\times$ 10$^{	-3	}$ &	-4.780 	$\times$ 10$^{	-3	}$\\
$^{	244	}$Pu	&	4.666 	&	0.237 	&	0.086 	&	-0.024 	&	15.410 	&	15.321 	&	8.021 	$\times$ 10$^{	-6	}$ &	5.570 	$\times$ 10$^{	-3	}$ &	-5.540 	$\times$ 10$^{	-3	}$\\
$^{	236	}$Am	&	6.255 	&	0.206 	&	0.116 	&	0.013 	&	6.732 	&	6.240 	&	7.910 	$\times$ 10$^{	-6	}$ &	2.900 	$\times$ 10$^{	-3	}$ &	-2.900 	$\times$ 10$^{	-3	}$\\
$^{	240	}$Am	&	5.469 	&	0.226 	&	0.108 	&	-0.009 	&	10.983 	&	10.433 	&	7.897 	$\times$ 10$^{	-6	}$ &	3.910 	$\times$ 10$^{	-3	}$ &	-3.900 	$\times$ 10$^{	-3	}$\\
$^{	233	}$Cm	&	7.475 	&	0.195 	&	0.114 	&	0.022 	&	2.107 	&	1.646 	&	7.894 	$\times$ 10$^{	-6	}$ &	1.980 	$\times$ 10$^{	-3	}$ &	-1.980 	$\times$ 10$^{	-3	}$\\
$^{	234	}$Cm	&	7.365 	&	0.195 	&	0.114 	&	0.022 	&	2.285 	&	2.032 	&	7.920 	$\times$ 10$^{	-6	}$ &	2.060 	$\times$ 10$^{	-3	}$ &	-2.060 	$\times$ 10$^{	-3	}$\\
$^{	236	}$Cm	&	7.067 	&	0.206 	&	0.116 	&	0.013 	&	3.351 	&	3.131 	&	7.932 	$\times$ 10$^{	-6	}$ &	2.270 	$\times$ 10$^{	-3	}$ &	-2.270 	$\times$ 10$^{	-3	}$\\
$^{	238	}$Cm	&	6.670 	&	0.216 	&	0.119 	&	0.004 	&	5.314 	&	4.734 	&	7.922 	$\times$ 10$^{	-6	}$ &	2.590 	$\times$ 10$^{	-3	}$ &	-2.590 	$\times$ 10$^{	-3	}$\\
$^{	240	}$Cm	&	6.398 	&	0.215 	&	0.106 	&	0.001 	&	6.419 	&	6.041 	&	7.974 	$\times$ 10$^{	-6	}$ &	2.840 	$\times$ 10$^{	-3	}$ &	-2.840 	$\times$ 10$^{	-3	}$\\
$^{	242	}$Cm	&	6.216 	&	0.226 	&	0.095 	&	-0.012 	&	7.148 	&	6.955 	&	8.003 	$\times$ 10$^{	-6	}$ &	3.040 	$\times$ 10$^{	-3	}$ &	-3.040 	$\times$ 10$^{	-3	}$\\
$^{	244	}$Cm	&	5.902 	&	0.237 	&	0.086 	&	-0.024 	&	8.757 	&	8.571 	&	8.002 	$\times$ 10$^{	-6	}$ &	3.420 	$\times$ 10$^{	-3	}$ &	-3.410 	$\times$ 10$^{	-3	}$\\
$^{	246	}$Cm	&	5.475 	&	0.237 	&	0.073 	&	-0.027 	&	11.172 	&	11.029 	&	8.015 	$\times$ 10$^{	-6	}$ &	4.030 	$\times$ 10$^{	-3	}$ &	-4.020 	$\times$ 10$^{	-3	}$\\
$^{	248	}$Cm	&	5.162 	&	0.237 	&	0.061 	&	-0.030 	&	13.079 	&	13.025 	&	8.045 	$\times$ 10$^{	-6	}$ &	4.600 	$\times$ 10$^{	-3	}$ &	-4.590 	$\times$ 10$^{	-3	}$\\
$^{	238	}$Cf	&	8.133 	&	0.205 	&	0.103 	&	0.010 	&	-0.076 	&	0.164 	&	7.980 	$\times$ 10$^{	-6	}$ &	1.720 	$\times$ 10$^{	-3	}$ &	-1.730 	$\times$ 10$^{	-3	}$\\
$^{	239	}$Cf	&	7.765 	&	0.215 	&	0.106 	&	0.001 	&	1.632 	&	1.372 	&	7.930 	$\times$ 10$^{	-6	}$ &	1.880 	$\times$ 10$^{	-3	}$ &	-1.880 	$\times$ 10$^{	-3	}$\\
$^{	240	}$Cf	&	7.711 	&	0.215 	&	0.106 	&	0.001 	&	1.612 	&	1.543 	&	7.966 	$\times$ 10$^{	-6	}$ &	1.950 	$\times$ 10$^{	-3	}$ &	-1.950 	$\times$ 10$^{	-3	}$\\
$^{	241	}$Cf	&	7.455 	&	0.215 	&	0.106 	&	0.001 	&	2.970 	&	2.460 	&	7.964 	$\times$ 10$^{	-6	}$ &	2.100 	$\times$ 10$^{	-3	}$ &	-2.100 	$\times$ 10$^{	-3	}$\\
$^{	242	}$Cf	&	7.517 	&	0.226 	&	0.095 	&	-0.012 	&	2.534 	&	2.288 	&	7.995 	$\times$ 10$^{	-6	}$ &	2.070 	$\times$ 10$^{	-3	}$ &	-2.070 	$\times$ 10$^{	-3	}$\\
$^{	244	}$Cf	&	7.329 	&	0.237 	&	0.085 	&	-0.014 	&	3.190 	&	2.975 	&	8.024 	$\times$ 10$^{	-6	}$ &	2.200 	$\times$ 10$^{	-3	}$ &	-2.200 	$\times$ 10$^{	-3	}$\\
$^{	245	}$Cf	&	7.258 	&	0.237 	&	0.086 	&	-0.024 	&	3.884 	&	3.264 	&	8.049 	$\times$ 10$^{	-6	}$ &	2.260 	$\times$ 10$^{	-3	}$ &	-2.260 	$\times$ 10$^{	-3	}$\\
$^{	246	}$Cf	&	6.862 	&	0.237 	&	0.086 	&	-0.024 	&	5.109 	&	4.871 	&	8.019 	$\times$ 10$^{	-6	}$ &	2.550 	$\times$ 10$^{	-3	}$ &	-2.550 	$\times$ 10$^{	-3	}$\\
$^{	248	}$Cf	&	6.361 	&	0.249 	&	0.063 	&	-0.029 	&	7.460 	&	7.195 	&	7.985 	$\times$ 10$^{	-6	}$ &	3.000 	$\times$ 10$^{	-3	}$ &	-3.000 	$\times$ 10$^{	-3	}$\\
$^{	250	}$Cf	&	6.128 	&	0.249 	&	0.051 	&	-0.032 	&	8.616 	&	8.381 	&	8.027 	$\times$ 10$^{	-6	}$ &	3.270 	$\times$ 10$^{	-3	}$ &	-3.270 	$\times$ 10$^{	-3	}$\\
$^{	252	}$Cf	&	6.217 	&	0.250 	&	0.039 	&	-0.035 	&	7.935 	&	7.937 	&	8.121 	$\times$ 10$^{	-6	}$ &	3.210 	$\times$ 10$^{	-3	}$ &	-3.210 	$\times$ 10$^{	-3	}$\\
$^{	253	}$Cf	&	6.078 	&	0.238 	&	0.036 	&	-0.025 	&	8.690 	&	8.686 	&	8.180 	$\times$ 10$^{	-6	}$ &	3.380 	$\times$ 10$^{	-3	}$ &	-3.380 	$\times$ 10$^{	-3	}$\\
$^{	254	}$Cf	&	5.926 	&	0.250 	&	0.026 	&	-0.027 	&	9.224 	&	9.459 	&	8.152 	$\times$ 10$^{	-6	}$ &	3.590 	$\times$ 10$^{	-3	}$ &	-3.590 	$\times$ 10$^{	-3	}$\\
$^{	241	}$Es	&	8.255 	&	0.226 	&	0.095 	&	-0.001 	&	0.708 	&	0.099 	&	7.967 	$\times$ 10$^{	-6	}$ &	1.710 	$\times$ 10$^{	-3	}$ &	-1.710 	$\times$ 10$^{	-3	}$\\
$^{	243	}$Es	&	8.075 	&	0.226 	&	0.095 	&	-0.012 	&	1.555 	&	0.702 	&	8.019 	$\times$ 10$^{	-6	}$ &	1.800 	$\times$ 10$^{	-3	}$ &	-1.810 	$\times$ 10$^{	-3	}$\\
$^{	245	}$Es	&	7.865 	&	0.237 	&	0.086 	&	-0.024 	&	2.128 	&	1.426 	&	8.039 	$\times$ 10$^{	-6	}$ &	1.920 	$\times$ 10$^{	-3	}$ &	-1.930 	$\times$ 10$^{	-3	}$\\
$^{	246	}$Es	&	7.495 	&	0.237 	&	0.086 	&	-0.024 	&	3.657 	&	2.750 	&	8.014 	$\times$ 10$^{	-6	}$ &	2.130 	$\times$ 10$^{	-3	}$ &	-2.130 	$\times$ 10$^{	-3	}$\\
$^{	247	}$Es	&	7.443 	&	0.237 	&	0.073 	&	-0.027 	&	3.591 	&	3.005 	&	8.052 	$\times$ 10$^{	-6	}$ &	2.170 	$\times$ 10$^{	-3	}$ &	-2.170 	$\times$ 10$^{	-3	}$\\
$^{	251	}$Es	&	6.597 	&	0.249 	&	0.051 	&	-0.032 	&	7.359 	&	6.541 	&	8.034 	$\times$ 10$^{	-6	}$ &	2.830 	$\times$ 10$^{	-3	}$ &	-2.830 	$\times$ 10$^{	-3	}$\\
$^{	253	}$Es	&	6.739 	&	0.250 	&	0.039 	&	-0.035 	&	6.248 	&	5.903 	&	8.139 	$\times$ 10$^{	-6	}$ &	2.740 	$\times$ 10$^{	-3	}$ &	-2.740 	$\times$ 10$^{	-3	}$\\
$^{	255	}$Es	&	6.400 	&	0.250 	&	0.026 	&	-0.027 	&	7.631 	&	7.486 	&	8.161 	$\times$ 10$^{	-6	}$ &	3.090 	$\times$ 10$^{	-3	}$ &	-3.090 	$\times$ 10$^{	-3	}$\\
$^{	244	}$Fm	&	8.546 	&	0.237 	&	0.085 	&	-0.014 	&	-0.506 	&	-0.436 	&	8.004 	$\times$ 10$^{	-6	}$ &	1.620 	$\times$ 10$^{	-3	}$ &	-1.620 	$\times$ 10$^{	-3	}$\\
$^{	246	}$Fm	&	8.379 	&	0.237 	&	0.073 	&	-0.027 	&	0.218 	&	0.154 	&	8.061 	$\times$ 10$^{	-6	}$ &	1.690 	$\times$ 10$^{	-3	}$ &	-1.700 	$\times$ 10$^{	-3	}$\\
$^{	248	}$Fm	&	7.995 	&	0.249 	&	0.063 	&	-0.029 	&	1.538 	&	1.386 	&	8.043 	$\times$ 10$^{	-6	}$ &	1.890 	$\times$ 10$^{	-3	}$ &	-1.890 	$\times$ 10$^{	-3	}$\\
$^{	250	}$Fm	&	7.557 	&	0.249 	&	0.051 	&	-0.032 	&	3.270 	&	2.984 	&	8.048 	$\times$ 10$^{	-6	}$ &	2.130 	$\times$ 10$^{	-3	}$ &	-2.140 	$\times$ 10$^{	-3	}$\\
$^{	252	}$Fm	&	7.154 	&	0.250 	&	0.039 	&	-0.035 	&	4.961 	&	4.579 	&	8.054 	$\times$ 10$^{	-6	}$ &	2.410 	$\times$ 10$^{	-3	}$ &	-2.410 	$\times$ 10$^{	-3	}$\\
$^{	254	}$Fm	&	7.307 	&	0.250 	&	0.027 	&	-0.037 	&	4.067 	&	3.962 	&	8.164 	$\times$ 10$^{	-6	}$ &	2.340 	$\times$ 10$^{	-3	}$ &	-2.340 	$\times$ 10$^{	-3	}$\\
$^{	256	}$Fm	&	7.025 	&	0.251 	&	0.014 	&	-0.030 	&	5.064 	&	5.116 	&	8.191 	$\times$ 10$^{	-6	}$ &	2.560 	$\times$ 10$^{	-3	}$ &	-2.560 	$\times$ 10$^{	-3	}$\\
$^{	251	}$No	&	8.755 	&	0.249 	&	0.051 	&	-0.032 	&	-0.017 	&	-0.310 	&	8.072 	$\times$ 10$^{	-6	}$ &	1.600 	$\times$ 10$^{	-3	}$ &	-1.610 	$\times$ 10$^{	-3	}$\\
$^{	252	}$No	&	8.548 	&	0.249 	&	0.051 	&	-0.032 	&	0.562 	&	0.313 	&	8.076 	$\times$ 10$^{	-6	}$ &	1.690 	$\times$ 10$^{	-3	}$ &	-1.690 	$\times$ 10$^{	-3	}$\\
$^{	254	}$No	&	8.226 	&	0.250 	&	0.039 	&	-0.035 	&	1.755 	&	1.370 	&	8.097 	$\times$ 10$^{	-6	}$ &	1.840 	$\times$ 10$^{	-3	}$ &	-1.850 	$\times$ 10$^{	-3	}$\\
$^{	256	}$No	&	8.581 	&	0.250 	&	0.027 	&	-0.037 	&	0.466 	&	0.216 	&	8.249 	$\times$ 10$^{	-6	}$ &	1.710 	$\times$ 10$^{	-3	}$ &	-1.720 	$\times$ 10$^{	-3	}$\\
$^{	252	}$Lr	&	9.165 	&	0.249 	&	0.051 	&	-0.032 	&	-0.424 	&	-1.174 	&	8.071 	$\times$ 10$^{	-6	}$ &	1.470 	$\times$ 10$^{	-3	}$ &	-1.480 	$\times$ 10$^{	-3	}$\\
$^{	253	}$Lr	&	8.915 	&	0.250 	&	0.039 	&	-0.035 	&	-0.154 	&	-0.421 	&	8.063 	$\times$ 10$^{	-6	}$ &	1.550 	$\times$ 10$^{	-3	}$ &	-1.550 	$\times$ 10$^{	-3	}$\\
$^{	255	}$Lr	&	8.502 	&	0.250 	&	0.027 	&	-0.037 	&	1.494 	&	0.864 	&	8.068 	$\times$ 10$^{	-6	}$ &	1.730 	$\times$ 10$^{	-3	}$ &	-1.740 	$\times$ 10$^{	-3	}$\\
$^{	259	}$Lr	&	8.575 	&	0.252 	&	0.002 	&	-0.033 	&	0.899 	&	0.609 	&	8.239 	$\times$ 10$^{	-6	}$ &	1.740 	$\times$ 10$^{	-3	}$ &	-1.750 	$\times$ 10$^{	-3	}$\\
\hline
\label{T1}
\end{longtable*}

\vspace{-35pt} 
\twocolumngrid   
\bibliographystyle{apsrev4-1}
\bibliography{1st}

\begin{thebibliography}{83}%
\makeatletter
\providecommand \@ifxundefined [1]{%
 \@ifx{#1\undefined}
}%
\providecommand \@ifnum [1]{%
 \ifnum #1\expandafter \@firstoftwo
 \else \expandafter \@secondoftwo
 \fi
}%
\providecommand \@ifx [1]{%
 \ifx #1\expandafter \@firstoftwo
 \else \expandafter \@secondoftwo
 \fi
}%
\providecommand \natexlab [1]{#1}%
\providecommand \enquote  [1]{``#1''}%
\providecommand \bibnamefont  [1]{#1}%
\providecommand \bibfnamefont [1]{#1}%
\providecommand \citenamefont [1]{#1}%
\providecommand \href@noop [0]{\@secondoftwo}%
\providecommand \href [0]{\begingroup \@sanitize@url \@href}%
\providecommand \@href[1]{\@@startlink{#1}\@@href}%
\providecommand \@@href[1]{\endgroup#1\@@endlink}%
\providecommand \@sanitize@url [0]{\catcode `\\12\catcode `\$12\catcode
  `\&12\catcode `\#12\catcode `\^12\catcode `\_12\catcode `\%12\relax}%
\providecommand \@@startlink[1]{}%
\providecommand \@@endlink[0]{}%
\providecommand \url  [0]{\begingroup\@sanitize@url \@url }%
\providecommand \@url [1]{\endgroup\@href {#1}{\urlprefix }}%
\providecommand \urlprefix  [0]{URL }%
\providecommand \Eprint [0]{\href }%
\providecommand \doibase [0]{http://dx.doi.org/}%
\providecommand \selectlanguage [0]{\@gobble}%
\providecommand \bibinfo  [0]{\@secondoftwo}%
\providecommand \bibfield  [0]{\@secondoftwo}%
\providecommand \translation [1]{[#1]}%
\providecommand \BibitemOpen [0]{}%
\providecommand \bibitemStop [0]{}%
\providecommand \bibitemNoStop [0]{.\EOS\space}%
\providecommand \EOS [0]{\spacefactor3000\relax}%
\providecommand \BibitemShut  [1]{\csname bibitem#1\endcsname}%
\let\auto@bib@innerbib\@empty
\bibitem [{\citenamefont {Hofmann}\ and\ \citenamefont
  {M\"unzenberg}(2000)}]{RevModPhys.72.733}%
  \BibitemOpen
  \bibfield  {author} {\bibinfo {author} {\bibfnamefont {S.}~\bibnamefont
  {Hofmann}}\ and\ \bibinfo {author} {\bibfnamefont {G.}~\bibnamefont
  {M\"unzenberg}},\ }\href {\doibase 10.1103/RevModPhys.72.733} {\bibfield
  {journal} {\bibinfo  {journal} {Rev. Mod. Phys.}\ }\textbf {\bibinfo {volume}
  {72}},\ \bibinfo {pages} {733} (\bibinfo {year} {2000})}\BibitemShut
  {NoStop}%
\bibitem [{\citenamefont {Wang}\ \emph {et~al.}(2022)\citenamefont {Wang},
  \citenamefont {Bai},\ and\ \citenamefont {Ren}}]{PhysRevC.105.024327}%
  \BibitemOpen
  \bibfield  {author} {\bibinfo {author} {\bibfnamefont {Z.}~\bibnamefont
  {Wang}}, \bibinfo {author} {\bibfnamefont {D.}~\bibnamefont {Bai}}, \ and\
  \bibinfo {author} {\bibfnamefont {Z.}~\bibnamefont {Ren}},\ }\href {\doibase
  10.1103/PhysRevC.105.024327} {\bibfield  {journal} {\bibinfo  {journal}
  {Phys. Rev. C}\ }\textbf {\bibinfo {volume} {105}},\ \bibinfo {pages}
  {024327} (\bibinfo {year} {2022})}\BibitemShut {NoStop}%
\bibitem [{\citenamefont {Sun}\ \emph {et~al.}(2016)\citenamefont {Sun},
  \citenamefont {Guo},\ and\ \citenamefont {Li}}]{PhysRevC.93.034316}%
  \BibitemOpen
  \bibfield  {author} {\bibinfo {author} {\bibfnamefont {X.~D.}\ \bibnamefont
  {Sun}}, \bibinfo {author} {\bibfnamefont {P.}~\bibnamefont {Guo}}, \ and\
  \bibinfo {author} {\bibfnamefont {X.~H.}\ \bibnamefont {Li}},\ }\href
  {\doibase 10.1103/PhysRevC.93.034316} {\bibfield  {journal} {\bibinfo
  {journal} {Phys. Rev. C}\ }\textbf {\bibinfo {volume} {93}},\ \bibinfo
  {pages} {034316} (\bibinfo {year} {2016})}\BibitemShut {NoStop}%
\bibitem [{\citenamefont {Deng}\ \emph {et~al.}(2020)\citenamefont {Deng},
  \citenamefont {Zhang},\ and\ \citenamefont {Royer}}]{PhysRevC.101.034307}%
  \BibitemOpen
  \bibfield  {author} {\bibinfo {author} {\bibfnamefont {J.~G.}\ \bibnamefont
  {Deng}}, \bibinfo {author} {\bibfnamefont {H.~F.}\ \bibnamefont {Zhang}}, \
  and\ \bibinfo {author} {\bibfnamefont {G.}~\bibnamefont {Royer}},\ }\href
  {\doibase 10.1103/PhysRevC.101.034307} {\bibfield  {journal} {\bibinfo
  {journal} {Phys. Rev. C}\ }\textbf {\bibinfo {volume} {101}},\ \bibinfo
  {pages} {034307} (\bibinfo {year} {2020})}\BibitemShut {NoStop}%
\bibitem [{\citenamefont {Luo}\ \emph {et~al.}(2025)\citenamefont {Luo},
  \citenamefont {Xu}, \citenamefont {Li}, \citenamefont {Wang}, \citenamefont
  {Zhang}, \citenamefont {Deng}, \citenamefont {Zhang},\ and\ \citenamefont
  {Ma}}]{PhysRevC.111.034330}%
  \BibitemOpen
  \bibfield  {author} {\bibinfo {author} {\bibfnamefont {J.}~\bibnamefont
  {Luo}}, \bibinfo {author} {\bibfnamefont {Y.}~\bibnamefont {Xu}}, \bibinfo
  {author} {\bibfnamefont {X.}~\bibnamefont {Li}}, \bibinfo {author}
  {\bibfnamefont {J.}~\bibnamefont {Wang}}, \bibinfo {author} {\bibfnamefont
  {Y.}~\bibnamefont {Zhang}}, \bibinfo {author} {\bibfnamefont
  {J.}~\bibnamefont {Deng}}, \bibinfo {author} {\bibfnamefont {F.}~\bibnamefont
  {Zhang}}, \ and\ \bibinfo {author} {\bibfnamefont {N.}~\bibnamefont {Ma}},\
  }\href {\doibase 10.1103/PhysRevC.111.034330} {\bibfield  {journal} {\bibinfo
   {journal} {Phys. Rev. C}\ }\textbf {\bibinfo {volume} {111}},\ \bibinfo
  {pages} {034330} (\bibinfo {year} {2025})}\BibitemShut {NoStop}%
\bibitem [{\citenamefont {Deng}\ \emph {et~al.}(2025)\citenamefont {Deng},
  \citenamefont {Li}, \citenamefont {Cheng}, \citenamefont {Mu},\ and\
  \citenamefont {Zhang}}]{Deng_2025}%
  \BibitemOpen
  \bibfield  {author} {\bibinfo {author} {\bibfnamefont {J.~G.}\ \bibnamefont
  {Deng}}, \bibinfo {author} {\bibfnamefont {J.~X.}\ \bibnamefont {Li}},
  \bibinfo {author} {\bibfnamefont {J.~H.}\ \bibnamefont {Cheng}}, \bibinfo
  {author} {\bibfnamefont {L.}~\bibnamefont {Mu}}, \ and\ \bibinfo {author}
  {\bibfnamefont {H.~F.}\ \bibnamefont {Zhang}},\ }\href {\doibase
  10.1088/1674-1137/adb2fa} {\bibfield  {journal} {\bibinfo  {journal} {Chin.
  Phys. C}\ }\textbf {\bibinfo {volume} {49}},\ \bibinfo {pages} {054104}
  (\bibinfo {year} {2025})}\BibitemShut {NoStop}%
\bibitem [{\citenamefont {Deng}\ \emph {et~al.}(2024)\citenamefont {Deng},
  \citenamefont {Cheng}, \citenamefont {Bao},\ and\ \citenamefont
  {Zhang}}]{Deng_2024}%
  \BibitemOpen
  \bibfield  {author} {\bibinfo {author} {\bibfnamefont {J.~G.}\ \bibnamefont
  {Deng}}, \bibinfo {author} {\bibfnamefont {J.~H.}\ \bibnamefont {Cheng}},
  \bibinfo {author} {\bibfnamefont {X.~J.}\ \bibnamefont {Bao}}, \ and\
  \bibinfo {author} {\bibfnamefont {H.~F.}\ \bibnamefont {Zhang}},\ }\href
  {\doibase 10.1088/1674-1137/ad30ef} {\bibfield  {journal} {\bibinfo
  {journal} {Chin. Phys. C}\ }\textbf {\bibinfo {volume} {48}},\ \bibinfo
  {pages} {064101} (\bibinfo {year} {2024})}\BibitemShut {NoStop}%
\bibitem [{\citenamefont {Dzyublik}(2014)}]{PhysRevC.90.054619}%
  \BibitemOpen
  \bibfield  {author} {\bibinfo {author} {\bibfnamefont {A.~Y.}\ \bibnamefont
  {Dzyublik}},\ }\href {\doibase 10.1103/PhysRevC.90.054619} {\bibfield
  {journal} {\bibinfo  {journal} {Phys. Rev. C}\ }\textbf {\bibinfo {volume}
  {90}},\ \bibinfo {pages} {054619} (\bibinfo {year} {2014})}\BibitemShut
  {NoStop}%
\bibitem [{\citenamefont {Wan}\ \emph {et~al.}(2015)\citenamefont {Wan},
  \citenamefont {Xu},\ and\ \citenamefont {Ren}}]{PhysRevC.92.024301}%
  \BibitemOpen
  \bibfield  {author} {\bibinfo {author} {\bibfnamefont {N.}~\bibnamefont
  {Wan}}, \bibinfo {author} {\bibfnamefont {C.}~\bibnamefont {Xu}}, \ and\
  \bibinfo {author} {\bibfnamefont {Z.~Z.}\ \bibnamefont {Ren}},\ }\href
  {\doibase 10.1103/PhysRevC.92.024301} {\bibfield  {journal} {\bibinfo
  {journal} {Phys. Rev. C}\ }\textbf {\bibinfo {volume} {92}},\ \bibinfo
  {pages} {024301} (\bibinfo {year} {2015})}\BibitemShut {NoStop}%
\bibitem [{\citenamefont {Wan}\ \emph {et~al.}(2016)\citenamefont {Wan},
  \citenamefont {Xu},\ and\ \citenamefont {Ren}}]{wan2016alpha}%
  \BibitemOpen
  \bibfield  {author} {\bibinfo {author} {\bibfnamefont {N.}~\bibnamefont
  {Wan}}, \bibinfo {author} {\bibfnamefont {C.}~\bibnamefont {Xu}}, \ and\
  \bibinfo {author} {\bibfnamefont {Z.~Z.}\ \bibnamefont {Ren}},\ }\href
  {\doibase https://doi.org/10.1007/s41365-016-0150-2} {\bibfield  {journal}
  {\bibinfo  {journal} {Nucl. Sci. Tech.}\ }\textbf {\bibinfo {volume} {27}},\
  \bibinfo {pages} {149} (\bibinfo {year} {2016})}\BibitemShut {NoStop}%
\bibitem [{\citenamefont {Karpeshin}\ \emph {et~al.}(2022)\citenamefont
  {Karpeshin}, \citenamefont {Trzhaskovskaya},\ and\ \citenamefont
  {Vitushkin}}]{PhysRevC.105.024307}%
  \BibitemOpen
  \bibfield  {author} {\bibinfo {author} {\bibfnamefont {F.~F.}\ \bibnamefont
  {Karpeshin}}, \bibinfo {author} {\bibfnamefont {M.~B.}\ \bibnamefont
  {Trzhaskovskaya}}, \ and\ \bibinfo {author} {\bibfnamefont {L.~F.}\
  \bibnamefont {Vitushkin}},\ }\href {\doibase 10.1103/PhysRevC.105.024307}
  {\bibfield  {journal} {\bibinfo  {journal} {Phys. Rev. C}\ }\textbf {\bibinfo
  {volume} {105}},\ \bibinfo {pages} {024307} (\bibinfo {year}
  {2022})}\BibitemShut {NoStop}%
\bibitem [{\citenamefont {Chwaszczewski}\ and\ \citenamefont
  {Słowiński}(2003)}]{CHWASZCZEWSKI200387}%
  \BibitemOpen
  \bibfield  {author} {\bibinfo {author} {\bibfnamefont {S.}~\bibnamefont
  {Chwaszczewski}}\ and\ \bibinfo {author} {\bibfnamefont {B.}~\bibnamefont
  {Słowiński}},\ }\href {\doibase
  https://doi.org/10.1016/S0306-2619(03)00022-9} {\bibfield  {journal}
  {\bibinfo  {journal} {Applied Energy}\ }\textbf {\bibinfo {volume} {75}},\
  \bibinfo {pages} {87} (\bibinfo {year} {2003})}\BibitemShut {NoStop}%
\bibitem [{\citenamefont {Gudowski}(2000)}]{GUDOWSKI2000169c}%
  \BibitemOpen
  \bibfield  {author} {\bibinfo {author} {\bibfnamefont {W.}~\bibnamefont
  {Gudowski}},\ }\href {\doibase https://doi.org/10.1016/S0375-9474(99)00585-0}
  {\bibfield  {journal} {\bibinfo  {journal} {Nucl. Phys. A}\ }\textbf
  {\bibinfo {volume} {663}},\ \bibinfo {pages} {169c} (\bibinfo {year}
  {2000})}\BibitemShut {NoStop}%
\bibitem [{\citenamefont {Kurniawan}\ \emph {et~al.}(2022)\citenamefont
  {Kurniawan}, \citenamefont {Othman}, \citenamefont {Singh}, \citenamefont
  {Avtar},\ and\ \citenamefont {$et\ al.$}}]{KURNIAWAN2022108736}%
  \BibitemOpen
  \bibfield  {author} {\bibinfo {author} {\bibfnamefont {T.~A.}\ \bibnamefont
  {Kurniawan}}, \bibinfo {author} {\bibfnamefont {M.~H.~D.}\ \bibnamefont
  {Othman}}, \bibinfo {author} {\bibfnamefont {D.}~\bibnamefont {Singh}},
  \bibinfo {author} {\bibfnamefont {R.}~\bibnamefont {Avtar}}, \ and\ \bibinfo
  {author} {\bibnamefont {$et\ al.$}},\ }\href {\doibase
  https://doi.org/10.1016/j.anucene.2021.108736} {\bibfield  {journal}
  {\bibinfo  {journal} {Annals of Nuclear Energy}\ }\textbf {\bibinfo {volume}
  {166}},\ \bibinfo {pages} {108736} (\bibinfo {year} {2022})}\BibitemShut
  {NoStop}%
\bibitem [{\citenamefont {Krause}\ \emph {et~al.}(2012)\citenamefont {Krause},
  \citenamefont {Rogers}, \citenamefont {Fischbach},\ and\ \citenamefont
  {\textit{et al.}}}]{KRAUSE201251}%
  \BibitemOpen
  \bibfield  {author} {\bibinfo {author} {\bibfnamefont {D.~E.}\ \bibnamefont
  {Krause}}, \bibinfo {author} {\bibfnamefont {B.~A.}\ \bibnamefont {Rogers}},
  \bibinfo {author} {\bibfnamefont {E.}~\bibnamefont {Fischbach}}, \ and\
  \bibinfo {author} {\bibnamefont {\textit{et al.}}},\ }\href {\doibase
  https://doi.org/10.1016/j.astropartphys.2012.05.002} {\bibfield  {journal}
  {\bibinfo  {journal} {Astropart. Phys.}\ }\textbf {\bibinfo {volume} {36}},\
  \bibinfo {pages} {51} (\bibinfo {year} {2012})}\BibitemShut {NoStop}%
\bibitem [{\citenamefont {Prelas}\ \emph {et~al.}(2014)\citenamefont {Prelas},
  \citenamefont {Weaver}, \citenamefont {Watermann}, \citenamefont {Lukosi},\
  and\ \citenamefont {$et\ al.$}}]{prelas2014review}%
  \BibitemOpen
  \bibfield  {author} {\bibinfo {author} {\bibfnamefont {M.~A.}\ \bibnamefont
  {Prelas}}, \bibinfo {author} {\bibfnamefont {C.~L.}\ \bibnamefont {Weaver}},
  \bibinfo {author} {\bibfnamefont {M.~L.}\ \bibnamefont {Watermann}}, \bibinfo
  {author} {\bibfnamefont {E.~D.}\ \bibnamefont {Lukosi}}, \ and\ \bibinfo
  {author} {\bibnamefont {$et\ al.$}},\ }\href {\doibase
  https://doi.org/10.1016/j.pnucene.2014.04.007} {\bibfield  {journal}
  {\bibinfo  {journal} {Pro. Nucl. Energy}\ }\textbf {\bibinfo {volume} {75}},\
  \bibinfo {pages} {117} (\bibinfo {year} {2014})}\BibitemShut {NoStop}%
\bibitem [{\citenamefont {Andrade}\ \emph {et~al.}(2023)\citenamefont
  {Andrade}, \citenamefont {Pereira},\ and\ \citenamefont
  {Velasquez}}]{doi:10.1504/IJNEST.2023.135375}%
  \BibitemOpen
  \bibfield  {author} {\bibinfo {author} {\bibfnamefont {K.}~\bibnamefont
  {Andrade}}, \bibinfo {author} {\bibfnamefont {C.}~\bibnamefont {Pereira}}, \
  and\ \bibinfo {author} {\bibfnamefont {C.~E.}\ \bibnamefont {Velasquez}},\
  }\href {\doibase 10.1504/IJNEST.2023.135375} {\bibfield  {journal} {\bibinfo
  {journal} {Int. Jour. Nucl. Energy Sci. Tech.}\ }\textbf {\bibinfo {volume}
  {16}},\ \bibinfo {pages} {194} (\bibinfo {year} {2023})}\BibitemShut
  {NoStop}%
\bibitem [{\citenamefont {Cowan}\ and\ \citenamefont
  {Sneden}(2006)}]{NatureCo}%
  \BibitemOpen
  \bibfield  {author} {\bibinfo {author} {\bibfnamefont {J.}~\bibnamefont
  {Cowan}}\ and\ \bibinfo {author} {\bibfnamefont {C.}~\bibnamefont {Sneden}},\
  }\href {\doibase 10.1038/nature04807} {\bibfield  {journal} {\bibinfo
  {journal} {Nature}\ }\textbf {\bibinfo {volume} {400}},\ \bibinfo {pages}
  {1151} (\bibinfo {year} {2006})}\BibitemShut {NoStop}%
\bibitem [{\citenamefont {Liolios}(2003)}]{PhysRevC.68.015804}%
  \BibitemOpen
  \bibfield  {author} {\bibinfo {author} {\bibfnamefont {T.~E.}\ \bibnamefont
  {Liolios}},\ }\href {\doibase 10.1103/PhysRevC.68.015804} {\bibfield
  {journal} {\bibinfo  {journal} {Phys. Rev. C}\ }\textbf {\bibinfo {volume}
  {68}},\ \bibinfo {pages} {015804} (\bibinfo {year} {2003})}\BibitemShut
  {NoStop}%
\bibitem [{\citenamefont {Arcones}\ and\ \citenamefont
  {Thielemann}(2023)}]{OrAstr}%
  \BibitemOpen
  \bibfield  {author} {\bibinfo {author} {\bibfnamefont {A.}~\bibnamefont
  {Arcones}}\ and\ \bibinfo {author} {\bibfnamefont {F.~K.}\ \bibnamefont
  {Thielemann}},\ }\href {\doibase https://doi.org/10.1007/s00159-022-00146-x}
  {\bibfield  {journal} {\bibinfo  {journal} {Astron. Astrophys. Rev.}\
  }\textbf {\bibinfo {volume} {31}},\ \bibinfo {pages} {1} (\bibinfo {year}
  {2023})}\BibitemShut {NoStop}%
\bibitem [{\citenamefont {Oganessian}\ and\ \citenamefont
  {Utyonkov}(2015)}]{Oganessian_2015}%
  \BibitemOpen
  \bibfield  {author} {\bibinfo {author} {\bibfnamefont {Y.~T.}\ \bibnamefont
  {Oganessian}}\ and\ \bibinfo {author} {\bibfnamefont {V.~K.}\ \bibnamefont
  {Utyonkov}},\ }\href {\doibase 10.1088/0034-4885/78/3/036301} {\bibfield
  {journal} {\bibinfo  {journal} {Rep. Prog. Phys.}\ }\textbf {\bibinfo
  {volume} {78}},\ \bibinfo {pages} {036301} (\bibinfo {year}
  {2015})}\BibitemShut {NoStop}%
\bibitem [{\citenamefont {Oganessian}\ \emph {et~al.}(2017)\citenamefont
  {Oganessian}, \citenamefont {Sobiczewski},\ and\ \citenamefont
  {Ter-Akopian}}]{Oganessian_2017}%
  \BibitemOpen
  \bibfield  {author} {\bibinfo {author} {\bibfnamefont {Y.~T.}\ \bibnamefont
  {Oganessian}}, \bibinfo {author} {\bibfnamefont {A.}~\bibnamefont
  {Sobiczewski}}, \ and\ \bibinfo {author} {\bibfnamefont {G.~M.}\ \bibnamefont
  {Ter-Akopian}},\ }\href {\doibase 10.1088/1402-4896/aa53c1} {\bibfield
  {journal} {\bibinfo  {journal} {Phys. Scr.}\ }\textbf {\bibinfo {volume}
  {92}},\ \bibinfo {pages} {023003} (\bibinfo {year} {2017})}\BibitemShut
  {NoStop}%
\bibitem [{\citenamefont {Martins}(2021)}]{inbook}%
  \BibitemOpen
  \bibfield  {author} {\bibinfo {author} {\bibfnamefont {R.}~\bibnamefont
  {Martins}},\ }\enquote {\bibinfo {title} {Becquerel's experimental
  mistakes},}\ \ (\bibinfo  {publisher} {Quamcumque Editum},\ \bibinfo {year}
  {2021})\ pp.\ \bibinfo {pages} {107--165}\BibitemShut {NoStop}%
\bibitem [{\citenamefont {Zhou}(2011)}]{Zhou_2011}%
  \BibitemOpen
  \bibfield  {author} {\bibinfo {author} {\bibfnamefont {S.~H.}\ \bibnamefont
  {Zhou}},\ }\href {\doibase 10.1088/1674-1137/35/5/008} {\bibfield  {journal}
  {\bibinfo  {journal} {Chin. Phys. C}\ }\textbf {\bibinfo {volume} {35}},\
  \bibinfo {pages} {449} (\bibinfo {year} {2011})}\BibitemShut {NoStop}%
\bibitem [{\citenamefont {Fu}\ \emph {et~al.}(2021)\citenamefont {Fu},
  \citenamefont {Zhang},\ and\ \citenamefont {Ma}}]{Fucb}%
  \BibitemOpen
  \bibfield  {author} {\bibinfo {author} {\bibfnamefont {C.}~\bibnamefont
  {Fu}}, \bibinfo {author} {\bibfnamefont {G.}~\bibnamefont {Zhang}}, \ and\
  \bibinfo {author} {\bibfnamefont {Y.}~\bibnamefont {Ma}},\ }\href {\doibase
  10.1063/5.0059405} {\bibfield  {journal} {\bibinfo  {journal} {Matter Radiat.
  Extremes}\ }\textbf {\bibinfo {volume} {7}},\ \bibinfo {pages} {024201}
  (\bibinfo {year} {2021})}\BibitemShut {NoStop}%
\bibitem [{\citenamefont {Strickland}\ and\ \citenamefont
  {Mourou}(1985)}]{STRICKLAND1985447}%
  \BibitemOpen
  \bibfield  {author} {\bibinfo {author} {\bibfnamefont {D.}~\bibnamefont
  {Strickland}}\ and\ \bibinfo {author} {\bibfnamefont {G.}~\bibnamefont
  {Mourou}},\ }\href {\doibase https://doi.org/10.1016/0030-4018(85)90151-8}
  {\bibfield  {journal} {\bibinfo  {journal} {Opt. Commun.}\ }\textbf {\bibinfo
  {volume} {55}},\ \bibinfo {pages} {447} (\bibinfo {year} {1985})}\BibitemShut
  {NoStop}%
\bibitem [{\citenamefont {Mourou}\ \emph {et~al.}(2006)\citenamefont {Mourou},
  \citenamefont {Tajima},\ and\ \citenamefont {Bulanov}}]{RevModPhys.78.309}%
  \BibitemOpen
  \bibfield  {author} {\bibinfo {author} {\bibfnamefont {G.~A.}\ \bibnamefont
  {Mourou}}, \bibinfo {author} {\bibfnamefont {T.}~\bibnamefont {Tajima}}, \
  and\ \bibinfo {author} {\bibfnamefont {S.~V.}\ \bibnamefont {Bulanov}},\
  }\href {\doibase 10.1103/RevModPhys.78.309} {\bibfield  {journal} {\bibinfo
  {journal} {Rev. Mod. Phys.}\ }\textbf {\bibinfo {volume} {78}},\ \bibinfo
  {pages} {309} (\bibinfo {year} {2006})}\BibitemShut {NoStop}%
\bibitem [{\citenamefont {Yoon}\ \emph {et~al.}(2021)\citenamefont {Yoon},
  \citenamefont {Kim}, \citenamefont {Choi}, \citenamefont {Sung},
  \citenamefont {Lee}, \citenamefont {Lee},\ and\ \citenamefont
  {Nam}}]{Yoon:21}%
  \BibitemOpen
  \bibfield  {author} {\bibinfo {author} {\bibfnamefont {J.~W.}\ \bibnamefont
  {Yoon}}, \bibinfo {author} {\bibfnamefont {Y.~G.}\ \bibnamefont {Kim}},
  \bibinfo {author} {\bibfnamefont {I.~W.}\ \bibnamefont {Choi}}, \bibinfo
  {author} {\bibfnamefont {J.~H.}\ \bibnamefont {Sung}}, \bibinfo {author}
  {\bibfnamefont {H.~W.}\ \bibnamefont {Lee}}, \bibinfo {author} {\bibfnamefont
  {S.~K.}\ \bibnamefont {Lee}}, \ and\ \bibinfo {author} {\bibfnamefont
  {C.~H.}\ \bibnamefont {Nam}},\ }\href {\doibase 10.1364/OPTICA.420520}
  {\bibfield  {journal} {\bibinfo  {journal} {Optica}\ }\textbf {\bibinfo
  {volume} {8}},\ \bibinfo {pages} {630} (\bibinfo {year} {2021})}\BibitemShut
  {NoStop}%
\bibitem [{\citenamefont {Danson}\ \emph {et~al.}(2019)\citenamefont {Danson},
  \citenamefont {Haefner}, \citenamefont {Bromage},\ and\ \citenamefont
  {\textit{et al.}}}]{Colin2}%
  \BibitemOpen
  \bibfield  {author} {\bibinfo {author} {\bibfnamefont {C.}~\bibnamefont
  {Danson}}, \bibinfo {author} {\bibfnamefont {C.}~\bibnamefont {Haefner}},
  \bibinfo {author} {\bibfnamefont {J.}~\bibnamefont {Bromage}}, \ and\
  \bibinfo {author} {\bibnamefont {\textit{et al.}}},\ }\href {\doibase
  10.1017/hpl.2019.36} {\bibfield  {journal} {\bibinfo  {journal} {High Power
  Laser Sci.}\ }\textbf {\bibinfo {volume} {7}},\ \bibinfo {pages} {e54}
  (\bibinfo {year} {2019})}\BibitemShut {NoStop}%
\bibitem [{\citenamefont {Radier}\ \emph {et~al.}(2022)\citenamefont {Radier},
  \citenamefont {Chalus}, \citenamefont {Charbonneau},\ and\ \citenamefont
  {\textit{et al.}}}]{Radier}%
  \BibitemOpen
  \bibfield  {author} {\bibinfo {author} {\bibfnamefont {C.}~\bibnamefont
  {Radier}}, \bibinfo {author} {\bibfnamefont {O.}~\bibnamefont {Chalus}},
  \bibinfo {author} {\bibfnamefont {M.}~\bibnamefont {Charbonneau}}, \ and\
  \bibinfo {author} {\bibnamefont {\textit{et al.}}},\ }\href {\doibase
  https://doi.org/10.1017/hpl.2022.11} {\bibfield  {journal} {\bibinfo
  {journal} {High Power Laser Sci.}\ }\textbf {\bibinfo {volume} {10}},\
  \bibinfo {pages} {e21} (\bibinfo {year} {2022})}\BibitemShut {NoStop}%
\bibitem [{\citenamefont {Li}\ \emph {et~al.}(2025)\citenamefont {Li},
  \citenamefont {Chen}, \citenamefont {Chen}, \citenamefont {Liu},\ and\
  \citenamefont {\textit{et al.}}}]{Yu-Tong}%
  \BibitemOpen
  \bibfield  {author} {\bibinfo {author} {\bibfnamefont {Y.~T.}\ \bibnamefont
  {Li}}, \bibinfo {author} {\bibfnamefont {L.~M.}\ \bibnamefont {Chen}},
  \bibinfo {author} {\bibfnamefont {M.}~\bibnamefont {Chen}}, \bibinfo {author}
  {\bibfnamefont {F.}~\bibnamefont {Liu}}, \ and\ \bibinfo {author}
  {\bibnamefont {\textit{et al.}}},\ }\href {\doibase 10.1017/hpl.2024.69}
  {\bibfield  {journal} {\bibinfo  {journal} {High Power Laser Sci.}\ }\textbf
  {\bibinfo {volume} {13}},\ \bibinfo {pages} {e12} (\bibinfo {year}
  {2025})}\BibitemShut {NoStop}%
\bibitem [{\citenamefont {Negoita}\ \emph {et~al.}(2022)\citenamefont
  {Negoita}, \citenamefont {Roth}, \citenamefont {Thirolf}, \citenamefont
  {Tudisco},\ and\ \citenamefont {\textit{et al.}}}]{ELI}%
  \BibitemOpen
  \bibfield  {author} {\bibinfo {author} {\bibfnamefont {F.}~\bibnamefont
  {Negoita}}, \bibinfo {author} {\bibfnamefont {M.}~\bibnamefont {Roth}},
  \bibinfo {author} {\bibfnamefont {P.~G.}\ \bibnamefont {Thirolf}}, \bibinfo
  {author} {\bibfnamefont {S.}~\bibnamefont {Tudisco}}, \ and\ \bibinfo
  {author} {\bibnamefont {\textit{et al.}}},\ }\href {\doibase
  https://doi.org/10.48550/arXiv.2201.01068} {\bibfield  {journal} {\bibinfo
  {journal} {Arxiv}\ }\textbf {\bibinfo {volume} {201.01068}},\ \bibinfo
  {pages} {.01068} (\bibinfo {year} {2022})}\BibitemShut {NoStop}%
\bibitem [{\citenamefont {Calin}(2023)}]{10.1117/12.2671369}%
  \BibitemOpen
  \bibfield  {author} {\bibinfo {author} {\bibfnamefont {A.~U.}\ \bibnamefont
  {Calin}},\ }in\ \href {\doibase 10.1117/12.2671369} {\emph {\bibinfo
  {booktitle} {Research Using Extreme Light: Entering New Frontiers with
  Petawatt-Class Lasers V}}},\ Vol.\ \bibinfo {volume} {12580},\ \bibinfo
  {organization} {International Society for Optics and Photonics}\ (\bibinfo
  {publisher} {SPIE},\ \bibinfo {year} {2023})\ p.\ \bibinfo {pages}
  {1258004}\BibitemShut {NoStop}%
\bibitem [{\citenamefont {Guo}\ \emph {et~al.}(2024)\citenamefont {Guo},
  \citenamefont {Fu},\ and\ \citenamefont {Bonasera}}]{GB}%
  \BibitemOpen
  \bibfield  {author} {\bibinfo {author} {\bibfnamefont {B.}~\bibnamefont
  {Guo}}, \bibinfo {author} {\bibfnamefont {C.~B.}\ \bibnamefont {Fu}}, \ and\
  \bibinfo {author} {\bibfnamefont {A.}~\bibnamefont {Bonasera}},\ }\href
  {\doibase 10.3389/fphy.2024.1503516} {\bibfield  {journal} {\bibinfo
  {journal} {Front. Phys.}\ }\textbf {\bibinfo {volume} {12}},\ \bibinfo
  {pages} {1503516} (\bibinfo {year} {2024})}\BibitemShut {NoStop}%
\bibitem [{\citenamefont {Xiao}\ \emph {et~al.}()\citenamefont {Xiao},
  \citenamefont {Cheng}, \citenamefont {Xu}, \citenamefont {Zou}, \citenamefont
  {Ren}, \citenamefont {Dzyublik},\ and\ \citenamefont {Yu}}]{Xiaoppnp}%
  \BibitemOpen
  \bibfield  {author} {\bibinfo {author} {\bibfnamefont {Q.}~\bibnamefont
  {Xiao}}, \bibinfo {author} {\bibfnamefont {J.~H.}\ \bibnamefont {Cheng}},
  \bibinfo {author} {\bibfnamefont {Y.~Y.}\ \bibnamefont {Xu}}, \bibinfo
  {author} {\bibfnamefont {Y.~T.}\ \bibnamefont {Zou}}, \bibinfo {author}
  {\bibfnamefont {Z.~Z.}\ \bibnamefont {Ren}}, \bibinfo {author} {\bibfnamefont
  {A.~Y.}\ \bibnamefont {Dzyublik}}, \ and\ \bibinfo {author} {\bibfnamefont
  {T.~P.}\ \bibnamefont {Yu}},\ }\href@noop {} {\bibfield  {journal} {\bibinfo
  {journal} {Rep. Prog. Phys.}\ }\textbf {\bibinfo {volume}
  {submitted}}}\BibitemShut {NoStop}%
\bibitem [{\citenamefont {Qi}\ \emph {et~al.}(2019)\citenamefont {Qi},
  \citenamefont {Li}, \citenamefont {Xu}, \citenamefont {Fu},\ and\
  \citenamefont {Wang}}]{PhysRevC99}%
  \BibitemOpen
  \bibfield  {author} {\bibinfo {author} {\bibfnamefont {J.~T.}\ \bibnamefont
  {Qi}}, \bibinfo {author} {\bibfnamefont {T.}~\bibnamefont {Li}}, \bibinfo
  {author} {\bibfnamefont {R.~H.}\ \bibnamefont {Xu}}, \bibinfo {author}
  {\bibfnamefont {L.~B.}\ \bibnamefont {Fu}}, \ and\ \bibinfo {author}
  {\bibfnamefont {X.}~\bibnamefont {Wang}},\ }\href {\doibase
  10.1103/PhysRevC.99.044610} {\bibfield  {journal} {\bibinfo  {journal} {Phys.
  Rev. C}\ }\textbf {\bibinfo {volume} {99}},\ \bibinfo {pages} {044610}
  (\bibinfo {year} {2019})}\BibitemShut {NoStop}%
\bibitem [{\citenamefont {Qi}\ \emph {et~al.}(2020)\citenamefont {Qi},
  \citenamefont {Fu},\ and\ \citenamefont {Wang}}]{PhysRevC102}%
  \BibitemOpen
  \bibfield  {author} {\bibinfo {author} {\bibfnamefont {J.~T.}\ \bibnamefont
  {Qi}}, \bibinfo {author} {\bibfnamefont {L.~B.}\ \bibnamefont {Fu}}, \ and\
  \bibinfo {author} {\bibfnamefont {X.}~\bibnamefont {Wang}},\ }\href {\doibase
  10.1103/PhysRevC.102.064629} {\bibfield  {journal} {\bibinfo  {journal}
  {Phys. Rev. C}\ }\textbf {\bibinfo {volume} {102}},\ \bibinfo {pages}
  {064629} (\bibinfo {year} {2020})}\BibitemShut {NoStop}%
\bibitem [{\citenamefont {Bai}\ \emph {et~al.}(2018)\citenamefont {Bai},
  \citenamefont {Deng},\ and\ \citenamefont {Ren}}]{BAI201823}%
  \BibitemOpen
  \bibfield  {author} {\bibinfo {author} {\bibfnamefont {D.}~\bibnamefont
  {Bai}}, \bibinfo {author} {\bibfnamefont {D.~M.}\ \bibnamefont {Deng}}, \
  and\ \bibinfo {author} {\bibfnamefont {Z.}~\bibnamefont {Ren}},\ }\href
  {\doibase https://doi.org/10.1016/j.nuclphysa.2018.05.004} {\bibfield
  {journal} {\bibinfo  {journal} {Nucl. Phys. A}\ }\textbf {\bibinfo {volume}
  {976}},\ \bibinfo {pages} {23} (\bibinfo {year} {2018})}\BibitemShut
  {NoStop}%
\bibitem [{\citenamefont {Mi$\c{S}$icu}\ and\ \citenamefont
  {Rizea}(2013)}]{Misicu2013}%
  \BibitemOpen
  \bibfield  {author} {\bibinfo {author} {\bibfnamefont {C.}~\bibnamefont
  {Mi$\c{S}$icu}}\ and\ \bibinfo {author} {\bibfnamefont {M.}~\bibnamefont
  {Rizea}},\ }\href {\doibase 10.1088/0954-3899/40/9/095101} {\bibfield
  {journal} {\bibinfo  {journal} {J. Phys. G: Nucl. Part. Phys.}\ }\textbf
  {\bibinfo {volume} {40}},\ \bibinfo {pages} {095101} (\bibinfo {year}
  {2013})}\BibitemShut {NoStop}%
\bibitem [{\citenamefont {Mi$\c{S}$icu}\ and\ \citenamefont
  {Rizea}(2016)}]{Misicurizea}%
  \BibitemOpen
  \bibfield  {author} {\bibinfo {author} {\bibfnamefont {C.}~\bibnamefont
  {Mi$\c{S}$icu}}\ and\ \bibinfo {author} {\bibfnamefont {M.}~\bibnamefont
  {Rizea}},\ }\href {\doibase doi:10.1515/phys-2016-0001} {\bibfield  {journal}
  {\bibinfo  {journal} {Open Phys.}\ }\textbf {\bibinfo {volume} {14}},\
  \bibinfo {pages} {81} (\bibinfo {year} {2016})}\BibitemShut {NoStop}%
\bibitem [{\citenamefont {Rehman}\ \emph {et~al.}(2022)\citenamefont {Rehman},
  \citenamefont {Shabbir}, \citenamefont {Shafiq},\ and\ \citenamefont
  {\textit{et al.}}}]{Po212}%
  \BibitemOpen
  \bibfield  {author} {\bibinfo {author} {\bibfnamefont {Z.~U.}\ \bibnamefont
  {Rehman}}, \bibinfo {author} {\bibfnamefont {N.}~\bibnamefont {Shabbir}},
  \bibinfo {author} {\bibfnamefont {S.}~\bibnamefont {Shafiq}}, \ and\ \bibinfo
  {author} {\bibnamefont {\textit{et al.}}},\ }\href {\doibase
  https://doi.org/10.1140/epja/s10050-022-00715-9} {\bibfield  {journal}
  {\bibinfo  {journal} {Eur. Phys. J. A}\ }\textbf {\bibinfo {volume} {58}},\
  \bibinfo {pages} {60} (\bibinfo {year} {2022})}\BibitemShut {NoStop}%
\bibitem [{\citenamefont {Cheng}\ \emph {et~al.}(2024)\citenamefont {Cheng},
  \citenamefont {Zhang}, \citenamefont {Xiao}, \citenamefont {Deng},\ and\
  \citenamefont {Yu}}]{Cheng123}%
  \BibitemOpen
  \bibfield  {author} {\bibinfo {author} {\bibfnamefont {J.~H.}\ \bibnamefont
  {Cheng}}, \bibinfo {author} {\bibfnamefont {W.~Y.}\ \bibnamefont {Zhang}},
  \bibinfo {author} {\bibfnamefont {Q.}~\bibnamefont {Xiao}}, \bibinfo {author}
  {\bibfnamefont {J.~G.}\ \bibnamefont {Deng}}, \ and\ \bibinfo {author}
  {\bibfnamefont {T.~P.}\ \bibnamefont {Yu}},\ }\href {\doibase
  https://doi.org/10.1016/j.physletb.2023.138322} {\bibfield  {journal}
  {\bibinfo  {journal} {Phys. Lett. B}\ }\textbf {\bibinfo {volume} {848}},\
  \bibinfo {pages} {138322} (\bibinfo {year} {2024})}\BibitemShut {NoStop}%
\bibitem [{\citenamefont {Xiao}\ \emph {et~al.}(2024)\citenamefont {Xiao},
  \citenamefont {Cheng}, \citenamefont {Xu}, \citenamefont {Zou},\ and\
  \citenamefont {Yu}}]{XqGamow}%
  \BibitemOpen
  \bibfield  {author} {\bibinfo {author} {\bibfnamefont {Q.}~\bibnamefont
  {Xiao}}, \bibinfo {author} {\bibfnamefont {J.~H.}\ \bibnamefont {Cheng}},
  \bibinfo {author} {\bibfnamefont {Y.~Y.}\ \bibnamefont {Xu}}, \bibinfo
  {author} {\bibfnamefont {Y.~T.}\ \bibnamefont {Zou}}, \ and\ \bibinfo
  {author} {\bibfnamefont {T.~P.}\ \bibnamefont {Yu}},\ }\href {\doibase
  https://doi.org/10.1007/s41365-024-01371-y} {\bibfield  {journal} {\bibinfo
  {journal} {Nucl. Sci. Tech.}\ }\textbf {\bibinfo {volume} {35}},\ \bibinfo
  {pages} {27} (\bibinfo {year} {2024})}\BibitemShut {NoStop}%
\bibitem [{\citenamefont {Cheng}\ \emph {et~al.}(2025)\citenamefont {Cheng},
  \citenamefont {Xiao}, \citenamefont {Deng}, \citenamefont {Xu}, \citenamefont
  {Zou},\ and\ \citenamefont {Yu}}]{CHENGoddA}%
  \BibitemOpen
  \bibfield  {author} {\bibinfo {author} {\bibfnamefont {J.~H.}\ \bibnamefont
  {Cheng}}, \bibinfo {author} {\bibfnamefont {Q.}~\bibnamefont {Xiao}},
  \bibinfo {author} {\bibfnamefont {J.~G.}\ \bibnamefont {Deng}}, \bibinfo
  {author} {\bibfnamefont {Y.~Y.}\ \bibnamefont {Xu}}, \bibinfo {author}
  {\bibfnamefont {Y.~T.}\ \bibnamefont {Zou}}, \ and\ \bibinfo {author}
  {\bibfnamefont {T.~P.}\ \bibnamefont {Yu}},\ }\href {\doibase
  https://doi.org/10.1007/s41365-024-01610-2} {\bibfield  {journal} {\bibinfo
  {journal} {Nucl. Sci. Tech.}\ }\textbf {\bibinfo {volume} {36}},\ \bibinfo
  {pages} {69} (\bibinfo {year} {2025})}\BibitemShut {NoStop}%
\bibitem [{\citenamefont {Wang}\ \emph {et~al.}(2025)\citenamefont {Wang},
  \citenamefont {Huang}, \citenamefont {Gao}, \citenamefont {Huang},
  \citenamefont {Xiao}, \citenamefont {Zhu},\ and\ \citenamefont
  {Su}}]{wang2025alpha}%
  \BibitemOpen
  \bibfield  {author} {\bibinfo {author} {\bibfnamefont {H.}~\bibnamefont
  {Wang}}, \bibinfo {author} {\bibfnamefont {Y.-G.}\ \bibnamefont {Huang}},
  \bibinfo {author} {\bibfnamefont {Z.-P.}\ \bibnamefont {Gao}}, \bibinfo
  {author} {\bibfnamefont {J.-L.}\ \bibnamefont {Huang}}, \bibinfo {author}
  {\bibfnamefont {E.-X.}\ \bibnamefont {Xiao}}, \bibinfo {author}
  {\bibfnamefont {L.}~\bibnamefont {Zhu}}, \ and\ \bibinfo {author}
  {\bibfnamefont {J.}~\bibnamefont {Su}},\ }\href {\doibase
  https://doi.org/10.1007/s41365-025-01753-w} {\bibfield  {journal} {\bibinfo
  {journal} {Nucl. Sci. Tech.}\ }\textbf {\bibinfo {volume} {36}},\ \bibinfo
  {pages} {1} (\bibinfo {year} {2025})}\BibitemShut {NoStop}%
\bibitem [{\citenamefont {Wu}\ \emph {et~al.}(2025)\citenamefont {Wu},
  \citenamefont {Fan}, \citenamefont {Ye}, \citenamefont {Gao}, \citenamefont
  {Yu},\ and\ \citenamefont {Liu}}]{PhysRevC1103}%
  \BibitemOpen
  \bibfield  {author} {\bibinfo {author} {\bibfnamefont {B.~B.}\ \bibnamefont
  {Wu}}, \bibinfo {author} {\bibfnamefont {Z.~F.}\ \bibnamefont {Fan}},
  \bibinfo {author} {\bibfnamefont {D.~F.}\ \bibnamefont {Ye}}, \bibinfo
  {author} {\bibfnamefont {C.~Z.}\ \bibnamefont {Gao}}, \bibinfo {author}
  {\bibfnamefont {C.~X.}\ \bibnamefont {Yu}}, \ and\ \bibinfo {author}
  {\bibfnamefont {J.}~\bibnamefont {Liu}},\ }\href {\doibase
  10.1103/PhysRevC.111.034602} {\bibfield  {journal} {\bibinfo  {journal}
  {Phys. Rev. C}\ }\textbf {\bibinfo {volume} {111}},\ \bibinfo {pages}
  {034602} (\bibinfo {year} {2025})}\BibitemShut {NoStop}%
\bibitem [{\citenamefont {Mi$\c{S}$icu}\ and\ \citenamefont
  {Rizea}(2019)}]{Misicu_2019}%
  \BibitemOpen
  \bibfield  {author} {\bibinfo {author} {\bibfnamefont {C.}~\bibnamefont
  {Mi$\c{S}$icu}}\ and\ \bibinfo {author} {\bibfnamefont {M.}~\bibnamefont
  {Rizea}},\ }\href {\doibase 10.1088/1361-6471/ab1d7c} {\bibfield  {journal}
  {\bibinfo  {journal} {Jour. Phys. G: Nucl. and Part. Phys.}\ }\textbf
  {\bibinfo {volume} {46}},\ \bibinfo {pages} {115106} (\bibinfo {year}
  {2019})}\BibitemShut {NoStop}%
\bibitem [{\citenamefont {Cheng}\ \emph {et~al.}(2022)\citenamefont {Cheng},
  \citenamefont {Li},\ and\ \citenamefont {Yu}}]{PhysRevC.105.024312}%
  \BibitemOpen
  \bibfield  {author} {\bibinfo {author} {\bibfnamefont {J.~H.}\ \bibnamefont
  {Cheng}}, \bibinfo {author} {\bibfnamefont {Y.}~\bibnamefont {Li}}, \ and\
  \bibinfo {author} {\bibfnamefont {T.~P.}\ \bibnamefont {Yu}},\ }\href
  {\doibase 10.1103/PhysRevC.105.024312} {\bibfield  {journal} {\bibinfo
  {journal} {Phys. Rev. C}\ }\textbf {\bibinfo {volume} {105}},\ \bibinfo
  {pages} {024312} (\bibinfo {year} {2022})}\BibitemShut {NoStop}%
\bibitem [{\citenamefont {Wu}\ and\ \citenamefont
  {Liu}(2022)}]{PhysRevC.106.064610}%
  \BibitemOpen
  \bibfield  {author} {\bibinfo {author} {\bibfnamefont {B.~B.}\ \bibnamefont
  {Wu}}\ and\ \bibinfo {author} {\bibfnamefont {J.}~\bibnamefont {Liu}},\
  }\href {\doibase 10.1103/PhysRevC.106.064610} {\bibfield  {journal} {\bibinfo
   {journal} {Phys. Rev. C}\ }\textbf {\bibinfo {volume} {106}},\ \bibinfo
  {pages} {064610} (\bibinfo {year} {2022})}\BibitemShut {NoStop}%
\bibitem [{\citenamefont {Zou}\ \emph {et~al.}(2024)\citenamefont {Zou},
  \citenamefont {Cheng}, \citenamefont {Xu},\ and\ \citenamefont {$et\
  al.$}}]{Zou_2024}%
  \BibitemOpen
  \bibfield  {author} {\bibinfo {author} {\bibfnamefont {Y.}~\bibnamefont
  {Zou}}, \bibinfo {author} {\bibfnamefont {J.~H.}\ \bibnamefont {Cheng}},
  \bibinfo {author} {\bibfnamefont {Y.~Y.}\ \bibnamefont {Xu}}, \ and\ \bibinfo
  {author} {\bibnamefont {$et\ al.$}},\ }\href {\doibase
  10.1088/1361-6471/ad2691} {\bibfield  {journal} {\bibinfo  {journal} {Jour.
  Phys. G: Nucl. and Part. Phys.}\ }\textbf {\bibinfo {volume} {51}},\ \bibinfo
  {pages} {045103} (\bibinfo {year} {2024})}\BibitemShut {NoStop}%
\bibitem [{\citenamefont {Liao}\ \emph {et~al.}(2025)\citenamefont {Liao},
  \citenamefont {Fan}, \citenamefont {Liu}, \citenamefont {He},\ and\
  \citenamefont {$et\ al.$}}]{w6wq-mj9b}%
  \BibitemOpen
  \bibfield  {author} {\bibinfo {author} {\bibfnamefont {L.~J.}\ \bibnamefont
  {Liao}}, \bibinfo {author} {\bibfnamefont {Y.}~\bibnamefont {Fan}}, \bibinfo
  {author} {\bibfnamefont {X.}~\bibnamefont {Liu}}, \bibinfo {author}
  {\bibfnamefont {B.}~\bibnamefont {He}}, \ and\ \bibinfo {author}
  {\bibnamefont {$et\ al.$}},\ }\href {\doibase 10.1103/w6wq-mj9b} {\bibfield
  {journal} {\bibinfo  {journal} {Phys. Rev. C}\ }\textbf {\bibinfo {volume}
  {112}},\ \bibinfo {pages} {014327} (\bibinfo {year} {2025})}\BibitemShut
  {NoStop}%
\bibitem [{\citenamefont {Wang}(2020)}]{PhysRevC.102.011601}%
  \BibitemOpen
  \bibfield  {author} {\bibinfo {author} {\bibfnamefont {X.}~\bibnamefont
  {Wang}},\ }\href {\doibase 10.1103/PhysRevC.102.011601} {\bibfield  {journal}
  {\bibinfo  {journal} {Phys. Rev. C}\ }\textbf {\bibinfo {volume} {102}},\
  \bibinfo {pages} {011601} (\bibinfo {year} {2020})}\BibitemShut {NoStop}%
\bibitem [{\citenamefont {Wu}\ \emph {et~al.}(2022)\citenamefont {Wu},
  \citenamefont {Duan},\ and\ \citenamefont {Liu}}]{PhysRevC.105.064615}%
  \BibitemOpen
  \bibfield  {author} {\bibinfo {author} {\bibfnamefont {B.}~\bibnamefont
  {Wu}}, \bibinfo {author} {\bibfnamefont {H.}~\bibnamefont {Duan}}, \ and\
  \bibinfo {author} {\bibfnamefont {J.}~\bibnamefont {Liu}},\ }\href {\doibase
  10.1103/PhysRevC.105.064615} {\bibfield  {journal} {\bibinfo  {journal}
  {Phys. Rev. C}\ }\textbf {\bibinfo {volume} {105}},\ \bibinfo {pages}
  {064615} (\bibinfo {year} {2022})}\BibitemShut {NoStop}%
\bibitem [{\citenamefont {Spohr}\ \emph {et~al.}(2023)\citenamefont {Spohr},
  \citenamefont {Doria}, \citenamefont {Baran},\ and\ \citenamefont {\textit{et
  al.}}}]{EPJA93Mo}%
  \BibitemOpen
  \bibfield  {author} {\bibinfo {author} {\bibfnamefont {K.~M.}\ \bibnamefont
  {Spohr}}, \bibinfo {author} {\bibfnamefont {D.}~\bibnamefont {Doria}},
  \bibinfo {author} {\bibfnamefont {V.}~\bibnamefont {Baran}}, \ and\ \bibinfo
  {author} {\bibnamefont {\textit{et al.}}},\ }\href {\doibase
  https://doi.org/10.1140/epja/s10050-023-01160-y} {\bibfield  {journal}
  {\bibinfo  {journal} {Eur. Phys. J. A}\ }\textbf {\bibinfo {volume} {59}},\
  \bibinfo {pages} {281} (\bibinfo {year} {2023})}\BibitemShut {NoStop}%
\bibitem [{\citenamefont {Qi}\ \emph {et~al.}(2023)\citenamefont {Qi},
  \citenamefont {Zhang},\ and\ \citenamefont {Wang}}]{PhysRevLett.130.112501}%
  \BibitemOpen
  \bibfield  {author} {\bibinfo {author} {\bibfnamefont {J.~T.}\ \bibnamefont
  {Qi}}, \bibinfo {author} {\bibfnamefont {H.~X.}\ \bibnamefont {Zhang}}, \
  and\ \bibinfo {author} {\bibfnamefont {X.}~\bibnamefont {Wang}},\ }\href
  {\doibase 10.1103/PhysRevLett.130.112501} {\bibfield  {journal} {\bibinfo
  {journal} {Phys. Rev. Lett.}\ }\textbf {\bibinfo {volume} {130}},\ \bibinfo
  {pages} {112501} (\bibinfo {year} {2023})}\BibitemShut {NoStop}%
\bibitem [{\citenamefont {Qi}\ \emph {et~al.}(2024)\citenamefont {Qi},
  \citenamefont {Liu},\ and\ \citenamefont {Wang}}]{PhysRevC.110.L051601}%
  \BibitemOpen
  \bibfield  {author} {\bibinfo {author} {\bibfnamefont {J.~T.}\ \bibnamefont
  {Qi}}, \bibinfo {author} {\bibfnamefont {B.~Q.}\ \bibnamefont {Liu}}, \ and\
  \bibinfo {author} {\bibfnamefont {X.}~\bibnamefont {Wang}},\ }\href {\doibase
  10.1103/PhysRevC.110.L051601} {\bibfield  {journal} {\bibinfo  {journal}
  {Phys. Rev. C}\ }\textbf {\bibinfo {volume} {110}},\ \bibinfo {pages}
  {L051601} (\bibinfo {year} {2024})}\BibitemShut {NoStop}%
\bibitem [{\citenamefont {Ma}\ \emph {et~al.}(2024)\citenamefont {Ma},
  \citenamefont {Wang}, \citenamefont {Yang},\ and\ \citenamefont {\textit{et
  al.}}}]{Mzg}%
  \BibitemOpen
  \bibfield  {author} {\bibinfo {author} {\bibfnamefont {Z.~G.}\ \bibnamefont
  {Ma}}, \bibinfo {author} {\bibfnamefont {Y.~M.}\ \bibnamefont {Wang}},
  \bibinfo {author} {\bibfnamefont {Y.}~\bibnamefont {Yang}}, \ and\ \bibinfo
  {author} {\bibnamefont {\textit{et al.}}},\ }\href {\doibase
  10.1063/5.0212163} {\bibfield  {journal} {\bibinfo  {journal} {Matter Radiat.
  Extremes}\ }\textbf {\bibinfo {volume} {9}},\ \bibinfo {pages} {055201}
  (\bibinfo {year} {2024})}\BibitemShut {NoStop}%
\bibitem [{\citenamefont {Feng}\ \emph {et~al.}(2022)\citenamefont {Feng},
  \citenamefont {Wang}, \citenamefont {Fu}, \citenamefont {Chen},\ and\
  \citenamefont {\textit{et al.}}}]{Feng}%
  \BibitemOpen
  \bibfield  {author} {\bibinfo {author} {\bibfnamefont {J.}~\bibnamefont
  {Feng}}, \bibinfo {author} {\bibfnamefont {W.~Z.}\ \bibnamefont {Wang}},
  \bibinfo {author} {\bibfnamefont {C.~B.}\ \bibnamefont {Fu}}, \bibinfo
  {author} {\bibfnamefont {L.~M.}\ \bibnamefont {Chen}}, \ and\ \bibinfo
  {author} {\bibnamefont {\textit{et al.}}},\ }\href {\doibase
  10.1103/PhysRevLett.128.052501} {\bibfield  {journal} {\bibinfo  {journal}
  {Phys. Rev. Lett.}\ }\textbf {\bibinfo {volume} {128}},\ \bibinfo {pages}
  {052501} (\bibinfo {year} {2022})}\BibitemShut {NoStop}%
\bibitem [{\citenamefont {Shvyd\'ko}\ \emph {et~al.}(2023)\citenamefont
  {Shvyd\'ko}, \citenamefont {R{\"o}hlsberger}, \citenamefont {Kocharovskaya},\
  and\ \citenamefont {$et al.$}}]{45Sc}%
  \BibitemOpen
  \bibfield  {author} {\bibinfo {author} {\bibfnamefont {Y.}~\bibnamefont
  {Shvyd\'ko}}, \bibinfo {author} {\bibfnamefont {R.}~\bibnamefont
  {R{\"o}hlsberger}}, \bibinfo {author} {\bibfnamefont {O.}~\bibnamefont
  {Kocharovskaya}}, \ and\ \bibinfo {author} {\bibnamefont {$et al.$}},\ }\href
  {\doibase https://doi.org/10.1038/s41586-023-06491-w} {\bibfield  {journal}
  {\bibinfo  {journal} {Nature}\ ,\ \bibinfo {pages} {471–475}} (\bibinfo
  {year} {2023})}\BibitemShut {NoStop}%
\bibitem [{\citenamefont {Bidkar}\ \emph {et~al.}(2024)\citenamefont {Bidkar},
  \citenamefont {Zerefa}, \citenamefont {Yadav}, \citenamefont {VanBrocklin},\
  and\ \citenamefont {Flavell}}]{NY1}%
  \BibitemOpen
  \bibfield  {author} {\bibinfo {author} {\bibfnamefont {A.~P.}\ \bibnamefont
  {Bidkar}}, \bibinfo {author} {\bibfnamefont {L.}~\bibnamefont {Zerefa}},
  \bibinfo {author} {\bibfnamefont {S.}~\bibnamefont {Yadav}}, \bibinfo
  {author} {\bibfnamefont {H.~F.}\ \bibnamefont {VanBrocklin}}, \ and\ \bibinfo
  {author} {\bibfnamefont {R.~R.}\ \bibnamefont {Flavell}},\ }\href {\doibase
  10.7150/thno.96403} {\bibfield  {journal} {\bibinfo  {journal}
  {Theranostics}\ }\textbf {\bibinfo {volume} {14}},\ \bibinfo {pages} {2969}
  (\bibinfo {year} {2024})}\BibitemShut {NoStop}%
\bibitem [{\citenamefont {Poty}\ \emph {et~al.}(2018)\citenamefont {Poty},
  \citenamefont {Francesconi}, \citenamefont {McDevitt}, \citenamefont
  {Morris},\ and\ \citenamefont {Lewis}}]{NY2}%
  \BibitemOpen
  \bibfield  {author} {\bibinfo {author} {\bibfnamefont {S.}~\bibnamefont
  {Poty}}, \bibinfo {author} {\bibfnamefont {L.~C.}\ \bibnamefont
  {Francesconi}}, \bibinfo {author} {\bibfnamefont {M.~R.}\ \bibnamefont
  {McDevitt}}, \bibinfo {author} {\bibfnamefont {M.~J.}\ \bibnamefont
  {Morris}}, \ and\ \bibinfo {author} {\bibfnamefont {J.~S.}\ \bibnamefont
  {Lewis}},\ }\href {\doibase https://doi.org/10.2967/jnumed.116.186338}
  {\bibfield  {journal} {\bibinfo  {journal} {Jour. Nucl. Medi.}\ }\textbf
  {\bibinfo {volume} {59}},\ \bibinfo {pages} {878} (\bibinfo {year}
  {2018})}\BibitemShut {NoStop}%
\bibitem [{\citenamefont {Tavares}\ \emph {et~al.}(2005)\citenamefont
  {Tavares}, \citenamefont {Medeiros},\ and\ \citenamefont
  {Terranova}}]{Tavares_2005}%
  \BibitemOpen
  \bibfield  {author} {\bibinfo {author} {\bibfnamefont {O.~A.~P.}\
  \bibnamefont {Tavares}}, \bibinfo {author} {\bibfnamefont {E.~L.}\
  \bibnamefont {Medeiros}}, \ and\ \bibinfo {author} {\bibfnamefont {M.~L.}\
  \bibnamefont {Terranova}},\ }\href {\doibase 10.1088/0954-3899/31/2/005}
  {\bibfield  {journal} {\bibinfo  {journal} {Jour. Phys. G: Nucl. and Part.
  Phys.}\ }\textbf {\bibinfo {volume} {31}},\ \bibinfo {pages} {129} (\bibinfo
  {year} {2005})}\BibitemShut {NoStop}%
\bibitem [{\citenamefont {Medeiros}\ \emph {et~al.}(2006)\citenamefont
  {Medeiros}, \citenamefont {Rodrigues}, \citenamefont {Duarte},\ and\
  \citenamefont {Tavares}}]{Medeiros_2006}%
  \BibitemOpen
  \bibfield  {author} {\bibinfo {author} {\bibfnamefont {E.~L.}\ \bibnamefont
  {Medeiros}}, \bibinfo {author} {\bibfnamefont {M.~M.~N.}\ \bibnamefont
  {Rodrigues}}, \bibinfo {author} {\bibfnamefont {S.~B.}\ \bibnamefont
  {Duarte}}, \ and\ \bibinfo {author} {\bibfnamefont {O.~A.~P.}\ \bibnamefont
  {Tavares}},\ }\href {\doibase 10.1088/0954-3899/32/8/B01} {\bibfield
  {journal} {\bibinfo  {journal} {Jour. Phys. G: Nucl. and Part. Phys.}\
  }\textbf {\bibinfo {volume} {32}},\ \bibinfo {pages} {B23} (\bibinfo {year}
  {2006})}\BibitemShut {NoStop}%
\bibitem [{\citenamefont {Tavares}\ and\ \citenamefont
  {Medeiros}(2012)}]{Tavares_2012}%
  \BibitemOpen
  \bibfield  {author} {\bibinfo {author} {\bibfnamefont {O.~A.~P.}\
  \bibnamefont {Tavares}}\ and\ \bibinfo {author} {\bibfnamefont {E.~L.}\
  \bibnamefont {Medeiros}},\ }\href {\doibase 10.1088/0031-8949/86/01/015201}
  {\bibfield  {journal} {\bibinfo  {journal} {Phys. Scr.}\ }\textbf {\bibinfo
  {volume} {86}},\ \bibinfo {pages} {015201} (\bibinfo {year}
  {2012})}\BibitemShut {NoStop}%
\bibitem [{\citenamefont {Zou}\ \emph {et~al.}(2021)\citenamefont {Zou},
  \citenamefont {Pan}, \citenamefont {Liu},\ and\ \citenamefont {$et\
  al.$}}]{Zou_2021}%
  \BibitemOpen
  \bibfield  {author} {\bibinfo {author} {\bibfnamefont {Y.~T.}\ \bibnamefont
  {Zou}}, \bibinfo {author} {\bibfnamefont {X.}~\bibnamefont {Pan}}, \bibinfo
  {author} {\bibfnamefont {H.~M.}\ \bibnamefont {Liu}}, \ and\ \bibinfo
  {author} {\bibnamefont {$et\ al.$}},\ }\href {\doibase
  10.1088/1402-4896/abf795} {\bibfield  {journal} {\bibinfo  {journal} {Phys.
  Scr.}\ }\textbf {\bibinfo {volume} {96}},\ \bibinfo {pages} {075301}
  (\bibinfo {year} {2021})}\BibitemShut {NoStop}%
\bibitem [{\citenamefont {Qi}\ \emph {et~al.}(2025)\citenamefont {Qi},
  \citenamefont {Wang}, \citenamefont {Cui}, \citenamefont {Wang},\ and\
  \citenamefont {Gu}}]{Qi_2025}%
  \BibitemOpen
  \bibfield  {author} {\bibinfo {author} {\bibfnamefont {L.~Q.}\ \bibnamefont
  {Qi}}, \bibinfo {author} {\bibfnamefont {H.~M.}\ \bibnamefont {Wang}},
  \bibinfo {author} {\bibfnamefont {J.~P.}\ \bibnamefont {Cui}}, \bibinfo
  {author} {\bibfnamefont {Y.~Z.}\ \bibnamefont {Wang}}, \ and\ \bibinfo
  {author} {\bibfnamefont {J.~Z.}\ \bibnamefont {Gu}},\ }\href {\doibase
  10.1088/1674-1137/addcc8} {\bibfield  {journal} {\bibinfo  {journal} {Chin.
  Phys. C}\ }\textbf {\bibinfo {volume} {49}},\ \bibinfo {pages} {094102}
  (\bibinfo {year} {2025})}\BibitemShut {NoStop}%
\bibitem [{\citenamefont {Dong}\ \emph {et~al.}(2010)\citenamefont {Dong},
  \citenamefont {Zuo}, \citenamefont {Gu}, \citenamefont {Wang},\ and\
  \citenamefont {Peng}}]{PhysRevC.81.064309}%
  \BibitemOpen
  \bibfield  {author} {\bibinfo {author} {\bibfnamefont {J.}~\bibnamefont
  {Dong}}, \bibinfo {author} {\bibfnamefont {W.}~\bibnamefont {Zuo}}, \bibinfo
  {author} {\bibfnamefont {J.}~\bibnamefont {Gu}}, \bibinfo {author}
  {\bibfnamefont {Y.}~\bibnamefont {Wang}}, \ and\ \bibinfo {author}
  {\bibfnamefont {B.}~\bibnamefont {Peng}},\ }\href {\doibase
  10.1103/PhysRevC.81.064309} {\bibfield  {journal} {\bibinfo  {journal} {Phys.
  Rev. C}\ }\textbf {\bibinfo {volume} {81}},\ \bibinfo {pages} {064309}
  (\bibinfo {year} {2010})}\BibitemShut {NoStop}%
\bibitem [{\citenamefont {Zhang}\ \emph {et~al.}(2011)\citenamefont {Zhang},
  \citenamefont {Royer},\ and\ \citenamefont {Li}}]{PhysRevC.84.027303}%
  \BibitemOpen
  \bibfield  {author} {\bibinfo {author} {\bibfnamefont {H.~F.}\ \bibnamefont
  {Zhang}}, \bibinfo {author} {\bibfnamefont {G.}~\bibnamefont {Royer}}, \ and\
  \bibinfo {author} {\bibfnamefont {J.~Q.}\ \bibnamefont {Li}},\ }\href
  {\doibase 10.1103/PhysRevC.84.027303} {\bibfield  {journal} {\bibinfo
  {journal} {Phys. Rev. C}\ }\textbf {\bibinfo {volume} {84}},\ \bibinfo
  {pages} {027303} (\bibinfo {year} {2011})}\BibitemShut {NoStop}%
\bibitem [{\citenamefont {Liu}\ \emph {et~al.}(2020)\citenamefont {Liu},
  \citenamefont {Zou}, \citenamefont {Pan}, \citenamefont {Bao},\ and\
  \citenamefont {Li}}]{Liu_2020}%
  \BibitemOpen
  \bibfield  {author} {\bibinfo {author} {\bibfnamefont {H.~M.}\ \bibnamefont
  {Liu}}, \bibinfo {author} {\bibfnamefont {Y.~T.}\ \bibnamefont {Zou}},
  \bibinfo {author} {\bibfnamefont {X.}~\bibnamefont {Pan}}, \bibinfo {author}
  {\bibfnamefont {X.~J.}\ \bibnamefont {Bao}}, \ and\ \bibinfo {author}
  {\bibfnamefont {X.~H.}\ \bibnamefont {Li}},\ }\href {\doibase
  10.1088/1674-1137/44/9/094106} {\bibfield  {journal} {\bibinfo  {journal}
  {Chin. Phys. C}\ }\textbf {\bibinfo {volume} {44}},\ \bibinfo {pages}
  {094106} (\bibinfo {year} {2020})}\BibitemShut {NoStop}%
\bibitem [{\citenamefont {Myers}\ and\ \citenamefont {\ifmmode \acute{S}\else
  \'{S}\fi{}wia\ifmmode~\mbox{\c{}}\else
  \c{}\fi{}tecki}(2000)}]{PhysRevC.62.044610}%
  \BibitemOpen
  \bibfield  {author} {\bibinfo {author} {\bibfnamefont {W.~D.}\ \bibnamefont
  {Myers}}\ and\ \bibinfo {author} {\bibfnamefont {W.~J.}\ \bibnamefont
  {\ifmmode \acute{S}\else \'{S}\fi{}wia\ifmmode~\mbox{\c{}}\else
  \c{}\fi{}tecki}},\ }\href {\doibase 10.1103/PhysRevC.62.044610} {\bibfield
  {journal} {\bibinfo  {journal} {Phys. Rev. C}\ }\textbf {\bibinfo {volume}
  {62}},\ \bibinfo {pages} {044610} (\bibinfo {year} {2000})}\BibitemShut
  {NoStop}%
\bibitem [{\citenamefont {Xu}\ and\ \citenamefont
  {Ren}(2004)}]{PhysRevC.69.024614}%
  \BibitemOpen
  \bibfield  {author} {\bibinfo {author} {\bibfnamefont {C.}~\bibnamefont
  {Xu}}\ and\ \bibinfo {author} {\bibfnamefont {Z.}~\bibnamefont {Ren}},\
  }\href {\doibase 10.1103/PhysRevC.69.024614} {\bibfield  {journal} {\bibinfo
  {journal} {Phys. Rev. C}\ }\textbf {\bibinfo {volume} {69}},\ \bibinfo
  {pages} {024614} (\bibinfo {year} {2004})}\BibitemShut {NoStop}%
\bibitem [{\citenamefont {Takigawa}\ \emph {et~al.}(2000)\citenamefont
  {Takigawa}, \citenamefont {Rumin},\ and\ \citenamefont
  {Ihara}}]{PhysRevC.61.044607}%
  \BibitemOpen
  \bibfield  {author} {\bibinfo {author} {\bibfnamefont {N.}~\bibnamefont
  {Takigawa}}, \bibinfo {author} {\bibfnamefont {T.}~\bibnamefont {Rumin}}, \
  and\ \bibinfo {author} {\bibfnamefont {N.}~\bibnamefont {Ihara}},\ }\href
  {\doibase 10.1103/PhysRevC.61.044607} {\bibfield  {journal} {\bibinfo
  {journal} {Phys. Rev. C}\ }\textbf {\bibinfo {volume} {61}},\ \bibinfo
  {pages} {044607} (\bibinfo {year} {2000})}\BibitemShut {NoStop}%
\bibitem [{\citenamefont {Gui}\ \emph {et~al.}(2022)\citenamefont {Gui},
  \citenamefont {Liu}, \citenamefont {Wu}, \citenamefont {Chu}, \citenamefont
  {He},\ and\ \citenamefont {Li}}]{Gui_2022}%
  \BibitemOpen
  \bibfield  {author} {\bibinfo {author} {\bibfnamefont {H.~F.}\ \bibnamefont
  {Gui}}, \bibinfo {author} {\bibfnamefont {H.~M.}\ \bibnamefont {Liu}},
  \bibinfo {author} {\bibfnamefont {X.~J.}\ \bibnamefont {Wu}}, \bibinfo
  {author} {\bibfnamefont {P.~C.}\ \bibnamefont {Chu}}, \bibinfo {author}
  {\bibfnamefont {B.}~\bibnamefont {He}}, \ and\ \bibinfo {author}
  {\bibfnamefont {X.~H.}\ \bibnamefont {Li}},\ }\href {\doibase
  10.1088/1572-9494/ac6576} {\bibfield  {journal} {\bibinfo  {journal} {Commun.
  Theor. Phys.}\ }\textbf {\bibinfo {volume} {74}},\ \bibinfo {pages} {055301}
  (\bibinfo {year} {2022})}\BibitemShut {NoStop}%
\bibitem [{\citenamefont {Deng}\ \emph {et~al.}(2018)\citenamefont {Deng},
  \citenamefont {Zhao}, \citenamefont {Chu},\ and\ \citenamefont
  {Li}}]{PhysRevC.97.044322}%
  \BibitemOpen
  \bibfield  {author} {\bibinfo {author} {\bibfnamefont {J.~G.}\ \bibnamefont
  {Deng}}, \bibinfo {author} {\bibfnamefont {J.~C.}\ \bibnamefont {Zhao}},
  \bibinfo {author} {\bibfnamefont {P.~C.}\ \bibnamefont {Chu}}, \ and\
  \bibinfo {author} {\bibfnamefont {X.~H.}\ \bibnamefont {Li}},\ }\href
  {\doibase 10.1103/PhysRevC.97.044322} {\bibfield  {journal} {\bibinfo
  {journal} {Phys. Rev. C}\ }\textbf {\bibinfo {volume} {97}},\ \bibinfo
  {pages} {044322} (\bibinfo {year} {2018})}\BibitemShut {NoStop}%
\bibitem [{\citenamefont {Wang}\ \emph {et~al.}(2021)\citenamefont {Wang},
  \citenamefont {Huang}, \citenamefont {Kondev}, \citenamefont {Audi},\ and\
  \citenamefont {Naimi}}]{Wang_2021}%
  \BibitemOpen
  \bibfield  {author} {\bibinfo {author} {\bibfnamefont {M.}~\bibnamefont
  {Wang}}, \bibinfo {author} {\bibfnamefont {W.}~\bibnamefont {Huang}},
  \bibinfo {author} {\bibfnamefont {F.}~\bibnamefont {Kondev}}, \bibinfo
  {author} {\bibfnamefont {G.}~\bibnamefont {Audi}}, \ and\ \bibinfo {author}
  {\bibfnamefont {S.}~\bibnamefont {Naimi}},\ }\href {\doibase
  10.1088/1674-1137/abddaf} {\bibfield  {journal} {\bibinfo  {journal} {Chin.
  Phys. C}\ }\textbf {\bibinfo {volume} {45}},\ \bibinfo {pages} {030003}
  (\bibinfo {year} {2021})}\BibitemShut {NoStop}%
\bibitem [{\citenamefont {Kondev}\ \emph {et~al.}(2021)\citenamefont {Kondev},
  \citenamefont {Wang}, \citenamefont {Huang}, \citenamefont {Naimi},\ and\
  \citenamefont {Audi}}]{Kondev_2021}%
  \BibitemOpen
  \bibfield  {author} {\bibinfo {author} {\bibfnamefont {F.}~\bibnamefont
  {Kondev}}, \bibinfo {author} {\bibfnamefont {M.}~\bibnamefont {Wang}},
  \bibinfo {author} {\bibfnamefont {W.}~\bibnamefont {Huang}}, \bibinfo
  {author} {\bibfnamefont {S.}~\bibnamefont {Naimi}}, \ and\ \bibinfo {author}
  {\bibfnamefont {G.}~\bibnamefont {Audi}},\ }\href {\doibase
  10.1088/1674-1137/abddae} {\bibfield  {journal} {\bibinfo  {journal} {Chin.
  Phys. C}\ }\textbf {\bibinfo {volume} {45}},\ \bibinfo {pages} {030001}
  (\bibinfo {year} {2021})}\BibitemShut {NoStop}%
\bibitem [{\citenamefont {M{\"o}ller}\ \emph {et~al.}(2016)\citenamefont
  {M{\"o}ller}, \citenamefont {Sierk}, \citenamefont {Ichikawa},\ and\
  \citenamefont {Sagawa}}]{MOLLER20161}%
  \BibitemOpen
  \bibfield  {author} {\bibinfo {author} {\bibfnamefont {P.}~\bibnamefont
  {M{\"o}ller}}, \bibinfo {author} {\bibfnamefont {A.~J.}\ \bibnamefont
  {Sierk}}, \bibinfo {author} {\bibfnamefont {T.}~\bibnamefont {Ichikawa}}, \
  and\ \bibinfo {author} {\bibfnamefont {H.}~\bibnamefont {Sagawa}},\ }\href
  {\doibase https://doi.org/10.1016/j.adt.2015.10.002} {\bibfield  {journal}
  {\bibinfo  {journal} {At. Data Nucl. Data Tables}\ }\textbf {\bibinfo
  {volume} {109-110}},\ \bibinfo {pages} {1} (\bibinfo {year}
  {2016})}\BibitemShut {NoStop}%
\bibitem [{\citenamefont {Zhang}\ \emph {et~al.}(2025)\citenamefont {Zhang},
  \citenamefont {Wang}, \citenamefont {Ma},\ and\ \citenamefont {$et\
  al.$}}]{210Pa}%
  \BibitemOpen
  \bibfield  {author} {\bibinfo {author} {\bibfnamefont {M.~M.}\ \bibnamefont
  {Zhang}}, \bibinfo {author} {\bibfnamefont {J.~G.}\ \bibnamefont {Wang}},
  \bibinfo {author} {\bibfnamefont {L.}~\bibnamefont {Ma}}, \ and\ \bibinfo
  {author} {\bibnamefont {$et\ al.$}},\ }\href {\doibase
  https://doi.org/10.1038/s41467-025-60047-2} {\bibfield  {journal} {\bibinfo
  {journal} {Nat. Commun.}\ }\textbf {\bibinfo {volume} {16}},\ \bibinfo
  {pages} {5003} (\bibinfo {year} {2025})}\BibitemShut {NoStop}%
\bibitem [{\citenamefont {Brabec}\ \emph {et~al.}(1996)\citenamefont {Brabec},
  \citenamefont {Ivanov},\ and\ \citenamefont {Corkum}}]{PhysRevA.54.R2551}%
  \BibitemOpen
  \bibfield  {author} {\bibinfo {author} {\bibfnamefont {T.}~\bibnamefont
  {Brabec}}, \bibinfo {author} {\bibfnamefont {M.~Y.}\ \bibnamefont {Ivanov}},
  \ and\ \bibinfo {author} {\bibfnamefont {P.~B.}\ \bibnamefont {Corkum}},\
  }\href {\doibase 10.1103/PhysRevA.54.R2551} {\bibfield  {journal} {\bibinfo
  {journal} {Phys. Rev. A}\ }\textbf {\bibinfo {volume} {54}},\ \bibinfo
  {pages} {R2551} (\bibinfo {year} {1996})}\BibitemShut {NoStop}%
\bibitem [{\citenamefont {Chen}\ \emph {et~al.}(2000)\citenamefont {Chen},
  \citenamefont {Liu}, \citenamefont {Fu},\ and\ \citenamefont
  {Zheng}}]{PhysRevA.63.011404}%
  \BibitemOpen
  \bibfield  {author} {\bibinfo {author} {\bibfnamefont {J.}~\bibnamefont
  {Chen}}, \bibinfo {author} {\bibfnamefont {J.}~\bibnamefont {Liu}}, \bibinfo
  {author} {\bibfnamefont {L.~B.}\ \bibnamefont {Fu}}, \ and\ \bibinfo {author}
  {\bibfnamefont {W.~M.}\ \bibnamefont {Zheng}},\ }\href {\doibase
  10.1103/PhysRevA.63.011404} {\bibfield  {journal} {\bibinfo  {journal} {Phys.
  Rev. A}\ }\textbf {\bibinfo {volume} {63}},\ \bibinfo {pages} {011404}
  (\bibinfo {year} {2000})}\BibitemShut {NoStop}%
\bibitem [{\citenamefont {Tanaka}\ \emph {et~al.}(2020)\citenamefont {Tanaka},
  \citenamefont {Spohr}, \citenamefont {Balabanski}, \citenamefont
  {Balascuta},\ and\ \citenamefont {$et\ al.$}}]{MREL123}%
  \BibitemOpen
  \bibfield  {author} {\bibinfo {author} {\bibfnamefont {K.~A.}\ \bibnamefont
  {Tanaka}}, \bibinfo {author} {\bibfnamefont {K.~M.}\ \bibnamefont {Spohr}},
  \bibinfo {author} {\bibfnamefont {D.~L.}\ \bibnamefont {Balabanski}},
  \bibinfo {author} {\bibfnamefont {S.}~\bibnamefont {Balascuta}}, \ and\
  \bibinfo {author} {\bibnamefont {$et\ al.$}},\ }\href {\doibase
  10.1063/1.5093535} {\bibfield  {journal} {\bibinfo  {journal} {Matter Radia.
  Extremes}\ }\textbf {\bibinfo {volume} {5}},\ \bibinfo {pages} {024402}
  (\bibinfo {year} {2020})}\BibitemShut {NoStop}%
\bibitem [{\citenamefont {Bucurescu}\ and\ \citenamefont
  {Zamfir}(2013)}]{PhysRevC.87.054324}%
  \BibitemOpen
  \bibfield  {author} {\bibinfo {author} {\bibfnamefont {D.}~\bibnamefont
  {Bucurescu}}\ and\ \bibinfo {author} {\bibfnamefont {N.~V.}\ \bibnamefont
  {Zamfir}},\ }\href {\doibase 10.1103/PhysRevC.87.054324} {\bibfield
  {journal} {\bibinfo  {journal} {Phys. Rev. C}\ }\textbf {\bibinfo {volume}
  {87}},\ \bibinfo {pages} {054324} (\bibinfo {year} {2013})}\BibitemShut
  {NoStop}%
\bibitem [{\citenamefont {Yahya}\ \emph {et~al.}(2025)\citenamefont {Yahya},
  \citenamefont {Majekodunmi}, \citenamefont {van~der Ventel}, \citenamefont
  {Mustapha},\ and\ \citenamefont {Mukeru}}]{PhysRevC.111.024322}%
  \BibitemOpen
  \bibfield  {author} {\bibinfo {author} {\bibfnamefont {W.~A.}\ \bibnamefont
  {Yahya}}, \bibinfo {author} {\bibfnamefont {J.~T.}\ \bibnamefont
  {Majekodunmi}}, \bibinfo {author} {\bibfnamefont {S.~I.~B.}\ \bibnamefont
  {van~der Ventel}}, \bibinfo {author} {\bibfnamefont {H.~A.}\ \bibnamefont
  {Mustapha}}, \ and\ \bibinfo {author} {\bibfnamefont {B.}~\bibnamefont
  {Mukeru}},\ }\href {\doibase 10.1103/PhysRevC.111.024322} {\bibfield
  {journal} {\bibinfo  {journal} {Phys. Rev. C}\ }\textbf {\bibinfo {volume}
  {111}},\ \bibinfo {pages} {024322} (\bibinfo {year} {2025})}\BibitemShut
  {NoStop}%
\end{thebibliography}%


\begin{thebibliography}{81}%
\makeatletter
\providecommand \@ifxundefined [1]{%
 \@ifx{#1\undefined}
}%
\providecommand \@ifnum [1]{%
 \ifnum #1\expandafter \@firstoftwo
 \else \expandafter \@secondoftwo
 \fi
}%
\providecommand \@ifx [1]{%
 \ifx #1\expandafter \@firstoftwo
 \else \expandafter \@secondoftwo
 \fi
}%
\providecommand \natexlab [1]{#1}%
\providecommand \enquote  [1]{``#1''}%
\providecommand \bibnamefont  [1]{#1}%
\providecommand \bibfnamefont [1]{#1}%
\providecommand \citenamefont [1]{#1}%
\providecommand \href@noop [0]{\@secondoftwo}%
\providecommand \href [0]{\begingroup \@sanitize@url \@href}%
\providecommand \@href[1]{\@@startlink{#1}\@@href}%
\providecommand \@@href[1]{\endgroup#1\@@endlink}%
\providecommand \@sanitize@url [0]{\catcode `\\12\catcode `\$12\catcode
  `\&12\catcode `\#12\catcode `\^12\catcode `\_12\catcode `\%12\relax}%
\providecommand \@@startlink[1]{}%
\providecommand \@@endlink[0]{}%
\providecommand \url  [0]{\begingroup\@sanitize@url \@url }%
\providecommand \@url [1]{\endgroup\@href {#1}{\urlprefix }}%
\providecommand \urlprefix  [0]{URL }%
\providecommand \Eprint [0]{\href }%
\providecommand \doibase [0]{http://dx.doi.org/}%
\providecommand \selectlanguage [0]{\@gobble}%
\providecommand \bibinfo  [0]{\@secondoftwo}%
\providecommand \bibfield  [0]{\@secondoftwo}%
\providecommand \translation [1]{[#1]}%
\providecommand \BibitemOpen [0]{}%
\providecommand \bibitemStop [0]{}%
\providecommand \bibitemNoStop [0]{.\EOS\space}%
\providecommand \EOS [0]{\spacefactor3000\relax}%
\providecommand \BibitemShut  [1]{\csname bibitem#1\endcsname}%
\let\auto@bib@innerbib\@empty

\end{thebibliography}

\end{document}